\documentclass[fleqn,usenatbib]{mnras}

\usepackage{mathptmx}

\usepackage[T1]{fontenc}

\DeclareRobustCommand{\VAN}[3]{#2}
\let\VANthebibliography\thebibliography
\def\thebibliography{\DeclareRobustCommand{\VAN}[3]{##3}\VANthebibliography}

\usepackage{graphicx}	
\usepackage{amsmath}	
\usepackage{amssymb}	
\usepackage{xspace}
\usepackage{longtable}
\usepackage{subfig}
\usepackage{booktabs}
\usepackage{lscape}
\usepackage{siunitx}
\usepackage[symbol]{footmisc}
\usepackage[dvipsnames]{xcolor}
\usepackage{subcaption}

\newcommand{\tess}{\ensuremath{\emph{TESS}}\xspace}

\title[Cold and under-dense: TOI-6478\,b]{TOI-6478\,b: a cold under-dense Neptune transiting a fully convective M dwarf from the thick disc}

\author[M. G. Scott et al.]{
Madison G. Scott$^{1}$$^{\href{https://orcid.org/0009-0006-3846-4558}{\includegraphics[scale=0.5]{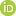}}}$, 
Amaury H.M.J. Triaud$^{1}$$^{\href{https://orcid.org/0000-0002-5510-8751}{\includegraphics[scale=0.5]{plots/orcid.jpg}}}$, 
Khalid Barkaoui$^{2,3,4}$$^{\href{https://orcid.org/0000-0003-1464-9276}{\includegraphics[scale=0.5]{plots/orcid.jpg}}}$,
Daniel Sebastian$^{1}$$^{\href{https://orcid.org/0000-0002-2214-9258}{\includegraphics[scale=0.5]{plots/orcid.jpg}}}$,
\newauthor
Adam J. Burgasser$^{5}$$^{\href{https://orcid.org/0000-0002-6523-9536}{\includegraphics[scale=0.5]{plots/orcid.jpg}}}$, 
Karen A. Collins$^{6}$$^{\href{https://orcid.org/0000-0001-6588-9574}{\includegraphics[scale=0.5]{plots/orcid.jpg}}}$,
Georgina Dransfield$^{7,8,1}$$^{\href{https://orcid.org/0000-0002-3937-630X}{\includegraphics[scale=0.5]{plots/orcid.jpg}}}$,
Coel Hellier$^{9}$$^{\href{https://orcid.org/0000-0002-3439-1439}{\includegraphics[scale=0.5]{plots/orcid.jpg}}}$,
Steve B. Howell$^{10}$$^{\href{https://orcid.org/0000-0002-2532-2853}{\includegraphics[scale=0.5]{plots/orcid.jpg}}}$,
\newauthor
Anjali A. A. Piette$^{1}$$^{\href{https://orcid.org/0000-0002-4487-5533}{\includegraphics[scale=0.5]{plots/orcid.jpg}}}$,
Benjamin V. Rackham$^{2}$$^{\href{https://orcid.org/0000-0002-3627-1676}{\includegraphics[scale=0.5]{plots/orcid.jpg}}}$,
Keivan G. Stassun$^{11}$$^{\href{https://orcid.org/0000-0002-3481-9052}{\includegraphics[scale=0.5]{plots/orcid.jpg}}}$,
Amalie Stokholm$^{1}$$^{\href{https://orcid.org/0000-0002-5496-365X}{\includegraphics[scale=0.5]{plots/orcid.jpg}}}$,
\newauthor
Mathilde Timmermans$^{1,2}$$^{\href{https://orcid.org/0009-0008-2214-5039}{\includegraphics[scale=0.5]{plots/orcid.jpg}}}$,
Cristilyn N. Watkins$^{6}$$^{\href{https://orcid.org/0000-0001-8621-6731}{\includegraphics[scale=0.5]{plots/orcid.jpg}}}$,
Michael Fausnaugh$^{12}$$^{\href{https://orcid.org/0000-0002-9113-7162}{\includegraphics[scale=0.5]{plots/orcid.jpg}}}$,
Akihiko Fukui$^{13,3}$$^{\href{https://orcid.org/0000-0002-4909-5763}{\includegraphics[scale=0.5]{plots/orcid.jpg}}}$,
\newauthor
Jon M. Jenkins$^{10}$$^{\href{https://orcid.org/0000-0002-4715-9460}{\includegraphics[scale=0.5]{plots/orcid.jpg}}}$,
Norio Narita$^{13,14,4}$$^{\href{https://orcid.org/0000-0001-8511-2981}{\includegraphics[scale=0.5]{plots/orcid.jpg}}}$,
George Ricker$^{15}$$^{\href{https://orcid.org/0000-0003-2058-6662}{\includegraphics[scale=0.5]{plots/orcid.jpg}}}$,
Emma Softich$^{5}$$^{\href{https://orcid.org/0000-0002-1420-1837}{\includegraphics[scale=0.5]{plots/orcid.jpg}}}$,
Richard P. Schwarz$^{6}$$^{\href{https://orcid.org/0000-0001-8227-1020}{\includegraphics[scale=0.5]{plots/orcid.jpg}}}$,
\newauthor
Sara Seager$^{15,16,17}$$^{\href{https://orcid.org/0000-0002-6892-6948}{\includegraphics[scale=0.5]{plots/orcid.jpg}}}$,
Avi Shporer$^{18}$$^{\href{https://orcid.org/0000-0002-1836-3120}{\includegraphics[scale=0.5]{plots/orcid.jpg}}}$,
Christopher Theissen$^{5}$$^{\href{https://orcid.org/0000-0002-9807-5435}{\includegraphics[scale=0.5]{plots/orcid.jpg}}}$,  
Joseph D. Twicken$^{19}$$^{\href{https://orcid.org/0000-0002-6778-7552}{\includegraphics[scale=0.5]{plots/orcid.jpg}}}$,
Joshua N. Winn$^{20}$,
\newauthor
David Watanabe$^{21}$$^{\href{https://orcid.org/0000-0002-3555-8464}{\includegraphics[scale=0.5]{plots/orcid.jpg}}}$
\\
$^{1}$School of Physics \& Astronomy, University of Birmingham, Edgbaston, Birmingham B15 2TT, United Kingdom\\
$^{2}$Astrobiology Research Unit, University of Li\`ege, All\'ee du 6 ao\^ut, 19, 4000 Li\`ege (Sart-Tilman), Belgium\\
$^{3}$Department of Earth, Atmospheric and Planetary Sciences, MIT, 77 Massachusetts Avenue, Cambridge, MA 02139, USA\\
$^{4}$Instituto de Astrof\'isica de Canarias (IAC), Calle V\'ia L\'actea s/n, 38200, La Laguna, Tenerife, Spain\\
$^{5}$Department of Astronomy \& Astrophysics, UC San Diego, 9500 Gilman Drive, La Jolla, CA 92093, USA \\
$^{6}$Center for Astrophysics \textbar \ Harvard \& Smithsonian, 60 Garden St, Cambridge, MA 02138, USA\\
$^{7}$Department of Astrophysics, University of Oxford, Denys Wilkinson Building, Keble Road, Oxford OX1 3RH, UK\\
$^{8}$Magdalen College, University of Oxford, Oxford OX1 4AU, UK\\
$^{9}$Astrophysics Group, Keele University, Staffordshire, ST5 5BG, UK\\
$^{10}$NASA Ames Research Center, Moffett Field, CA 94035, USA\\
$^{11}$Department of Physics \& Astronomy, Vanderbilt University, 6301 Stevenson Center Ln., Nashville, TN 37235, USA \\
$^{12}$Department of Physics \& Astronomy, Texas Tech University, Lubbock TX, 79409-1051, USA\\
$^{13}$Komaba Institute for Science, The University of Tokyo, 3-8-1 Komaba, Meguro, Tokyo 153-8902, Japan\\
$^{14}$Astrobiology Center, 2-21-1 Osawa, Mitaka, Tokyo 181-8588, Japan\\
$^{15}$Kavli Institute for Astrophysics and Space Research, Massachusetts Institute of Technology, Cambridge, MA 02139, USA\\
$^{16}$Departamento de Astrofísica, Universidad de La Laguna (ULL), 38206 La Laguna, Tenerife, Spain\\
$^{17}$Department of Astronomy, University of Maryland, College Park, MD 20742, USA\\
$^{18}$Department of Physics and Kavli Institute for Astrophysics and Space Research, Massachusetts Institute of Technology, Cambridge, MA 02139, USA\\
$^{19}$SETI Institute, Mountain View, CA 94043 USA/NASA Ames Research Center, Moffett Field, CA 94035 USA\\
$^{20}$Department of Astrophysical Sciences, Princeton University, Princeton, NJ 08544, USA\\
$^{21}$Planetary Discoveries, Valencia CA 91354, USA\\
}

\date{Accepted XXX. Received YYY; in original form ZZZ}

\pubyear{2024}

\begin{document}
\label{firstpage}
\pagerange{\pageref{firstpage}--\pageref{lastpage}}
\maketitle

\begin{abstract}
Growing numbers of exoplanet detections continue to reveal the diverse nature of planetary systems.
Planet formation around late-type M dwarfs is of particular interest. These systems provide practical laboratories to measure exoplanet occurrence rates for M dwarfs, thus testing how the outcomes of planet formation scale with host mass, and how they compare to Sun-like stars. Here, we report the discovery of TOI-6478\,b, a cold ($T_{\text{eq}}=204\,$K) Neptune-like planet orbiting an M5 star ($R_\star=0.234\pm0.012\,\text{R}_\odot$, $M_\star=0.230\pm0.007\,\text{M}_\odot$, $T_{\text{eff}}=3230\pm75\,$K) which is a member of the Milky Way's thick disc. We measure a planet radius of $R_b=4.6\pm0.24\,\text{R}_\oplus$ on a \textcolor{black}{$P_b=34.005019\pm0.000025\,$}d orbit. Using radial velocities, we calculate an upper mass limit of \textcolor{black}{$M_b\leq9.9\,\text{M}_\oplus$ ($M_b\leq0.6\,\text{M}_{\text{Nep}})$}, with $3\,\sigma$ confidence. TOI-6478\,b is a milestone planet in the study of cold, Neptune-like worlds. Thanks to its large atmospheric scale height, it is amenable to atmospheric characterisation with facilities such as JWST, and will provide an excellent probe of atmospheric chemistry in this cold regime. It is one of very few transiting exoplanets that orbit beyond their system's ice-line whose atmospheric chemical composition can be measured. Based on our current understanding of this planet, we estimate TOI-6478\,b's spectroscopic features (in transmission) can be $\sim2.5\times$ as high as the widely studied planet K2-18\,b. 
\end{abstract}




\section{Introduction}

Over 75\% of stars in our galaxy are classified as M dwarfs \citep{chabrier2003, henry2006, reyle2021}. Their slow evolution, complex spectra and low luminosities make accurate stellar characterisation tricky compared to solar-type stars; however, efforts by studies such as the EBLM project \citep[Eclipsing Binaries--Low Mass;][]{eblm2013, maxted2023} aim to populate the mass--radius parameter for these stars. More specifically, they investigate the low-mass end of this parameter space, i.e., fully convective M dwarfs \citep[$\leq0.35\,\text{M}_{\odot}$;][]{2010A&ARv..18...67T, 2018ApJS..237...21M}, where stellar evolution models likely under-estimate stellar radii \citep[see e.g.,][]{2008MNRAS.389..585C, 2013ApJ...776...87S, 2018AJ....155..225K, duck2023, swayne2024, davis2024}. Improving mass-radius relations for low-mass stars is crucial because planets discovered transiting these stars provide astronomers with a particularly good opportunity to make detailed atmospheric characterisations, for instance with the James Webb Space Telescope \citep[JWST;][]{jwstERS} \citep[see e.g.][]{benneke2019, madhu2020, kempton2023}. Planets transiting M dwarfs produce comparatively larger transit depths than similar planets transiting Sun-like stars. Similarly, transmission spectroscopy is a modulation of the transit depth by the transiting planet's atmosphere, and this modulation is also larger when the star is smaller, for a fixed planet size. Smaller stellar radii makes it also easier to perform follow-up observations using ground based facilities \citep[e.g.][]{Brown:2013, speculoos, triaud2021_smallstars}.

To date, the Transiting Exoplanet Survey Satellite \citep[\textit{TESS};][]{tess} has identified over 7000 exoplanet candidates. Of the \textcolor{black}{600} confirmed\footnote{NASA Exoplanet Archive, \textcolor{black}{January 2025} \url{https://exoplanetarchive.ipac.caltech.edu/}}, there are less than 30 Neptune-sized planets that orbit M dwarfs, and less that 10 orbiting late-type M dwarfs ($T_{\mathrm{eff}}\,\leq\,3300\mathrm{K}$).

These planets present a unique opportunity to probe Neptune-like planet formation around stars much different to the Sun. 

TOI-6478\,b is a cold, seemingly under-dense \textcolor{black}{($\leq0.56\,\text{g}\,\text{cm}^{-3}$)}, Neptune-sized planet orbiting a cool late-type, fully convective M dwarf. Similar to exoplanets such as TOI-620\,b \citep{2022AJ....163..269R}, this exoplanet appears to be low-density, indicative of a high atmospheric mass fraction.

A sub-category of these low-density exoplanets, namely "super-puffs", are defined to be those which have densities $\leq0.3\,\text{g}\,\text{cm}^{-3}$ \citep[e.g.][]{2011ApJS..197....7C, 2019arXiv191107355S}. It has been speculated that some super-puff planets could be explained as ringed exoplanets \citep{2020AJ....159..131P} or planets with hazy atmospheres \citep{gao&zhang2020}, which can explain their apparent inflated radii.

Under-dense exoplanets provide an exciting opportunity for observations with JWST because their large atmospheric scale heights enhance the detection of transmission spectroscopy features. While this is desirable for all kinds of exoplanets, it is especially compelling for planets that lie in the poorly populated region of parameter space of cold ($<250\,$K), short-period \textcolor{black}{($<50\,$d}) planets. More specifically, Neptune-sized planets within this temperature range likely have cool H$_2$/He-dominated atmospheres, more akin to the giant planets of our Solar System than most exoplanets discovered thus far \citep[e.g.][]{2010A&A...523A..26N,2010ARA&A..48..631S}. They therefore represent an important link between the exoplanet population and the planets in our Solar System. Cool atmospheres of planets beyond the ice-line are relatively under-studied because most orbit Sun-like stars and consequently have long orbital periods ($>100\,$d), making transmission spectroscopy challenging and impractical. Our paper highlights the remarkable opportunity that TOI-6478\,b presents for atmospheric and planet formation studies.

Another interesting property of the TOI-6478 system is in being part of the thick disc, which implies an age older than the Solar system. A recent population study reveals hints that under-dense planets appear more likely in older systems \textcolor{black}{\citep{weeks2024}}, a trend TOI-6478\,b appears to follow. 
\color{black}

This paper is organised as follows. In Section~\ref{sec:star}, we describe the observations and methods used to characterise the host star. Section~\ref{sec:tess} describes the conclusions we are able to draw from the available \textit{TESS} data, and Section~\ref{sec:followup} presents the photometric and radial-velocity ground-based follow-up observations used for the validation of TOI-6478\,b's planetary nature. Section~\ref{sec:analysis} describes the joint photometric and radial-velocity fitting, followed by a discussion and conclusion of these results in Section~\ref{sec:discussion}.

\section{Stellar Characterisation}
\label{sec:star}
TOI-6478 is a nearby \citep[$\rm 38\,pc$;][]{BJdist} M dwarf of spectral type M5, with an M3 co-moving companion. The planetary information will be derived using the host star's parameters, therefore we describe the characterisation of TOI-6478 in the following sections. All photometry and stellar parameters adopted for this work can be found in Table \ref{tab:starpar}.

\begin{table}
\centering
\caption{Stellar parameters adopted for this work.}
\begin{tabular}{@{}lp{25mm}p{30mm}@{}}
\toprule
{\bf Designations} & \multicolumn{2}{p{65mm}}{TOI-6478, TIC 332657786, 2MASS  J09595797-1609323, Gaia DR2 5673934548598976256, UCAC4 370-057370, WISE J095958.01-160934.4, LP 789-76B} \\ \midrule
{\bf Parameter} & {\bf Value}              & {\bf Source} \\ \midrule
T mag           & 13.0041$\pm$0.0082        & \cite{TICv8} \\
V mag           & 15.99$\pm$0.2        & \cite{TICv8} \\
G mag           & 14.313$\pm$0.00069      & \cite{gaiaDR3cat} \\
J mag           & 11.389$\pm$0.024          & \cite{2masscat} \\
H mag           & 10.91$\pm$0.022          & \cite{2masscat} \\
K mag           & 10.657$\pm$0.021         & \cite{2masscat} \\
W1 mag           & 10.488$\pm$0.022          & \cite{wisecat} \\
W2 mag           & 10.261$\pm$0.021          & \cite{wisecat} \\
W3 mag           & 10.14 $\pm$0.061            & \cite{wisecat} \\
W4 mag           & 9.105 $\pm$0.525       & \cite{wisecat} \\
Distance         & 38.61$\pm$0.1\,pc       & \cite{TICv8} \\
$\alpha$           & 09:59:58.03      & \cite{gaiaDR3cat} \\
$\delta$           & \text{-}16:09:35.26       & \cite{gaiaDR3cat} \\
$\mu_{\alpha}$           & $\rm 52.4\,mas\,yr^{-1}$      & \cite{gaiaDR3cat} \\
$\mu_{\delta}$           & $\rm \text{-}172.3\,mas\,yr^{-1}$      & \cite{gaiaDR3cat} \\
SpT (optical)             & M5  & This work \\
SpT (NIR)             & M4.0$\pm$0.5  & This work \\
$R_{\star}$     & 0.234$\pm$0.012 R$_{\odot}$ & This work \\
$M_{\star}$     & 0.230$\pm$ 0.007 M$_{\odot}$   & This work                  \\
${\rm T_{eff}}$ & 3230$\pm$75 K             & This work   \\
                 & 3193$\pm$1 K             & \cite{gaiaDR3cat}   \\
$\log g_\star$           & 5.016$\pm$0.005           & \cite{gaiaDR3cat}                  \\
$\rm [Fe/H]$    & $-$0.18$\pm$0.20 dex              & This work (opt. spectroscopy)         \\    
$\rm [Fe/H]$    & $-$0.53$\pm$0.13 dex              & This work (NIR spectroscopy)         \\    
                & $-$0.195$\pm$0.003 dex              & \cite{gaiaDR3cat}         \\    
 Age   & $\gtrsim$6--7~Gyr             & This work (spectroscopy)                \\ 
         
\bottomrule
\end{tabular}
\label{tab:starpar}
\end{table}

\subsection{Reconnaissance Spectroscopy}
\label{sec:recspec}

\subsubsection{Shane/Kast}
\label{sec:kast}

TOI-6478 and its brighter co-moving companion LP~789-76\,A (aka 2MASS J09595660$-$1609206, aka TIC 332657787, aka Gaia DR3 5673934617318453248; $G$ = 12.68, 23$\arcsec$ separation) were observed simultaneously with the Shane 3m Kast double spectrograph \citep{kastspectrograph} on 2024 April 9 (UT). Conditions were clear and windy with 1$\arcsec$ seeing, and we used the 1$\farcs$5 (3.5~pixel) slit rotated to include both sources. We used the 600/4310 grism on the blue channel to acquire 3750--5600~{\AA} spectra at a resolution of $\lambda/\Delta\lambda$ $\approx$ 1200, and the 600/7500 grating on the red channel to acquire 5800--9000~{\AA} spectra at a resolution of $\lambda/\Delta\lambda$ $\approx$ 1500. Total integrations of 1000~s were acquired in both channels, split into two exposures of 500~s each in the red, at an average airmass of 1.70. We observed the nearby G2~V star 43 Hya ($V$ = 7.62) immediately afterward at a similar airmass for telluric absorption correction, and the flux standard Feige~34 ($V$ = 11.2; \citealt{1990ApJ...358..344M,1990AJ.....99.1621O}) later in the night. Arclamps (HeHgCd in the blue, HeNeAr in the red), quartz flat field lamps, and bias frames were obtained at the start of the night for wavelength and pixel response calibration.

\begin{figure*}
    \centering
    \includegraphics[width=0.8\textwidth]{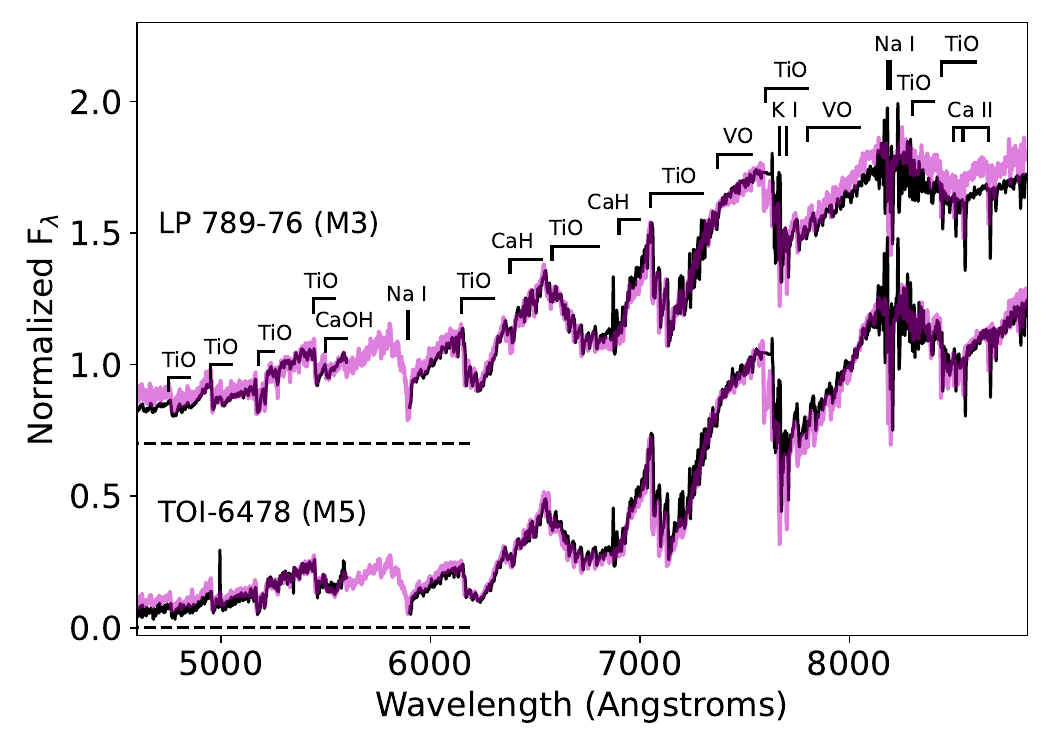}
    \caption{Kast spectra (black lines) of the co-moving companions LP~789-76 (top) and TOI-6478 (bottom) compared to best-fit M3 and M5 spectral templates from \citet[magenta lines]{2007AJ....133..531B}. All spectra are normalized at 7500~{\AA}, with the blue and red orders of Kast relatively scaled to match the spectral standard, and the spectrum of LP~789-76 is offset by 0.7 flux units for clarity (zeropoints are indicated by dashed lines). Key atomic and molecular spectral features are labeled.}
    \label{fig:kast}
\end{figure*}

Data were reduced using the {\tt kastredux} package\footnote{\url{https://github.com/aburgasser/kastredux}} following standard approaches for optical spectroscopic data reduction. In brief, we used the flat field lamp exposures to trace the illuminated orders and measure pixel response, the lamp exposures and bias frames to define a mask array, and the arc lamps to determine the wavelength scale with a precision of 0.19~{\AA} (13~km/s) in the blue and 0.23~{\AA} (10~km/s) in the red. Source spectra were traced along the tilted dispersion axes and extracted using a boxcar profile with a linear fit to the background. The red data were also corrected for cosmic ray contamination using an outlier rejection algorithm between the two exposures. Spectral flux density calibration was determined from band measurements of Feige~34 from \citet{1990ApJ...358..344M}, with a second-order correction made using the G2~V telluric calibrator and an empirical template from \citet{1998PASP..110..863P}. The G2~V spectrum was also used to correct for telluric absorption in the science targets. Our final data, shown in Figure~\ref{fig:kast}, have signal-to-noise (S/N) of 
22 at 5400~{\AA} and 121 at 7500~{\AA} for TOI-6478 and
96 at 5400~{\AA} and 281 at 7500~{\AA} for LP~789-76.

We used tools in {\tt kastredux} to characterize both spectra, which display the characteristic molecular bands and red optical slope of early-to-mid M dwarfs.
Comparison to SDSS spectral templates from \citet{2007AJ....133..531B} yield classifications of M5 for TOI~6478 and M3 for LP~789-76, which are 
consistent with index-based classifications based on the system of \citet{2003AJ....125.1598L}.
We see no evidence of H$\alpha$ or H$\beta$ emission indicative of magnetic activity in either source.
Indeed, TOI~6478 shows H$\alpha$ in absorption with an equivalent width EW = 0.37$\pm$0.07~{\AA}, suggesting a system activity age $\gtrsim$6--7~Gyr \citep{2008AJ....135..785W,2023MNRAS.526.4787R}. 
Such an old age is consistent with the kinematics of the system. The {\it Gaia\/} Catalog of Nearby Stars (GCN; \citealt{2021AA...649A...6G}) reports $[U,V,W]$ = [+2,$-$99,+36] for LP~789-76, giving the system a 99\% probability of being part of the Galactic thick disk based on the results of \citet{2003AA...410..527B}.

Analysis of the metallicity-sensitive features of TiO and CaH around 6800--7100~{\AA} yield discrepant $\zeta$ values of 0.85$\pm$0.01 and 1.01$\pm$0.01 for TOI~6478 and LP~789-76, respectively \citet{2013AJ....145..102L}. The former is close to the dwarf/subdwarf boundary defined by \citet{2007ApJ...669.1235L} suggesting that the system may be slightly metal-poor, with an estimate of [Fe/H] = $-$0.18$\pm$0.20 based on \citet{Mann2013}.
The latter is consistent with solar metallicity.
Both sources have atmosphere parameters reported in {\it Gaia\/} DR3, with metallicities that are consistent with our measurements and also discrepant with each other: 
[Fe/H] = $-$0.195$\pm$0.003 for TOI~6478 and 
$-$0.026$\pm$0.011 for LP~789-76. 
{\it Gaia\/} also reports effective temperatures 
of $T_\textup{eff}$ = 3193~K and 3428~K (with unrealistic precisions of 1~K),
respectively. The latter is consistent with 

the $T_{\rm eff}$ = 3456$\pm$115~K estimate of 
\citet{2023AJ....165..267H} using {\it Gaia\/} photometry and the empirical relationship of \citet{2013ApJS..208....9P}.

\subsubsection{IRTF/SpeX}
\label{sec:spex}

We gathered medium-resolution near-infrared spectra of TOI-6478 and its comoving companion LP 789-76 on \textcolor{black}{2024 May 10} (UT) using the SpeX spectrograph \citep{Rayner2003} on the 3.2-m NASA Infrared Telescope Facility (IRTF).
Conditions were clear, and seeing was 0$\farcs$7.
We used the short-wavelength cross-dispersed (SXD) mode with the $0\farcs3 \times 15''$ slit aligned to the parallactic angle.
This setup provides 0.80--2.42\,$\mu$m spectra at a resolving power of $R{\sim}2000$ with 2.5\,pixels per resolution element.
While nodding in an ABBA pattern, we collected six, 300-s integrations on TOI-6478 and six, 120-s integrations on LP 789-76.
We gathered a set of standard SXD flat-field and arc-lamp calibrations after each target, followed by six, 20-s integrations of the A0\,V standard HD\,86593.
We reduced the data with Spextool v4.1 \citep{Cushing2004}, following the approach of previous analyses \citep{Delrez2022, Ghachoui2023, Barkaoui2023}.
The final spectra of TOI-6478 and LP 789-76 have median per-pixel signal-to-noise ratios of 111 and 141, respectively.

\begin{figure}
    \centering
    \includegraphics[width=\linewidth]{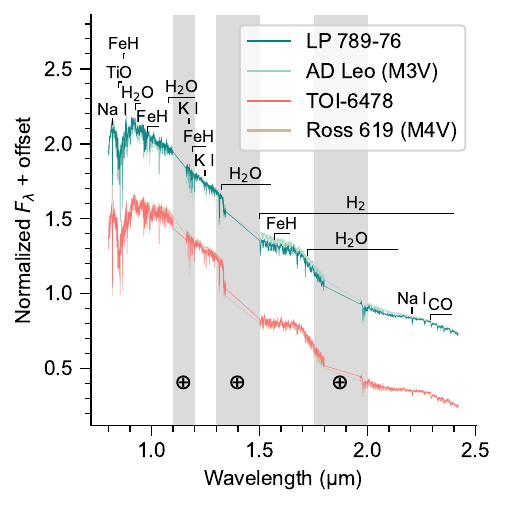}
    \caption{
    SpeX SXD spectra of TOI-6478 (coral) and LP 789-76 (teal) alongside their best-matching spectral standards.
    Prominent atomic and molecular spectral features of M dwarfs are highlighted.
    }
    \label{fig:spex}
\end{figure}

The SpeX SXD spectra of TOI-6478 and LP 789-76 are shown in Fig.\,\ref{fig:spex}.
As in previous SpeX analyses \citep[e.g.,][]{Triaud2023, Gillon2024, Timmermans2024}, we used the SpeX Prism Library Analysis Toolkit \citep[SPLAT, ][]{splat} to assign spectral types and estimate stellar 
metallicities.
Comparing the spectra to standard spectra of single stars in the IRTF Spectral Library \citep{Cushing2005, Rayner2009}, we find close matches to Ross 619 (M4V) and AD Leo (M3V) for TOI-6478 and LP 789-76, respectively.
We therefore adopt infrared spectral types of M4.0 $\pm$ 0.5 for TOI-6478 and M3.0 $\pm$ 0.5 for LP 789-76.
These are slightly earlier than and consistent with our optical spectral types, respectively.
Using the H2O--K2 index \citep{Rojas-Ayala2012} and \citet{Mann2013} relation, we estimate stellar iron abundances of $\mathrm{[Fe/H]} = -0.53 \pm 0.13$ for TOI-6478 and $\mathrm{[Fe/H]} = -0.48 \pm 0.12$ for LP 789-76, indicating a consistent, sub-solar metallicity for both stars.
For TOI-6478, this estimate is lower than our estimate based on the optical spectra but consistent with it at $1.5\sigma$.

\subsection{Spectral Energy Distribution}
\label{sec:sed}

\begin{figure}
    \centering
    \includegraphics[width=\linewidth,trim=80 75 50 50,clip]{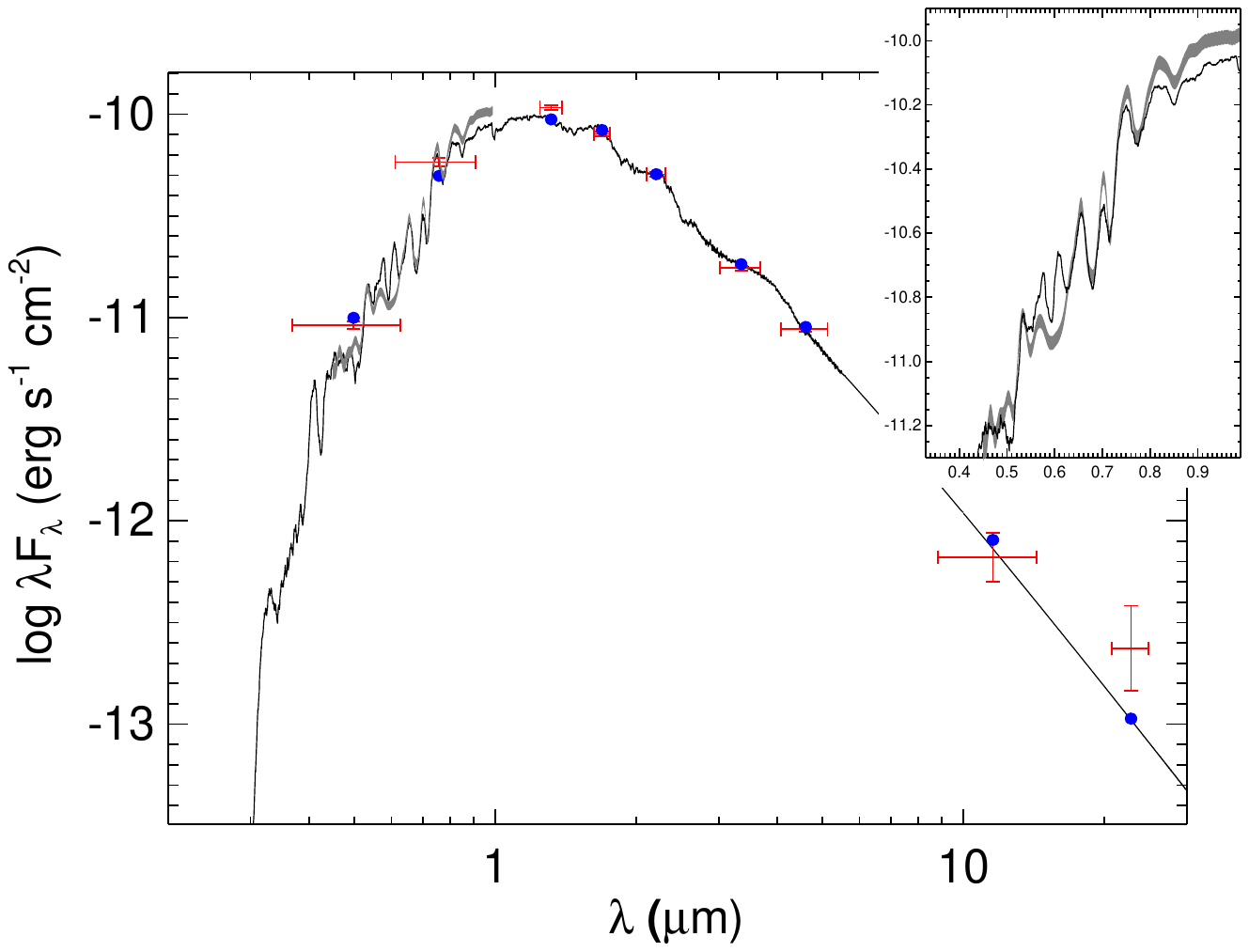}
\caption{\textcolor{black}{Spectral energy distribution of TOI-6478. Red symbols represent the observed photometric measurements, where the horizontal bars represent the effective width of the passband. Blue symbols are the model fluxes from the best-fit PHOENIX atmosphere model (black). The {\it Gaia\/} absolute flux-calibrated spectrophotometry is overlaid as the grey swathe (also shown in detail in the inset plot). }\label{fig:sed}}
\end{figure}

As an independent determination of the basic stellar parameters, we performed an analysis of the broadband spectral energy distribution (SED) of the star together with the {\it Gaia\/} DR3 parallax \citep[with no systematic offset applied; see, e.g.,][]{StassunTorres:2021}, in order to determine an empirical measurement of the stellar radius, following the procedures described in \citet{Stassun:2016,Stassun:2017,Stassun:2018}. We pulled the the $JHK_S$ magnitudes from {\it 2MASS}, the W1--W4 magnitudes from {\it WISE}, the $G_{\rm BP} G_{\rm RP}$ magnitudes from {\it Gaia}, as well as the {\it Gaia\/} spectrophotometry. Together, the available photometry spans the full stellar SED over the wavelength range 0.4--20~$\mu$m (see Figure~\ref{fig:sed}).  

We performed a fit using PHOENIX stellar atmosphere models \citep{Husser:2013}, fitting for the effective temperature ($T_{\rm eff}$) while adopting the metallicity ([Fe/H]) from the spectroscopic analysis. We fixed the extinction $A_V \equiv 0$ due to the close proximity of the system. The resulting fit (Figure~\ref{fig:sed}) has a best-fit $T_{\rm eff} = 3230 \pm 75$~K, with a reduced $\chi^2$ of 3.4 and consistent with the {\it Gaia\/} estimate. Integrating the model SED gives the bolometric flux at Earth, $F_{\rm bol} = 1.151 \pm 0.040 \times 10^{-10}$ erg~s$^{-1}$~cm$^{-2}$. Taking the $F_{\rm bol}$ and the {\it Gaia\/} parallax directly gives the bolometric luminosity, $L_{\rm bol} = 0.00534 \pm 0.00019$~L$_\odot$. The stellar radius then follows from the Stefan-Boltzmann relation, $R_\star = 0.234 \pm 0.012$~R$_\odot$. In addition, we can estimate the stellar mass from the empirical $M_K$ relations of \citet{Mann:2016}, giving $M_\star = 0.230 \pm 0.007$~M$_\odot$. 
These values are consistent with $T_{\rm eff}$, $L_{\rm bol}$, and $R_\star$ estimates from \citet{2023AJ....165..267H} based on {\it Gaia\/} data alone, while our mass is nearly 3$\sigma$ higher than the 0.203$\pm$0.006~M$_\odot$ reported in that study.

\subsection{Galactic orbit of TOI-6478} \label{sec:galactic_orbit}

The orbit of TOI-6478 in the Milky Way galaxy was computed from the 5D astrometric information and line-of-sight velocity from {\it Gaia\/} DR3 \citep{gaia2016, gaiadr3}.
We use the \textsc{Python} package \texttt{galpy} \citep{galpy} for this computation of the Galactic orbital properties. As a description of the Milky Way potential, we use the axisymmetric gravitational potential \texttt{McMillan2017}  \citep{mcmillan2017}. As for the Galactic location and velocity of the Sun, we assume  $(X_{\odot},Y_{\odot},Z_{\odot})=(8.2,0,0.0208)$~\si{\kilo pc} and  $(U_{\odot},V_{\odot},W_{\odot})=(11.1,12.24,7.25)$~\si{\kilo\metre\per\second} with a circular velocity of \SI{240}{\kilo\metre\per\second} \citep{schonrich2010,bovy2015,bennett2019,gravity2019}.

\begin{figure*}
    \centering
    \includegraphics{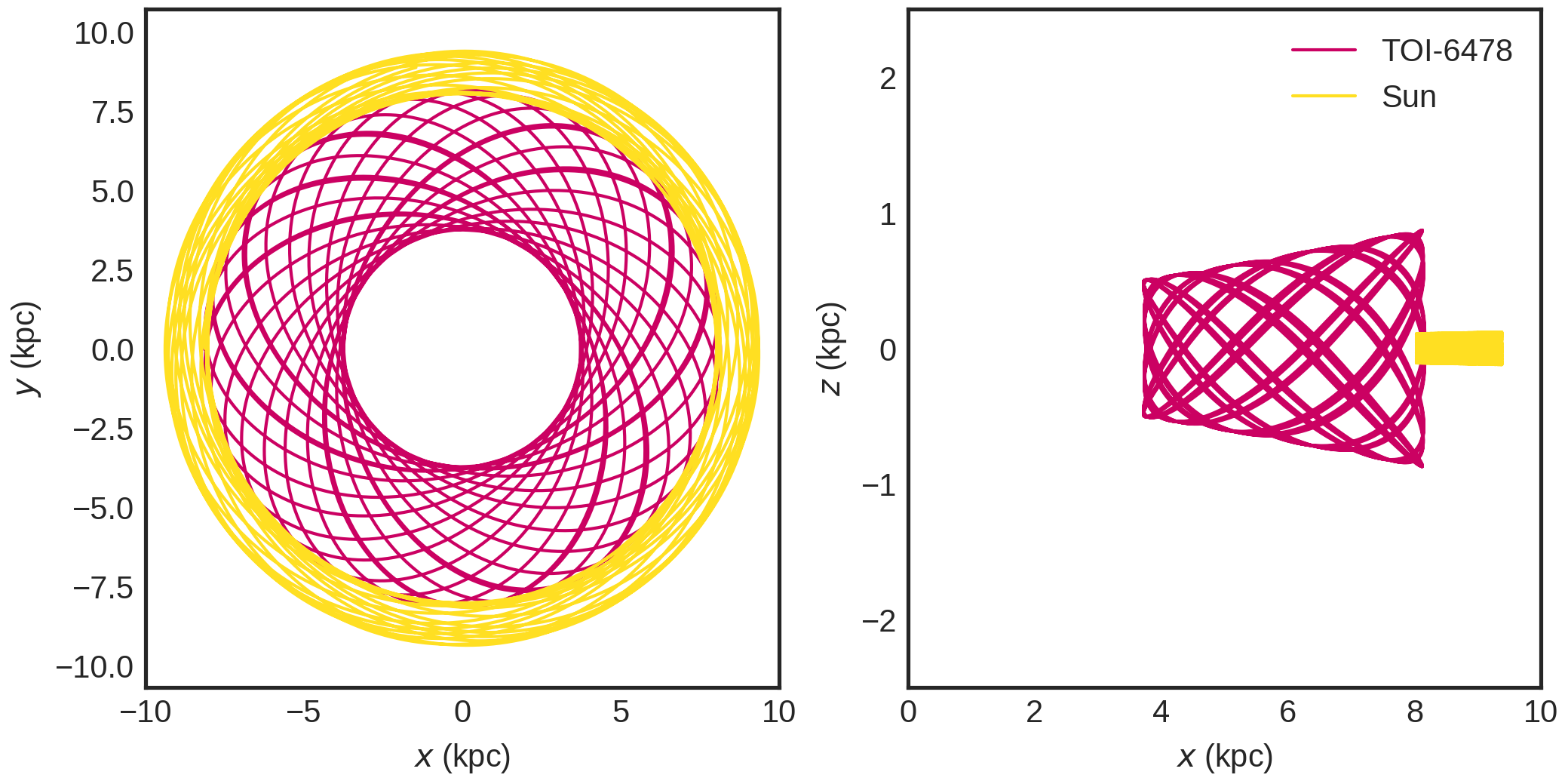}
    \caption{The integrated orbit of TOI-6478 within the Milky Way's potential over a $5$~Gyr period, depicted in Cartesian Galactocentric coordinates. The left panel presents a top-down view ($x, y$), while the right panel shows an edge-on perspective ($x, z$). For comparison, the Sun's integrated orbit is included in both views.}
    \label{fig:orbitintegration}
\end{figure*}

Fig.~\ref{fig:orbitintegration} shows the integration of the orbit of TOI-6478 in the Milky Way potential in Cartesian Galactocentric coordinates across a period of time of $5$~Gyr, highlighting the extent of the system's movement within the Milky Way.

It is interesting to note that this system spends time both in the denser inner part of the Galactic disk, at a minimum inner distance about a factor of 2 closer to the Galactic centre than the solar orbit, as well as significantly further away from the Galactic midplane than our Sun, reaching a maximum height of almost $1$~kpc over the course of its orbit.

\begin{figure}
    \centering
    \includegraphics[width=1\linewidth]{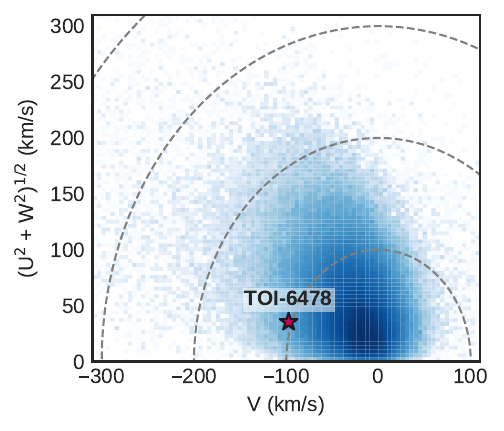}
    \caption{A Toomre diagram illustrating the Galactic velocity components perpendicular to the plane (vertical axis) versus those within the plane (horizontal axis). TOI-6478 is marked as a pink star. The background density map represents the number count of targets from {\it Gaia\/}~DR3 identified as single stars with reliable astrometric data and available line-of-sight velocities (\texttt{astrometric\_params\_solved = 95}, \texttt{rv\_nb\_transits > 0}, \texttt{ruwe < 1.4}, and \texttt{non\_single\_star = 0}). Darker regions indicate higher target densities, prominently highlighting the Galactic thin disk around ($0, 0$).}
    \label{fig:toomre}
\end{figure}

Fig.~\ref{fig:toomre} shows location of TOI-6478 in a so-called Toomre diagram. 
The diagram plots a star's velocity relative to the Sun in two components, along the horizontal axis it shows the component of a star's rotational velocity around in the plane of the Galaxy, while the vertical axis shows the combination of a star's velocity components perpendicular to the Galactic plane.
From this diagram, it is evident that different Galactic stellar populations such as the thin disk, thick disk, and halo show different characteristics, with stars in the Galactic thin disk -- like our Sun -- move similar to the Sun and thus the Galactic standard of rest ($(0, 0)$ in Fig.~\ref{fig:toomre}) and with stars belonging to the Galactic halo rotating significantly slower than the Sun ($\gtrsim 200$~\si{\kilo\metre\per\second}) but move with higher velocities in the dimensions perpendicular to the Galactic plane.

TOI-6478 is found to be in the region of velocity and orbital space typically occupied by the Galactic thick disk stars. 
This identification is in alignment with the sub-solar metallicity (see Table~\ref{tab:starpar}), which is also characteristic for these stars. As the thick disk formed early on in the history of the Galaxy \citep[$\gtrsim 9$~Gyr][]{spitoni2024}, this identification also agrees with the lower age bound estimated from the H$\alpha$ lines. 

The evolution of conditions for planet formation throughout the history of the Milky Way remains an area with many unresolved questions. \citet{zink2023} conducted a demographic study of exoplanets around FGK dwarf stars and found that close-in small planets are approximately $50$~\% less common around stars in the Galactic thick disk compared to those in the thin disk, emphasizing how the Galactic surroundings affect the possible exoplantary architecture. \citet{hallatt2024arXiv240809319H} attributed this difference to the more hostile conditions of the primordial thick disk, where the denser and more radiative environment likely led to the more efficient destruction of protoplanetary disks, thereby reducing the timescale available for planet formation.
TOI-6478 is an excellent candidate for future studies on the interior structure and atmospheric properties of planets in thick disk systems. These studies need to be extended to cover the abundant M dwarfs and with a greater sample we can explore variations in formation efficiency and evolutionary processes among planetary systems in different Galactic components.

\section{TESS data} \label{sec:tess}

\begin{figure*}
    \centering
    \includegraphics[width=\textwidth]{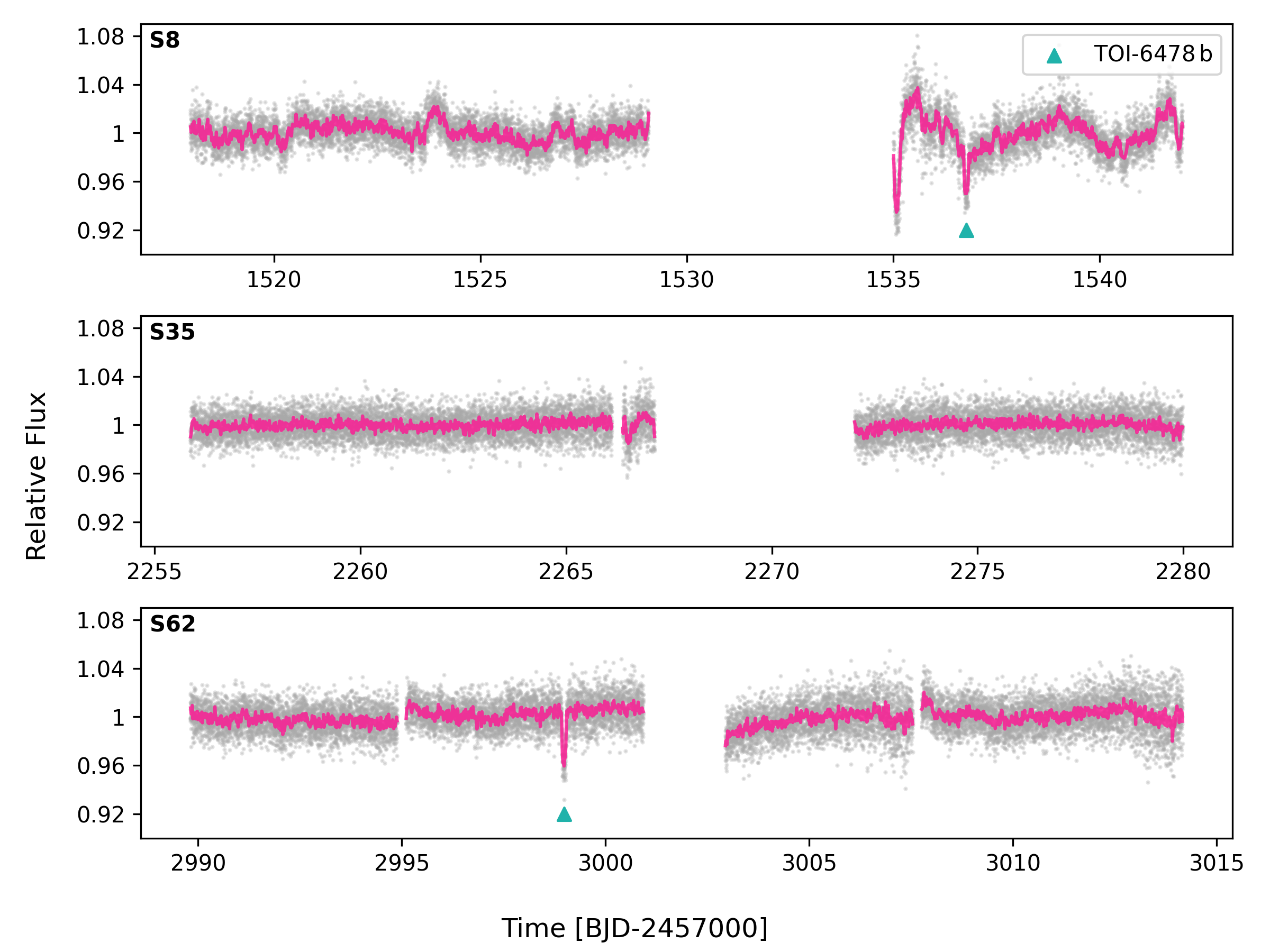}
    \caption{\textcolor{black}{The SPOC PDCSAP \textit{TESS} light curves for TOI-6478\,b, observed at 120s cadence. The star was observed in 3 sectors: Sector 8, Sector 35, and Sector 62. The grey points show the 120s data, and the pink shows this data binned to 20 mins. The transits detected in Sector 8 and Sector 62 are indicated by a green triangle.}}
    \label{fig:tess_transits}
\end{figure*}

\subsection{Determining the orbital period from \textit{TESS} duo-transits}

\textit{TESS} observed two transits throughout three sectors (S8, S35, S62) with a cadence of 120s. The data has been processed by the SPOC \citep[Science Processing Operations Centre;][\textcolor{black}{Figure~\ref{fig:tess_transits}}]{SPOC}. The signature of TOI-6478 b was detected first in a search of sectors 8, 35 and 62 by the FAINT search pipeline \citep{kunimoto2021, QLP2020a, QLP2020b} and was alerted by the TESS Science Office on 15 June 2023 \citep{guerrero2021} after the data validation reports were vetted. The TESS Science Processing Operations Center (SPOC) detected the transit signature in the same sectors shortly thereafter with a noise-compensating matched filter \citep{jenkins2002, jenkins2010, jenkins2020} and was fitted with an initial limb-darkened transit model \citep{li2019} and passed a suite of diagnostic tests \citep{twicken2018} including the difference image centroid test which located the host star within 3.0 +/- 2.5 arcsec of the transit source. 

Prior to the QLP faint transit search detection, and subsequent alert as TOI-6478.01, the transit signature was flagged as a planet candidate CTOI by the Planet Hunters \citep{eisner2021} on 15 May 2020 based only on the first transit. TOI-6478.01 was then alerted by the TESS Science Office as a planet candidate on 15 June 2023, as discussed above.

Transit 1 occurred at 1536.7668 BTJD (S8) and transit 2 occurred 1462.21 days later in S62 at 2998.9728 BTJD. \textcolor{black}{Based on the observed transit duration ($\sim$ 3 hours) and the available data (e.g. considering gaps where transits could fall), we predicted an orbital period of 34\,d. Longer orbital periods that were compatible with the data were ruled out based on an incompatibility with the duration. We initiated a ground-based follow-up photometric observation (see Section~\ref{subsec:ground}), on \textcolor{black}{2024 March 03} assuming the approximate 34\,d orbital period. A full transit was observed at a transit time of \textcolor{black}{3373.0354 BTJD}, thus confirming this orbital period.}

\subsection{Aperture photometry}

\begin{figure*}
    \centering
        \centering
        \includegraphics[width=\textwidth]{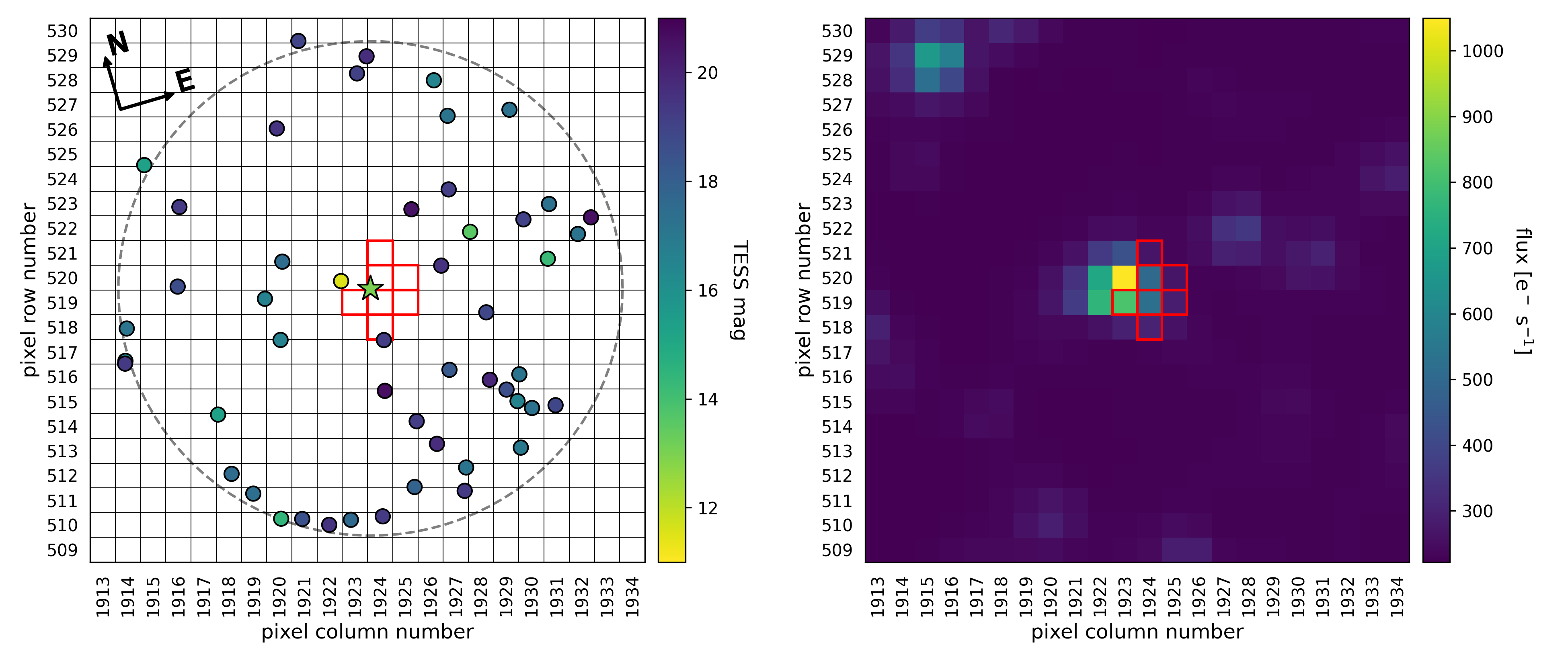}
        \label{fig:subfig1}
    
    \vspace{0.5cm} 
    
        \centering
        \includegraphics[width=\textwidth]{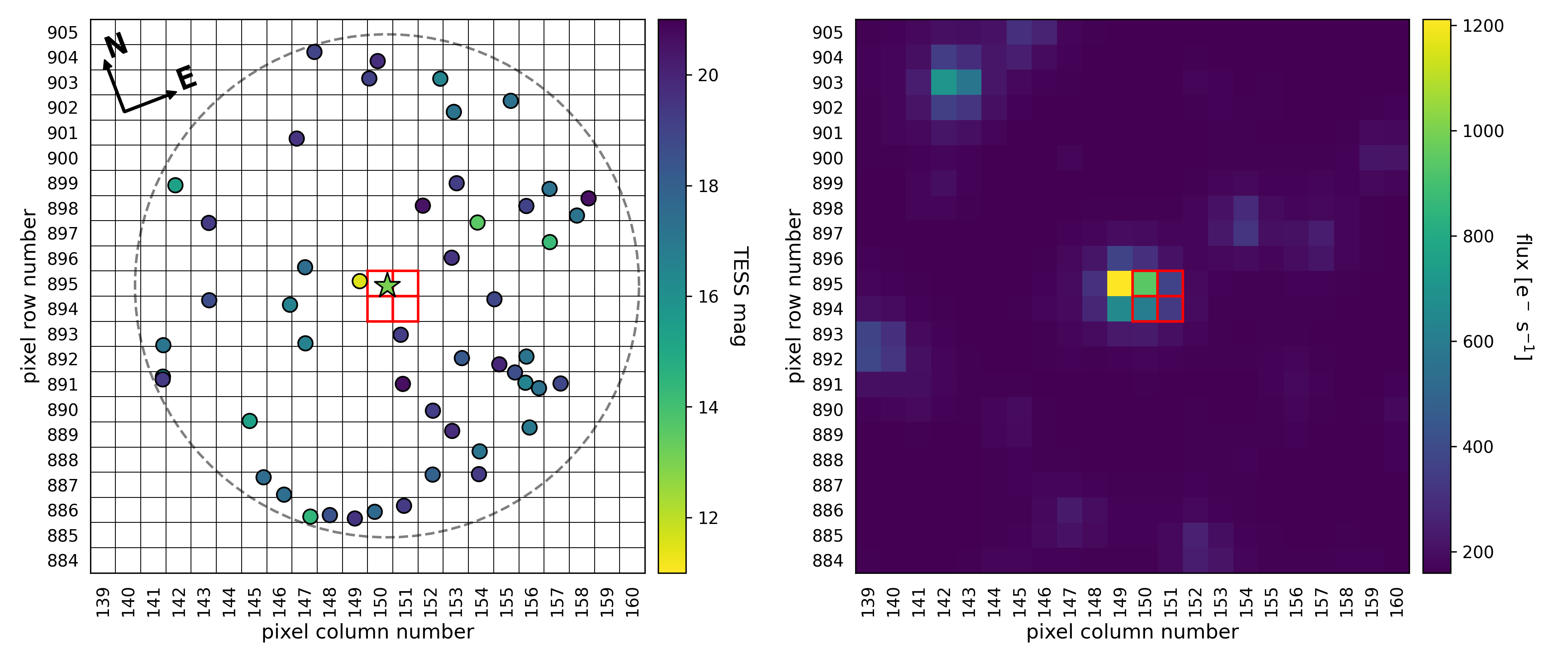}
        \label{fig:subfig2}
    
    \caption{\textit{Left}: Stars within the field of the target star TOI-6478 (depicted by the star), obtained from the \texttt{triceratops} pipeline for sector 8 (top) and sector 62 (bottom). The colours of each star correspond to their \textit{TESS magnitude}. The red shows the \textit{TESS} pipeline aperture. The yellow point is the brighter comoving companion to TOI-6478, LP~789-76. \textit{Right}: \textit{TESS} target pixel files with pipeline apertures for sector 8 (top) and sector 62 (bottom). }
    \label{fig:apertures}
\end{figure*}

Figure~\ref{fig:apertures} shows the target pixel files (TPFs) for S8 and S62 for TOI-6478\,b. Due to the large pixel size of \textit{TESS} (21" px$^{-1}$), it is not uncommon to have crowding within the aperture. Here, we see that our target is blended in the \textit{TESS} aperture with its brighter, co-moving companion LP 789-76 (23'' separation). The flux contribution from this star can cause the transits we observe in \textit{TESS} to be diluted in contrast to the transits we obtain from the ground (see Section~\ref{subsec:ground}) which typically have smaller apertures. The target pixel files (TPFs) with the associated apertures are shown in Figure~\ref{fig:apertures}. We also show the surrounding stars in the field (plot obtained from \texttt{triceratops}).

\section{Vetting and Validation}
\label{sec:followup}

This section describes the efforts taken to validate the planetary nature of TOI-6478\,b. We first describe the high-resolution imaging observations, followed by an outline of the ground-based photometric and radial velocity observations collected from the LCO and Gemini North observatories respectively. We conclude this section with a discussion of how these observations were used in the validation of TOI-6478\,b as a planetary candidate. All follow-up observations are summarised in Table \ref{tab:followup}.

\begin{table*}
\centering
\caption{Summary of ground-based follow-up observations carried out for TOI-6478\,b.}
\begin{tabular}{@{}ccccc@{}}
\midrule \midrule
\multicolumn{5}{c}{\textbf{Follow-up Observations}}                                              \\ \midrule \midrule
\multicolumn{5}{c}{\textbf{High Resolution Imaging}}                                             \\ 
\textbf{Observatory} & \textbf{Filter} & \textbf{Date}     & \textbf{Sensitivity Limit} & \textbf{Result}\\ \midrule
Gemini South      & $562~{\rm nm}$     & 2024 March 13 & $\Delta m=4.84$ at $0.5\arcsec$  & No sources detected   \\ 
Gemini South      & $832~{\rm nm}$     & 2024 March 13 & $\Delta m=6.97$ at $0.5\arcsec$  & No sources detected   \\\midrule
\multicolumn{5}{c}{\textbf{Photometric Follow-up}}                                               \\
\textbf{Observatory} & \textbf{Filter} & \textbf{Date}     & \textbf{Coverage} & \textbf{Result} \\ \midrule
LCO-SSO-2m0M    & \textcolor{black}{{\it Pan-STARRS-$z_s$}}     & 2024 March 03 & Full  & Detection\\ 
LCO-SSO-2m0M    & {\it Sloan-$i'$}     & 2024 March 03 & Full  & Detection\\ 
LCO-SSO-2m0M    & {\it Sloan-$r'$}     & 2024 March 03 & Full  & Detection\\ 
LCO-SSO-2m0M    & {\it Sloan-$g'$}     & 2024 March 03 & Full  & Detection\\ 
LCO-SSO-1m0    & {\it Pan-STARRS-zs}     & 2024 March 03 & Full  & Detection\\ 
LCO-SSO-1m0    & {\it $V$}     & 2024 March 03 & Full  & Detection \\\midrule
\multicolumn{5}{c}{\textbf{\color{black}Spectroscopic Observations}}                                               \\ 
\textbf{Instrument} & \textbf{Wavelength Range} & \textbf{Date}     & \textbf{Number of Spectra} & \textbf{Use}\\ \midrule
Shane/Kast & 375-560~\rm nm \& 580-900~\rm nm  & 2024 April 09 & \textcolor{black}{1} &Stellar characterisation \\
IRTF/SpeX & 800-2420~\rm nm & 2024 May 10 & \textcolor{black}{1} & Stellar characterisation \\
Gemini North/MAROON-X & 500-670~\rm nm \& 650-920~\rm nm & 2024 March 29 - 2024 April 23 & 10 & Mass upper limit \\
\hline

\end{tabular}
\label{tab:followup}
\end{table*}
\color{black}

\subsection{Archival imaging}

\begin{figure*}
    \centering
    \includegraphics[width=\linewidth]{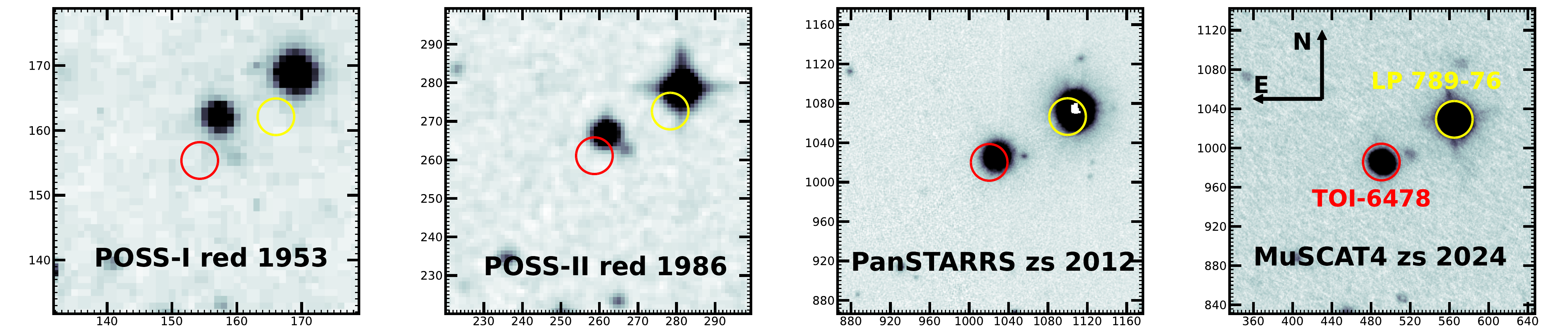}
    \caption{\textcolor{black}{Archival images of the field of view around TOI-6478. From left to right: image taken with the DSS/POSS-I in 1953, image taken with the DSS/POSS-II in 1986 in red, image taken with PanSTARRS in 2012 in the $z_s$-band, and the image taken with MuSCAT4 in 2024 in the $z_s$-band. The red and yellow circles depict the positions of TOI-6478 and LP~789-76 in the 2024 image, respectively.}}
    \label{fig:archival_imaging}
\end{figure*}

In order to investigate whether a background object is blending with the target star, we examine archival imaging of the field of view of TOI-6478. Given its high proper motion of 180 mas\,yr$^{-1}$ \citep{gaiaDR3cat}, this is able to be done through inspection of \textcolor{black} {DSS/POSS-I and DSS/POSS-II \citep{reid1991, lasker1996} images taken in red in 1953 and 1986 respectively}, and comparing to the $z_s$-band MuSCAT4 \citep{Narita:2020} observations taken in 2024, {\textcolor{black}{71 years later}} \textcolor{black}{(see Figure~\ref{fig:archival_imaging})}. We can thus conclude that at TOI-6478's current position, there is no background star that could be impacting our conclusions due to being the source of the observed transit events.

\subsection{High resolution imaging}
\label{sec:hires}

\begin{figure}
    \centering
    \includegraphics[width=\linewidth]{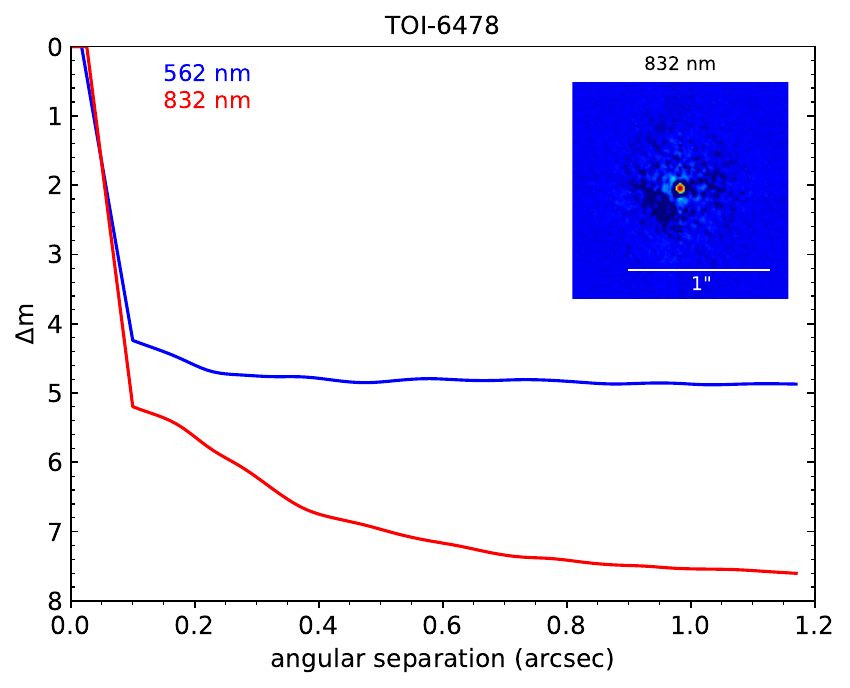}
    \caption{$5\sigma$ speckle imaging contrast curves obtained with Zorro for TOI-6478 in 562\,nm and 832\,nm bands on \textcolor{black}{2024 March 13}, as a function of angular separation. The reconstructed 832\,nm speckle image is shown in the top right inset with a 1" scale bar. No close companions are detected within the angular and brightness contrast levels.}

    \label{fig:hires}
\end{figure}

A critical observation to perform in order to rule out false positives such as background eclipsing binaries is that of obtaining high-resolution images of the exoplanet host star. While nearby M stars such as TOI-6478 have proper motions which allow historic sky imagery to eliminate confounding stars at the present day position of the exoplanet host, the spatial scale of such images as well as present day seeing limited images cannot rule out stellar companions. Inside of 0.5 arcsec (19 au at the distance of TOI-6478), essentially no technique other than high-resolution imaging allows for search and detection of very close bound companions. Optical high-resolution speckle offers angular resolutions and deep contrast levels from which M star stellar companions can readily be identified. Spatially close bound companions produce ``third-light” flux that can lead to underestimated planetary radii if not accounted for in the transit model \citep{2015ApJ...805...16C}, incorrect planet and star parameters \citep{2017AJ....154...66F, 2020ApJ...898...47F}, and can cause non-detections of small planets residing with the same exoplanetary system \citep{2021AJ....162...75L}. Thus, to search for very close bound companions, unresolved in TESS or other ground-based follow-up observations, we obtained high-resolution imaging speckle observations of TOI-6478.

TOI-6478 was observed on 2024 March 13 UT using the Zorro speckle instrument on the Gemini South 8-m telescope \citep{2021FrASS...8..138S}. Zorro provides simultaneous speckle imaging in two bands (562nm and 832nm) with output data products including a reconstructed image with robust contrast limits on companion detections \citep[e.g.][]{2016ApJ...829L...2H}. Due to the red nature of the M star TOI-6478, sixteen sets of $1000 \times 0.06$ second images were obtained and processed in our standard reduction pipeline \citep{2011AJ....142...19H}. Figure~\ref{fig:hires} shows our final 5$\sigma$ magnitude contrast curves and the 832 nm reconstructed speckle image. We find that, with the Zorro field of view, TOI-6478 is a single star with no companion brighter than 5-7.5 magnitudes below that of the target star from the Gemini 8-m telescope diffraction limit (20 mas) out to 1.2”. At the distance of TOI-6478 (d=38 pc) these angular limits correspond to spatial limits of 0.76  to 45 au. 
\textcolor{black}{
Another false positive scenario is that there is a bounded stellar companion below the Zorro detection sensitivity that is the source of the transit signal. This would mean that TOI-6478 has a wide binary companion that has a magnitude $\geq7.5\times$ fainter and an angular separation up to around 1 arcsec. Using the \citet{Baraffe2015} models in I-band (Zorro wavelength), assuming an old $>10\,$Gyr star (thick disc -- see Section~\ref{sec:galactic_orbit}), the mass of a body $7.5\times$ fainter than TOI-6478 is $\sim0.072\,\mathrm{M}_\odot$, corresponding to a brown dwarf just below the hydrogen burning limit. This magnitude difference results in a flux ratio of $f_{\mathrm{BD}}/f_{\mathrm{TOI-}6478}=0.001$. Therefore, if the transit event were to be on this brown dwarf companion, it would be diluted by 1000 due to the flux from TOI-6478, creating an unphysical transit depth ($\sim0.004\%$). The observed transit depth is already $\sim40\times$ too deep compared to the largest eclipse that could happen on the possible bounded brown dwarf companion. We thus conclude that even if there is a bounded companion, the transit cannot be happening on this companion. }

\subsection{Photometric follow-up}
\label{sec:photometry}

\subsubsection{LCOGT Photometry\label{subsec:ground}}

The \textit{TESS} pixel scale is $\sim 21\arcsec$ pixel$^{-1}$ and photometric apertures typically extend out to roughly 1 arcminute, generally causing multiple stars to blend in the \textit{TESS} photometric aperture. To determine the true source of the \textit{TESS} detection, we acquired ground-based time-series follow-up photometry of the field around TOI-6478 as part of the \textit{TESS} Follow-up Observing Program \citep[TFOP;][]{collins2019}\footnote{{\href{https://tess.mit.edu/followup}{https://tess.mit.edu/followup}}}. The on-target follow-up light curves are also used to place constraints on the transit depth and the \textit{TESS} ephemeris. We used the {\tt TESS Transit Finder}, which is a customized version of the {\tt Tapir} software package \citep{Jensen:2013}, to schedule our transit observations.

We observed a full transit of TOI-6478.01 on UTC 2024 March 03 simultaneously from 1.0\,m and 2.0\,m telescopes at the Las Cumbres Observatory Global Telescope (LCOGT) \citep{Brown:2013} node at Siding Spring Observatory near Coonabarabran, Australia (SSO). Alternating Johnson/Cousins V and Pan-STARRS $z_s$ filters were used from the 1\,m network node while simultaneous images were taken in Sloan $g'$, $r'$, $i'$, and \textcolor{black}{{\it Pan-STARRS-$z_s$}} bands from the 2\,m Faulkes Telescope South. The 1\,m telescope is equipped with a $4096\times4096$ SINISTRO camera having an image scale of $0\farcs389$ per pixel, resulting in a $26\arcmin\times26\arcmin$ field of view. The 2\,m telescope is equipped with the MuSCAT4 multi-band imager \citep{Narita:2020}.  All images were calibrated by the standard LCOGT {\tt BANZAI} pipeline \citep{McCully:2018}, and differential photometric data were extracted using {\tt AstroImageJ} \citep{Collins:2017}.  2.4”-4.7” We used circular photometric apertures with radii ranging from $2\farcs4$---$4\farcs7$ that excluded all of the flux from the nearest known neighbor in the Gaia DR3 catalog (Gaia DR3 5673934617318453248, LP 789-76) that is $22\farcs9$ northeast of TOI-6478. A $\sim$34 ppt event was detected on-target and confirms the 34 day period alias. All light curve data are available on the {\tt EXOFOP-TESS} website\footnote{\href{https://exofop.ipac.caltech.edu/tess/target.php?id=332657786}{https://exofop.ipac.caltech.edu/tess/target.php?id=332657786}} and are included in the global modelling described in Section~\ref{sec:analysis}.

\subsubsection{WASP}

\begin{figure}
    \includegraphics
    {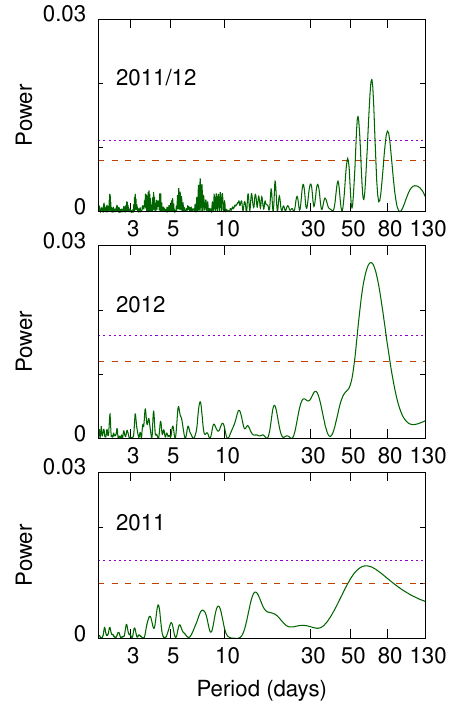}
 \caption{Periodograms of the WASP data for TOI-6478 and its co-moving companion from 2011 and 2012 and (top) both years combined. The dashed lines are at the estimated 10\%- and 1\%-likelihood false-alarm levels. There is a significant modulation at a period of 66 $\pm$ 4 d, though it is unclear which star it originates from.}
\label{fig:wasp}   
\end{figure}

The Wide Angle Search for Planets (WASP) project \citep{2006PASP..118.1407P} was a ground-based exoplanet transit survey with instruments in La Palma and South Africa. In 2011 and 2012 it covered the field of TOI-6478. The resulting data indicate a possible rotational modulation at 66 $\pm$ 4 days (see Fig.~\ref{fig:wasp}); however, as with \textit{TESS}, WASP has a large pixel size (13.7"), and therefore the extraction aperture contains both TOI-6478 and its co-moving companion LP 789-76. Due to LP 789-76 being the brighter star, it is likely that this modulation does not come from TOI-6478. We therefore do not include this in our analysis but note that it prompts further observations of these stars from other ground-based observatories in order to untangle the origin of this signal.

\subsection{Radial Velocity Follow-Up} \label{sec:rvs}
\subsubsection{MAROON-X}
TOI-6478 was observed with the MAROON-X spectrograph at the Gemini North Observatory in Hilo, Hawai'i \citep{maroonx2016, maroonx2020, maroonx2022}. The MAROON-X spectrograph is a high resolution radial velocity spectrograph which uses red-optical fibre feeding and covers a wavelength range of 500-920\,nm across two channels: blue (500-670\,nm) and red (650-920\,nm). It has a resolving power of $\sim85\,000$. 

We obtained 10 spectra with both the blue and red arms (645\,nm and 875\,nm respectively) over a period of 27 days (\textcolor{black}{2024 March 29 to 2024 April 23}) with 900s exposure time \textcolor{black}{(see Table~\ref{tab:rvs} in Appendix)}. The mean SNR for the blue and red arms were \textcolor{black}{21 and 52} respectively. These data were reduced, with RVs extracted by the MAROON-X team using \textcolor{black}{a specific version of} the publicly available python pipeline \textsc{serval} \citep[SpEctrum Radial Velocity AnaLyser;][]{serval} \textcolor{black}{which has been modified for use with MAROON-X data}, in which the analysis of the RVs is performed through a template matching code \citep[see e.g.][]{winters2022, kanodia2024, martioli2024}. This is favoured over other methods such as classical binary mask cross-correlation codes as it typically outperforms for M-dwarf stars. Correction for the main instrument drift is performed in the wavelength solution, and the H\;$\alpha$ line is observed in both channels. \textcolor{black}{It was noted that the 2024-04-01 09:22:47 (BJD 2460401.901) observation was flagged as `usable' rather than `pass', since the atmospheric dispersion compensator (ADC) was not following the telescope. In order to be more conservative with our analyses (see Sections~\ref{sec:analysis} and ~\ref{sec:discussion}), we choose to omit this data point.}

The cross-correlation function analysis confirmed the systemic velocity of TOI-6478 ($\sim 99.5\,{\rm km\,s^{-1}}$). We also measure the full-width-half-maximum (FWHM) of the average line profile, resulting in $5.4\pm0.15\,{\rm km\,s^{-1}}$ and $5.6\pm 0.05\,{\rm km\,s^{-1}}$ for the blue and red channels respectively. This corresponds to the expected FWHM of a non rotating M-dwarf at the resolution of MAROON-X. We can conclude that TOI-6478 is a slow rotating star with an upper limit of $v\,\sin i_{\star} \lesssim 2\,{\rm km\,s^{-1}}$.

\subsection{Statistical Validation}

\label{sec:validation}

We make use of \texttt{TRICERATOPS}\footnote{\url{ https://github.com/stevengiacalone/triceratops}} \citep{triceratops, 2020ascl.soft02004G}, a statistical validation package used to evaluate the false positive probability (FPP) of TOI-6478\,b. 

Upon providing the \textit{TESS} apertures for the target, \texttt{TRICERATOPS} calculates the amount of flux contributed from nearby stars, which enables it to assess whether the transit signal could be caused by an alternate scenario to that which is assumed, i.e., a transiting planet on the target star. Based on the aperture information, as well as the photometric data and high resolution speckle imaging constrast curves (Figure~\ref{fig:hires}), a range of scenarios for transiting planets (TPs) and eclipsing binaries (EBs) are considered and their relative probabilities calculated. The threshold for a planet to be considered validated is $\text{FPP}<0.015$.

\begin{figure}
    \centering
    \includegraphics[width=\linewidth]{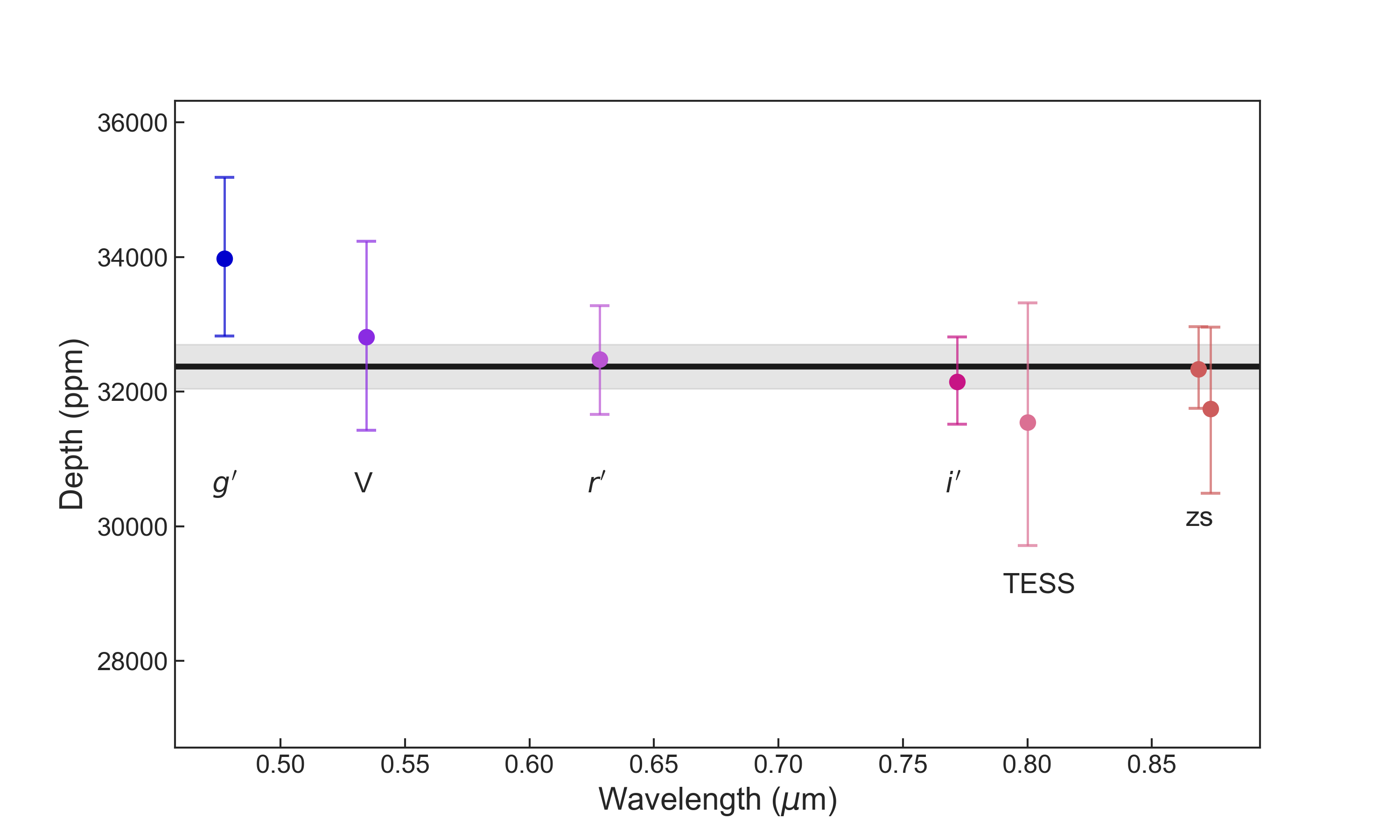}
    \caption{The measured transit depths for each transit vs wavelength, where for \textit{TESS} this is the average from the two transits. The grey band shows the weighted average of the transit depths. All bands agree to 1$\sigma$ except the blue band $g'$, which agrees to 1.3$\sigma$.} 
 
    \label{fig:chromaticity}
\end{figure}

Considering only the two transits from \textit{TESS}, we find a $\text{FPP}=0.6499$ and a nearby false positive probability (NFPP) of $\text{NFPP}=0.6498$. This reduces to $\text{FPP}=\text{NFPP}=0.47$ with the inclusion of the transit from LCO. We choose the light curve observed in the \textit{i'} band for the validation since the transit depth, while statistically consistent between  bands, is still wavelength dependent (see Figure~\ref{fig:chromaticity}), and the \textit{TESS} bandpass is centered on \textit{i'}. The FPP is rather large, likely due to the fact that our star is blended in the \textit{TESS} aperture with its brighter comoving companion TIC 332657787 (LP~789-76\,A), located $\sim\,23$" away. However, since the ground-based LCO data has a much smaller aperture situated only on the target star, the stars are resolved and we are able to confirm the transit is on target. Therefore, we adopt the procedure used in \citet{Timmermans2024} and remove the probability of a transiting planet around a nearby star, $P_{\text{NTP}}$. This places the FPP at $\text{FPP}=5.86\times 10^{-18} \pm 7.045\times 10^{-17}$, averaging from 20 iterations, which is well below the validated planet threshold.

\section{Global analysis}
\label{sec:analysis}

\begin{figure*}
    \centering
     \includegraphics[scale=0.45]
    {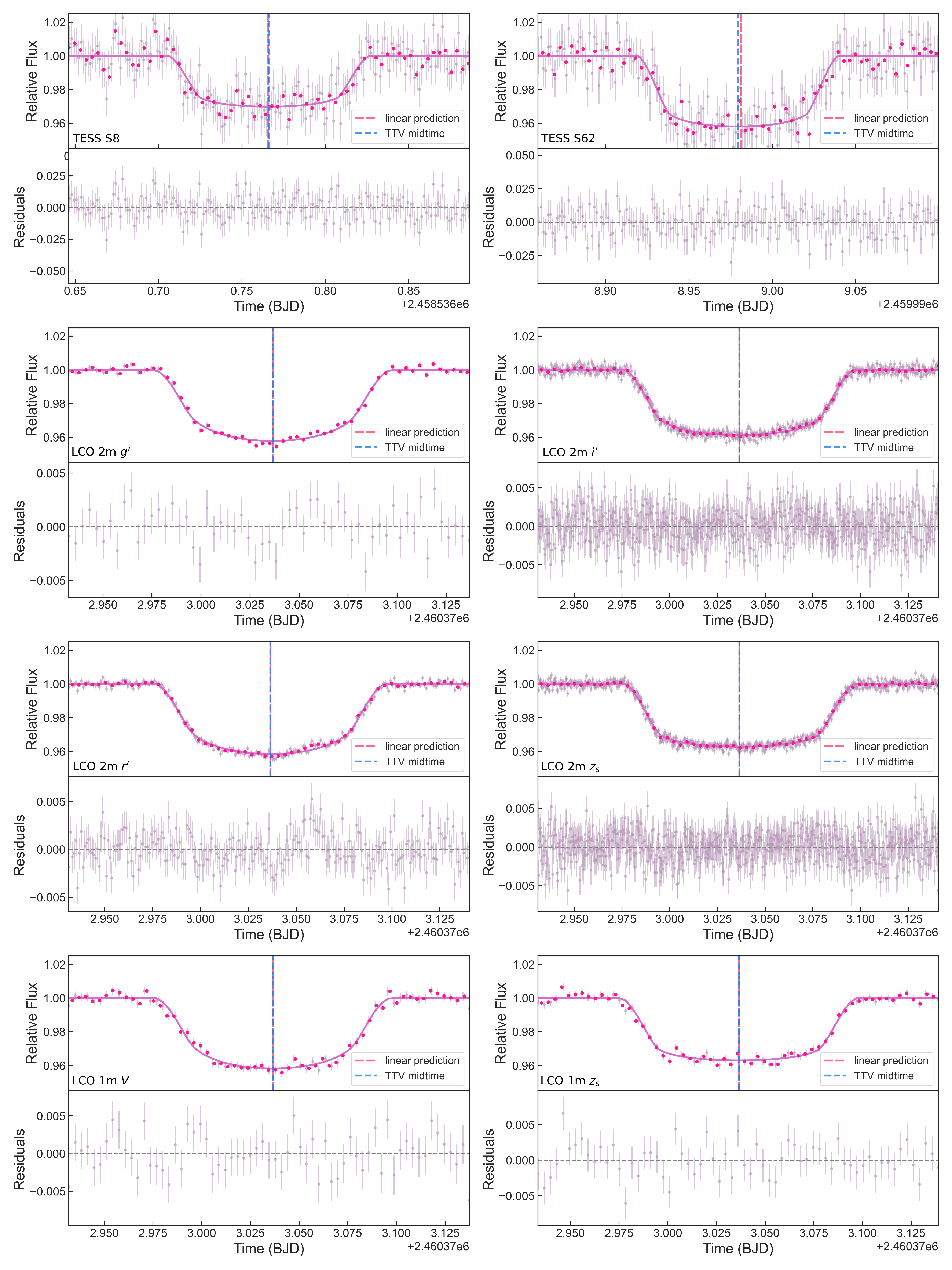}
    \caption{\textcolor{black}{Fitted \textit{TESS} (top 2 figures) and LCO transits (bottom 6 figures) obtained from the global photometric analysis using \textsc{allesfitter}. Top panels show the model and transit data with the baseline subtracted, as well as the transit timing variations (TTVs) and linear predicted mid-time. Bottom panels show the residuals from our fit. The presence of a starspot is inferred from the LCO transits ($\sim2460373.06 $\,BJD).} }
    \label{fig:allesfitter_results}
\end{figure*}

\begin{figure}
    \centering
    \includegraphics[width=\linewidth]{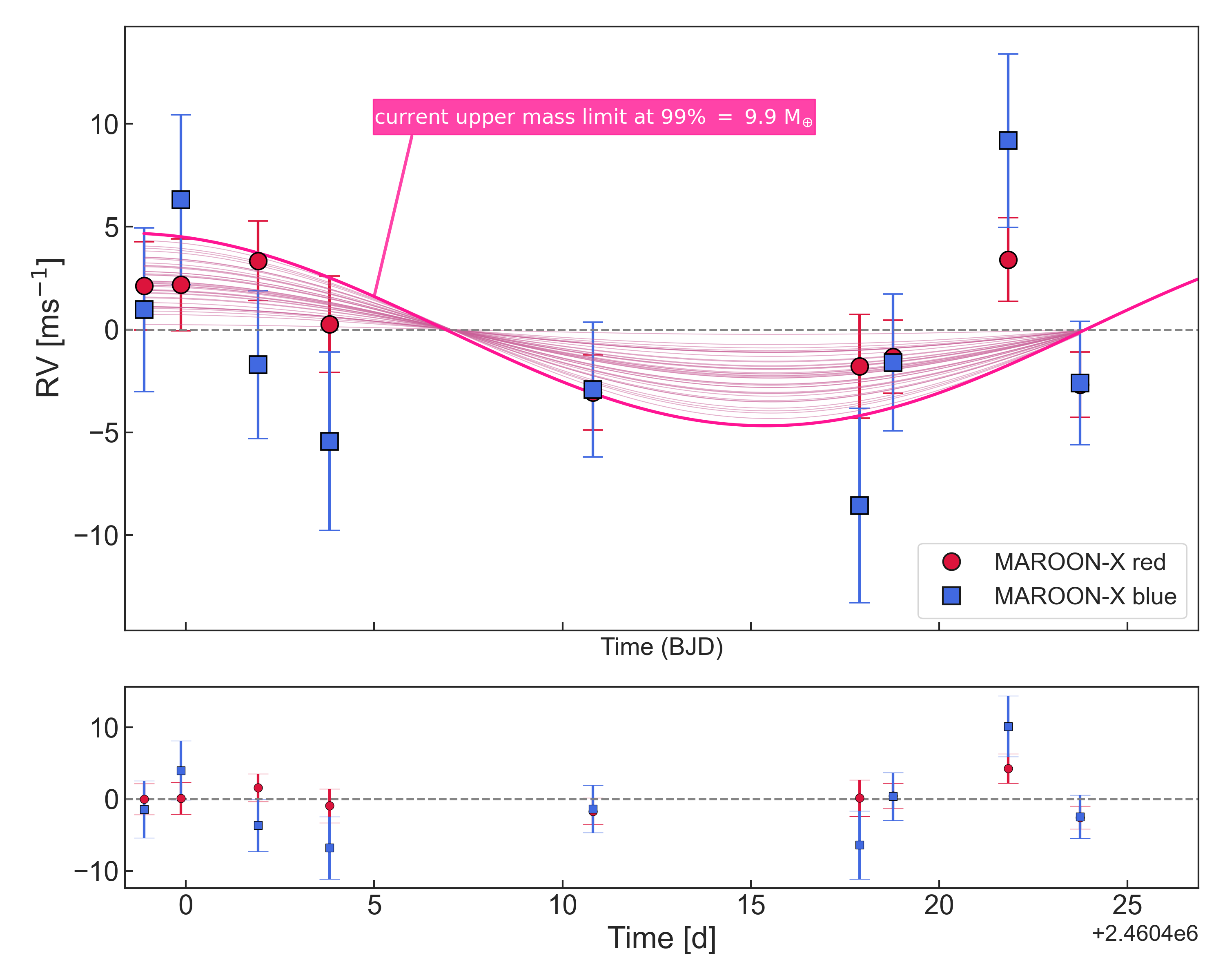}
    \caption{\textcolor{black}{Radial Velocity (RV) points measured with MAROON-X at the Gemini-North observatory. Nested sampling fit using both red- and blue-arm RV data achieved from results from global photometric analysis in order to obtain an upper limit on the semi-amplitude ($\leq4.68\,\mathrm{m}\,\mathrm{s}^{-1}$ -bright pink line) and thus the mass of TOI-6478\,b ($\leq9.9\,\mathrm{M}_\oplus$). Darker pink lines show a sample of the posterior models from the fit. Bottom panel shows the residuals of the fit.}}
    \label{fig:rvs}
\end{figure}

We perform a global photometric analysis of both the \textit{TESS} and LCO photometry using the flexible, publicly available \textsc{python} package \textsc{{allesfitter}} \citep{allesfitter-paper, allesfitter-code}. This inference-based package primarily makes use of the light curve generating \textsc{python} code \textsc{{ellc}} \citep{ellc}, and the one-dimension scalable Gaussian Process Regression code \textsc{{celerite}} \citep{celerite}. While both nested sampling (NS) and Markov Chain Monte Carlo (MCMC) algorithms can be used in \textsc{{allesfitter}} (with the packages \textsc{{dynesty}} \citep{dynesty} and \textsc{{emcee}} \citep{emcee} respectively) to obtain the best fit models, we select the NS method as this allows the Bayesian evidence to be calculated at each sampling step; this is particularly useful when wanting to compare different models, as from this we can calculate the Bayes Factor \citep{BayesFactor} and assess which model, if any, is favoured. Since we initially run both circular and eccentric fits, this provides a quantitative argument as to which model should be assumed for this planetary system based on the data available.

We include all 8 transits in our analysis of TOI-6478\,b---$2\times$\textit{TESS}, $6\times$LCO (same epoch)---however, we omit the RV data from our global fit. The \textit{TESS} data used throughout this analysis is the SPOC Presearch Data Conditioning Simple Aperture Photometry \citep[PDCSAP;][]{stumpe2012, stumpe2014, smith2012} 2-minute \textit{TESS} light curves (see Figure~\ref{fig:tess_transits}). 
Once the global photometric fit is finalised, we use the results to do an RV-only fit, with tight priors on all parameters except semi-amplitude, $K$. This is discussed later in this section.

Initially we run two fits with both circular and eccentric models, where the eccentricity is either fixed at 0 or allowed to vary. Following \citet{Triaud2011}, the eccentricity is reparameterised as $\sqrt{e_{\text{b}}} \cos\,\omega_{\text{b}}$ and $\sqrt{e_{\text{b}}} \sin\,\omega_{\text{b}}$, where $\omega_{\text{b}}$ is the argument of periastron. We calculate the logarithm of the Bayes Factor for the circular and eccentric cases to be 9964.3 and 9894.9 respectively. To significantly favour one scenario over another, it is often desired that $\Delta\log Z \geq 5$. We find $\Delta\log Z = 69.4$, therefore very clearly favouring the circular model over the eccentric. 

We use the stellar parameters described in Section~\ref{sec:star} and uniformly fit for all planetary parameters, namely $t_0$, $P$, $\cos i$, $R_p/R_\star$ and $(R_p + R_\star)/a$.

Since the \textit{TESS} apertures include a significant amount of light from stars other than the transit host \textcolor{black}{(see Figure~\ref{fig:apertures})}, the observed transit depth needs to be corrected. Therefore, we fit for a dilution factor for the two \textit{TESS} transits. This was the main motivation for fitting the \textit{TESS} transits as separate instruments in \textsc{{allesfitter}}, since it was likely they would have different dilution factors due to having different apertures. We fix the LCO dilution at 0 since there is no contamination due to the small aperture. As per the \textsc{{allesfitter}} documentation, the dilution is equivalent to $D_0 = 1-C$, where $C$ is the `crowdsap' value in the \textit{TESS} FITS file headers. This gives the ratio of the the flux from the target star to the total flux within the aperture. For S8 and S62 we obtain $C_{8}=0.46$ and $C_{62}=0.37$ respectively, however, we decide to freely fit the dilution with $\rm{U}[-1, 1]$ as this will also be informed by the uncontaminated LCO transits and we did not want to bias the fit if the crowdsap values are inaccurate.

For each photometric band, we calculate the quadratic limb darkening coefficients using the \textsc{python} package \textsc{\large{pyltdk}} \citep{pyldtk} and the \textsc{\large{phoenix}} stellar atmosphere library \citep{phoenix}. This code requires the $T_{\text{eff}}$, $\log$\,g and metallicity, z, which are listed in Table~\ref{tab:starpar}. This gives us the $u_1$ and $u_2$ limb darkening coefficients, which we reparameterise to $q_1$ and $q_2$ as in \citet{kippingldcs}. These are adopted as normal priors in our fit. We also make use of \textsc{\large{allesfitter}}'s `coupled\_with' function as for observations taken in the same photometric band we typically want the limb darkening coefficients to agree. We apply this to the \textit{TESS} light curves, but not to the two LCO transits taken in the $\text{z}_{\text{s}}$ band in order to show that there is still agreement with the derived $u_1$ and $u_2$ values.

As expected, the light curves show variations from red noise caused by stellar variability and instrumental systematics; we choose to model the baseline with a hybrid spline for all data sets. While often a Gaussian Process (GP) is used to account for this noise, this was not necessary since we model each transit separately, and hence there is no significant out-of-transit data. The hybrid spline works by automatically trying to describe the residuals with a smooth spline at every sampling step. This method does not require any parameters to be given as priors. We also use the `error scaling' function within \textsc{{allesfitter}} which accounts for the white noise in our data; we give this a wide prior and select the `sample` method, which samples as every other parameter.

We run a static NS within \textsc{{allesfitter}} with 500 live points and a tolerance threshold of 0.02. We take the median posterior values as the results from our fit, with the uncertainty estimated as the 1$\sigma$ confidence intervals. We then use these in an RV-only fit \textcolor{black}{as mentioned above}, again assuming a circular orbit. The results for the photometry-only and RV-only fits are shown in Figures~\ref{fig:allesfitter_results} and ~\ref{fig:rvs} respectively. All fitted and derived parameters from the circular, TTV fit are presented in Tables~\ref{tab:fitted_params} and ~\ref{tab:derived_params}.

\begin{table} 
    \centering
    \caption{\textcolor{black}{Fit parameters from global analysis of the photometry (see Section~\ref{sec:analysis}). The last column indicates whether the parameter was fixed for the nested sampler. The limb darkening parameters for the two \textit{TESS} transits are coupled.}}
    \begin{tabular}{c|c|c|c}
        parameter & value & unit & fit/fixed \\ 
        \hline 
        \multicolumn{4}{c}{\textit{Fitted parameters}} \\ 
        \hline 
        $R_b / R_\star$ & $0.18027\pm0.00088$ & $R_b/R_\star$& fit \\ 
        $(R_\star + R_b) / a_b$ & $0.01131_{-0.00013}^{+0.00011}$ & & fit \\ 
        $\cos{i_b}$ & $0.00178_{-0.00070}^{+0.00044}$ & & fit \\ 
        $T_{0;b}$ & $2459454.90101\pm0.00068$ & $\mathrm{BJD}$& fit \\ 
        $P_b$ & $34.005019\pm0.000025$ & $\mathrm{d}$& fit \\ 
        $\sqrt{e_b} \cos{\omega_b}$ & $0.0$ & & fixed \\ 
        $\sqrt{e_b} \sin{\omega_b}$ & $0.0$ & & fixed \\ 
        $D_\mathrm{0; TESS\,S8}$ & $0.182\pm0.048$ & & fit \\ 
        $D_\mathrm{0; TESS\,S62}$ & $-0.125\pm0.055$ & & fit \\ 
        $D_\mathrm{0; LCO\,gp2m}$ & $0.0$ & & fixed \\ 
        $D_\mathrm{0; LCO\,ip2m}$ & $0.0$ & & fixed \\ 
        $D_\mathrm{0; LCO\,rp2m}$ & $0.0$ & & fixed \\ 
        $D_\mathrm{0; LCO\,zs2m}$ & $0.0$ & & fixed \\ 
        $D_\mathrm{0; LCO\,V1m}$ & $0.0$ & & fixed \\ 
        $D_\mathrm{0; LCO\,zs1m}$ & $0.0$ & & fixed \\ 
        $q_{1; \mathrm{TESS\,S8}}$ & $0.256\pm0.048$ & & fit \\ 
        $q_{2; \mathrm{TESS\,S8}}$ & $0.287\pm0.049$ & & fit \\ 
        $q_{1; \mathrm{TESS\,S62}}$ & $0.2564481081075988$ & & coupled \\ 
        $q_{2; \mathrm{TESS\,S62}}$ & $0.2865715657990476$ & & coupled \\ 
        $q_{1; \mathrm{LCO\,gp2m}}$ & $0.732\pm0.043$ & & fit \\ 
        $q_{2; \mathrm{LCO\,gp2m}}$ & $0.359\pm0.037$ & & fit \\ 
        $q_{1; \mathrm{LCO\,ip2m}}$ & $0.382_{-0.030}^{+0.032}$ & & fit \\ 
        $q_{2; \mathrm{LCO\,ip2m}}$ & $0.338_{-0.033}^{+0.036}$ & & fit \\ 
        $q_{1; \mathrm{LCO\,rp2m}}$ & $0.641\pm0.038$ & & fit \\ 
        $q_{2; \mathrm{LCO\,rp2m}}$ & $0.378_{-0.030}^{+0.032}$ & & fit \\ 
        $q_{1; \mathrm{LCO\,zs2m}}$ & $0.279_{-0.028}^{+0.030}$ & & fit \\ 
        $q_{2; \mathrm{LCO\,zs2m}}$ & $0.278\pm0.039$ & & fit \\ 
        $q_{1; \mathrm{LCO\,V1m}}$ & $0.720\pm0.045$ & & fit \\ 
        $q_{2; \mathrm{LCO\,V1m}}$ & $0.345_{-0.038}^{+0.042}$ & & fit \\ 
        $q_{1; \mathrm{LCO\,zs1m}}$ & $0.254_{-0.038}^{+0.041}$ & & fit \\ 
        $q_{2; \mathrm{LCO\,zs1m}}$ & $0.275_{-0.045}^{+0.049}$ & & fit \\ 
        $\ln{\sigma_\mathrm{TESS}}$ & $-4.560\pm0.037$ & & fit \\ 
        $\ln{\sigma_\mathrm{TESS}}$ & $-4.512_{-0.035}^{+0.037}$ & & fit \\ 
        $\ln{\sigma_\mathrm{LCO}}$ & $-6.356\pm0.093$ & & fit \\ 
        $\ln{\sigma_\mathrm{LCO}}$ & $-6.181_{-0.028}^{+0.027}$ & & fit \\ 
        $\ln{\sigma_\mathrm{LCO}}$ & $-6.396\pm0.045$ & & fit \\ 
        $\ln{\sigma_\mathrm{LCO}}$ & $-6.260_{-0.026}^{+0.028}$ & & fit \\ 
        $\ln{\sigma_\mathrm{LCO}}$ & $-5.989_{-0.084}^{+0.089}$ & & fit \\ 
        $\ln{\sigma_\mathrm{LCO}}$ & $-6.068_{-0.088}^{+0.093}$ & & fit \\

    \end{tabular}
    \label{tab:fitted_params}
\end{table}

\begin{table*}
\caption{\textcolor{black}{Derived parameters for TOI-6478\,b from global analysis of the photometry. All depths are fractional.}}
    \begin{tabular}{c|c|c}
        Parameter & Value & Source \\ 
        \hline 
        \multicolumn{3}{c}{\textit{Derived parameters}} \\ 
        \hline 
        Host radius over semi-major axis b; $R_\star/a_\mathrm{b}$ & $0.009579_{-0.00010}^{+0.000088}$ & derived \\ 
        Semi-major axis b over host radius; $a_\mathrm{b}/R_\star$ & $104.40_{-0.95}^{+1.1}$ & derived \\ 
        Companion radius b over semi-major axis b; $R_\mathrm{b}/a_\mathrm{b}$ & $0.001727_{-0.000026}^{+0.000023}$ & derived \\ 
        Companion radius b; $R_\mathrm{b}$ ($\mathrm{R_{\oplus}}$) & $4.60\pm0.24$ & derived \\ 
        Companion radius b; $R_\mathrm{b}$ ($\mathrm{R_{jup}}$) & $0.411\pm0.021$ & derived \\ 
        Semi-major axis b; $a_\mathrm{b}$ ($\mathrm{R_{\odot}}$) & $24.4\pm1.3$ & derived \\ 
        Semi-major axis b; $a_\mathrm{b}$ (AU) & $0.1136\pm0.0060$ & derived \\ 
        Inclination b; $i_\mathrm{b}$ (deg) & $89.898_{-0.026}^{+0.040}$ & derived \\ 
        Impact parameter b; $b_\mathrm{tra;b}$ & $0.186_{-0.072}^{+0.044}$ & derived \\ 
        Total transit duration b; $T_\mathrm{tot;b}$ (h) & $2.900\pm0.011$ & derived \\ 
        Full-transit duration b; $T_\mathrm{full;b}$ (h) & $1.987\pm0.015$ & derived \\ 
        Host density from orbit b; $\rho_\mathrm{\star;b}$ (cgs) & $18.61_{-0.50}^{+0.60}$ & derived \\ 
        Equilibrium temperature b; $T_\mathrm{eq;b}$ (K) & $204.4\pm4.9$ & derived \\ 
        Transit depth (undil.) b; $\delta_\mathrm{tr; undil; b; TESS\,S8}$ & $0.0370_{-0.0029}^{+0.0031}$ & derived \\ 
        Transit depth (dil.) b; $\delta_\mathrm{tr; dil; b; TESS\,S8}$ & $0.0303_{-0.0018}^{+0.0016}$ & derived \\ 
        Transit depth (undil.) b; $\delta_\mathrm{tr; undil; b; TESS\,S62}$ & $0.0370_{-0.0023}^{+0.0025}$ & derived \\ 
        Transit depth (dil.) b; $\delta_\mathrm{tr; dil; b; TESS\,S62}$ & $0.0416_{-0.0018}^{+0.0019}$ & derived \\ 
        Transit depth (undil.) b; $\delta_\mathrm{tr; undil; b; LCO\,gp2m}$ & $0.04212_{-0.00052}^{+0.00056}$ & derived \\ 
        Transit depth (dil.) b; $\delta_\mathrm{tr; dil; b; LCO\,gp2m}$ & $0.04212_{-0.00052}^{+0.00056}$ & derived \\ 
        Transit depth (undil.) b; $\delta_\mathrm{tr; undil; b; LCO\,ip2m}$ & $0.03871_{-0.00033}^{+0.00029}$ & derived \\ 
        Transit depth (dil.) b; $\delta_\mathrm{tr; dil; b; LCO\,ip2m}$& $0.03871_{-0.00033}^{+0.00029}$ & derived \\ 
        Transit depth (undil.) b; $\delta_\mathrm{tr; undil; b; LCO\,rp2m}$ & $0.04159_{-0.00034}^{+0.00037}$ & derived \\ 
        Transit depth (dil.) b; $\delta_\mathrm{tr; dil; b; LCO\,rp2m}$ & $0.04159_{-0.00034}^{+0.00037}$ & derived \\ 
        Transit depth (undil.) b; $\delta_\mathrm{tr; undil; b; LCO\,zs2m}$ & $0.03722\pm0.00029$ & derived \\ 
        Transit depth (dil.) b; $\delta_\mathrm{tr; dil; b; LCO\,zs2m}$ & $0.03722\pm0.00029$ & derived \\ 
        Transit depth (undil.) b; $\delta_\mathrm{tr; undil; b; LCO\,V1m}$ & $0.04186\pm0.00063$ & derived \\ 
        Transit depth (dil.) b; $\delta_\mathrm{tr; dil; b; LCO\,V1m}$ & $0.04186\pm0.00063$ & derived \\ 
        Transit depth (undil.) b; $\delta_\mathrm{tr; undil; b; LCO\,zs1m}$ & $0.03695_{-0.00050}^{+0.00054}$ & derived \\ 
        Transit depth (dil.) b; $\delta_\mathrm{tr; dil; b; LCO\,zs1m}$ & $0.03695_{-0.00050}^{+0.00054}$ & derived \\ 
        Limb darkening; $u_\mathrm{1; TESS\,S8}$ & $0.287_{-0.052}^{+0.056}$ & derived \\ 
        Limb darkening; $u_\mathrm{2; TESS\,S8}$ & $0.213_{-0.053}^{+0.058}$ & derived \\ 
        Limb darkening; $u_\mathrm{1; LCO\,gp2m}$ & $0.614\pm0.058$ & derived \\ 
        Limb darkening; $u_\mathrm{2; LCO\,gp2m}$ & $0.240\pm0.067$ & derived \\ 
        Limb darkening; $u_\mathrm{1; LCO\,ip2m}$ & $0.418\pm0.035$ & derived \\ 
        Limb darkening; $u_\mathrm{2; LCO\,ip2m}$ & $0.200\pm0.049$ & derived \\ 
        Limb darkening; $u_\mathrm{1; LCO\,rp2m}$ & $0.606\pm0.040$ & derived \\ 
        Limb darkening; $u_\mathrm{2; LCO\,rp2m}$ & $0.195\pm0.055$ & derived \\ 
        Limb darkening; $u_\mathrm{1; LCO\,zs2m}$ & $0.293_{-0.034}^{+0.032}$ & derived \\ 
        Limb darkening; $u_\mathrm{2; LCO\,zs2m}$ & $0.234\pm0.050$ & derived \\ 
        Limb darkening; $u_\mathrm{1; LCO\,V1m}$ & $0.586\pm0.065$ & derived \\ 
        Limb darkening; $u_\mathrm{2; LCO\,V1m}$ & $0.262\pm0.072$ & derived \\ 
        Limb darkening; $u_\mathrm{1; LCO\,zs1m}$ & $0.276_{-0.044}^{+0.050}$ & derived \\ 
        Limb darkening; $u_\mathrm{2; LCO\,zs1m}$ & $0.226\pm0.054$ & derived \\ 
        Combined host density from all orbits; $\rho_\mathrm{\star; combined}$ (cgs) & $18.61_{-0.50}^{+0.60}$ & derived \\

    \end{tabular}
    \label{tab:derived_params}
\end{table*}

\section{Discussion and Conclusion}
\label{sec:discussion}

\begin{figure*}
    \centering
    \includegraphics[width=\textwidth]{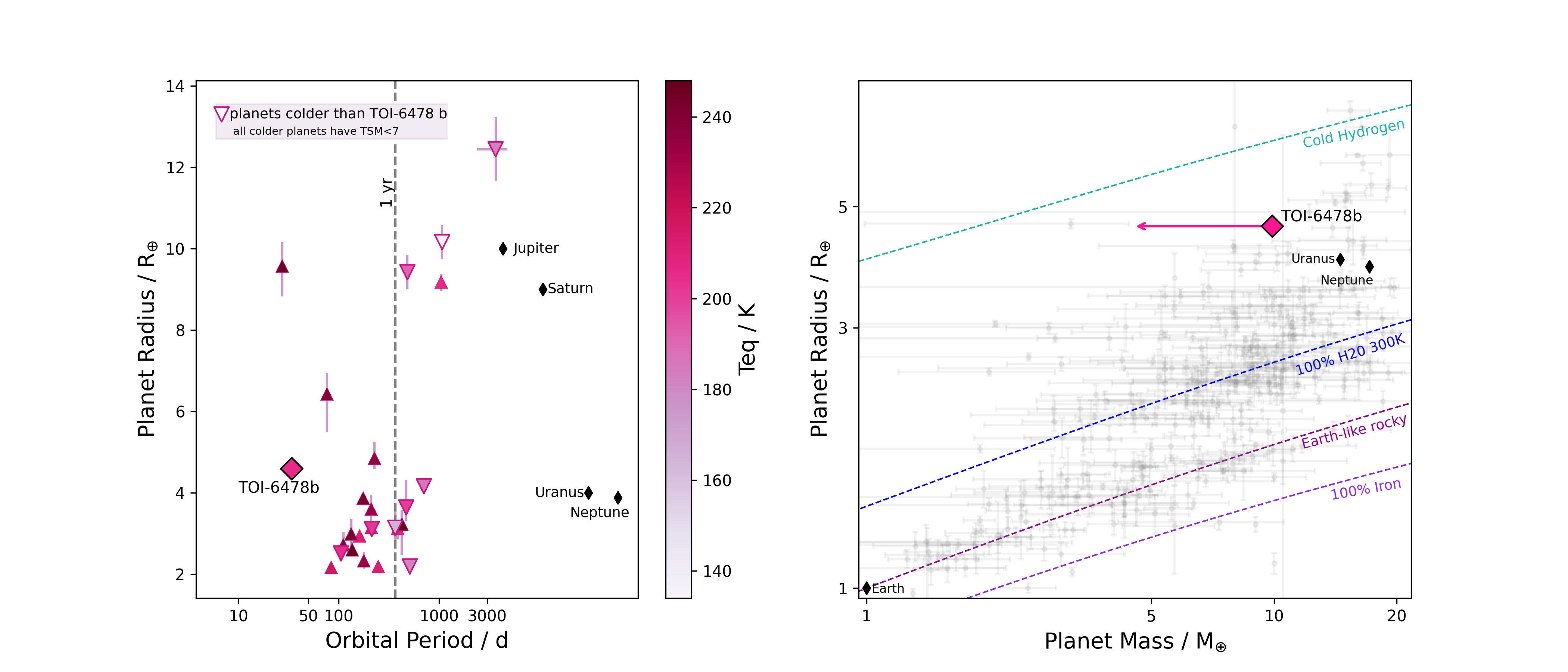}
    \caption{\textcolor{black}{Comparison of TOI-6478\,b (diamond) to the known exoplanet population. \textit{Left}: Planet radius vs orbital period for exoplanets with equilibrium temperatures (color bar) <\,250\,K and radii $>2\,\rm{R}_{\oplus}$. One year is highlighted by the dashed grey line. Of these 25 exoplanets, only 9 are colder than TOI-6478\,b, but all have orbital periods at least $3\times$ larger. Solar systems planets are shown as reference. \textit{Right}: Mass-radius diagram with key mass-radius relations highlighted with coloured dashed lines. TOI-6478\,b is shown as the pink diamond, placed at the upper-mass limit from the current RVs. Solar system planets (black diamonds) are shown as reference.}} 
    \label{fig:mass-rad}
\end{figure*}

\begin{figure}
    \centering
    \includegraphics[width=\linewidth]{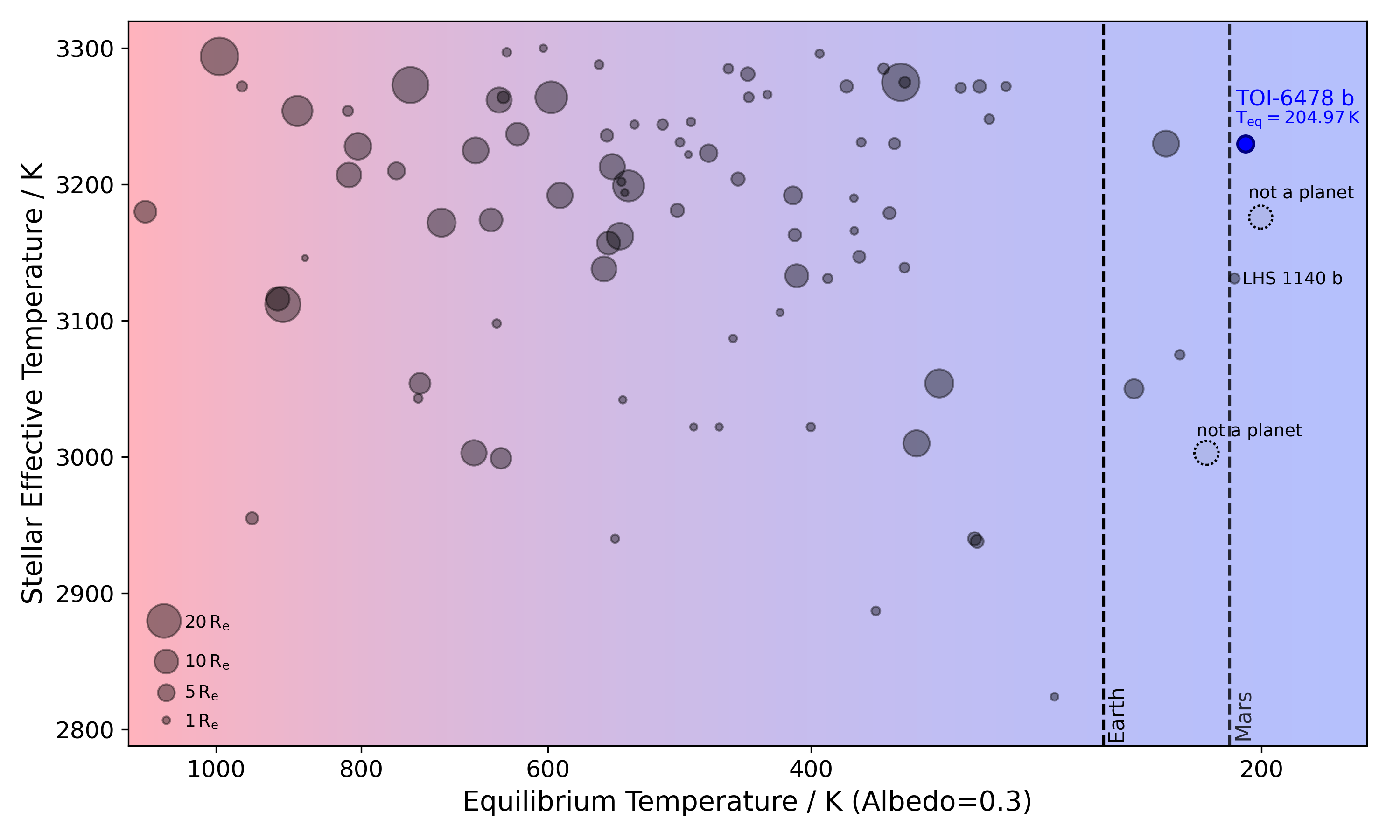}
    \caption{Effective temperature vs equilibrium temperature for all \textit{TESS} Objects of Interest (TOIs) for $T_{\text{eff}}\leq3300\,$K, with points as the relative sizes of the exoplanets. TOIs with non-planetary flags have been removed (False alarm (FA), false positive (FP), nearby eclipsing binary (NEB), background eclipsing binary (BEB), eclipsing binary (EB)). TOI-6478\,b is highlighted in blue. Equilibrium temperatures for Earth and Mars are depicted by the dashed vertical lines. TOIs known not to be planetary in nature (TOI-6508\,b, Barkaoui et al, in prep; TOI-5575\,b, Gan et al, in prep) but have non-updated flags, have been included but highlighted as dashed circles. Well-known rocky super-Earth LHS 1140\,b \citep{lhs1140b} is labelled. TOI-6478\,b is the coldest TOI amongst late-type M dwarfs.}
    \label{fig:tois_temps}
\end{figure}

From the circular fit using two transits from \textit{TESS}, and six same-epoch transits from LCO, we find that TOI-6478\,b hosts a $\sim$\,Neptune-sized planet of radius \textcolor{black}{$R_{\text{b}}=4.6\pm0.24\,\text{R}_\oplus$} ($\sim\,1.2\,\text{R}_{\text{Nep}}$) on a \textcolor{black}{$P_{\mathrm{b}}=34.005\pm0.000025$\,d} orbit. 

We compare the stellar host density, $\rho_\star$, calculated from both orbital dynamics (i.e., a combination of the transit duration, impact parameter and orbital period, assuming a circular orbit) and the stellar parameters (i.e., the mass and radius of the star); we find the derived stellar host density from our fit (which we do not use as a prior) to be \textcolor{black}{$18.6^{+0.6}_{-0.5}$  g\,cm$^{-3}$}, which is within \textcolor{black}{1.5$\sigma$} of the prior value $25.27\pm4.39\,$g\,cm$^{-3}$.

Figure~\ref{fig:chromaticity} shows the measured depths for each photometric band (corrected for limb darkening). These are obtained by doing a separate fit of the data where all bands have a uniform prior on the dilution factor (as for the \textit{TESS} bands in Section~\ref{sec:analysis}) except one ($i'$ here) as to act as a reference. This is to perform a chromaticity check on the measured depths across different wavelengths. We find that the transits are achromatic and agree to $1\sigma$ (except the bluest band $g'$, which agrees to $1.3\sigma$), which is indicative that the transiting events we observe are produced by a planet rather than a star in a binary orbit with TOI-6478, (or equally a nearby or background eclipsing binary).

We note the potential presence of a star spot in the transits from LCO at around \textcolor{black}{2460373.06 BJD} (see Figure~\ref{fig:allesfitter_results}). We removed the spot and redid the analysis as described in Section~\ref{sec:analysis} in order to test whether this was having an effect on the measured depth or duration of the transit. We find statistically consistent results to the original analysis and thus conclude this spot does not influence the fit. While the presence of a star spot presents an opportunity to measure spot properties on a late-type M-dwarf, it does have implications for the atmospheric characterisation and mass measurements of the planet \citep[see e.g.][]{2011IAUS..273..281B, 2015MNRAS.448.2546B, 2016MNRAS.459.3565V}. 

We fit the \textcolor{black}{9 RVs (BJD=2460401.901 data point omitted, see Section~\ref{sec:rvs})} with a NS using \textsc{{allesfitter}} as discussed in Section~\ref{sec:analysis} and from this calculate an upper limit on the semi-amplitude and thus an upper limit on the mass of the planet. \textcolor{black}{We include both the red- and blue-arm RVs in this fit}. From the mass-radius relations described in \citet{2017ApJ...834...17C}, the estimated mass of a $4.6\,\text{R}_\oplus$ planet is $19.2\,\text{M}_\oplus$, which would produce an RV signal of $9.07\,\text{m\,s}^{-1}$ assuming the orbital parameters from our fit. A signal of this amplitude is not seen in our RVs; instead we calculate the upper limit to be $K_\text{fit}+3\sigma$, where $K_\text{fit}$ is the is semi-amplitude result from the NS and $\sigma$ is the uncertainty on this measurement. From this we obtain \textcolor{black}{$K_\text{max}=4.7\,\text{m\,s}^{-1}$} which corresponds to a $3\sigma$ (99\%) upper mass limit of \textcolor{black}{$m_\text{max}=9.9\,\text{M}_\oplus$}. \textcolor{black}{Figure~\ref{fig:mass-rad} \textit{right} shows a mass-radius diagram for confirmed exoplanets (grey) along with key mass-radius relations \citep{zeng2016} (dashed lines) and Solar System planets (black). TOI-6478\,b (pink) is placed at its current upper mass limit, where the pink arrow indicates the range of masses we expect to be its true mass. With decreasing mass, TOI-6478\,b enters an underpopulated region of parameter space, that is of cold, under-dense Neptune-like exoplanets.}

Figure~\ref{fig:tois_temps} shows all the TOIs \citep[TESS Objects of Interest;][]{toi} for stars with $T_{\text{eff}}\leq3300\,$K (as of \textcolor{black}{2024 May 22}). We remove non-planet entries (e.g. false positives, eclipsing binaries etc) and calculate the equilibrium temperatures for each planet assuming an albedo of 0.3 (Earth). While we know this is likely not the albedo value for every planet, we adopt this here for consistency as this parameter is typically not available, especially for planet candidates. We find that TOI-6478\,b is the coldest planet in the TOI sample for late-type M dwarfs. While there is one TOI planet candidate colder - TOI-5575.01\footnote{\url{https://exofop.ipac.caltech.edu/tess/target.php?id=160162137}}, an approximately Jupiter-sized object candidate orbiting a slightly cooler ($T_{\text{eff}}=3176\pm157\,$K) star, we remove this from our Figure~\ref{fig:tois_temps} as observations have revealed since it is not a transiting exoplanet (Gan et al., in prep).

\begin{figure}
    \centering
    \includegraphics[width=\linewidth]{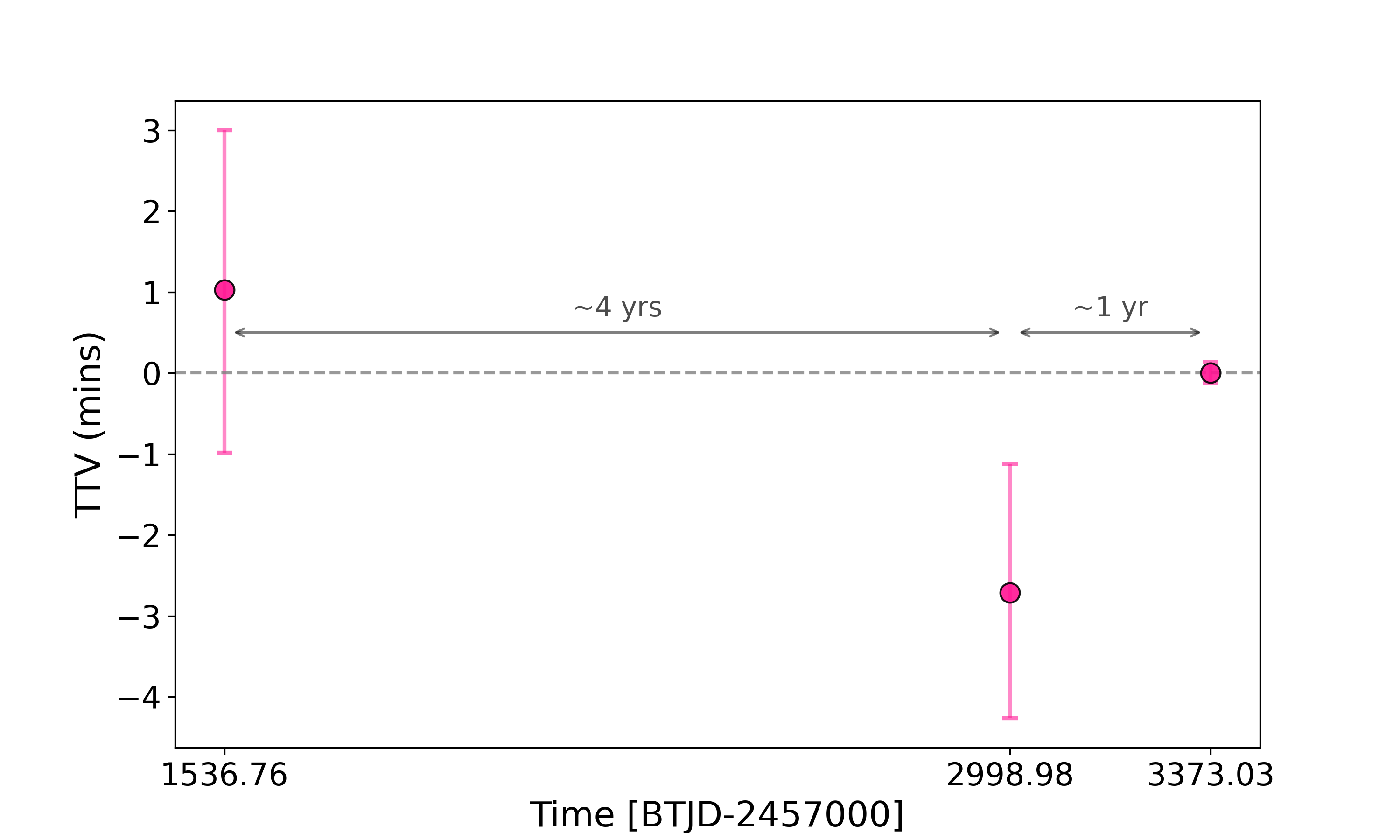}
    \caption{\textcolor{black}{Transit timing variations (TTVs) for TOI-6478\,b from the global photometric analysis of the three transits (\textit{TESS}$\times2$, LCOGT$\times1$ epoch), with approximate time between transits highlighted on the plot. No significant TTVs are detected. Nb: the error on the LCO transit TTV (third point) is small due to the fact that we fit 6 transits at this epoch (4x2m0, 2x1m0).}
    }
    \label{fig:ttvs}
\end{figure}

\textcolor{black}{The method discussed in Section~\ref{sec:analysis} assumes a constant orbital period. In order to check from transit timing variations (TTVs), we repeat same process but now allowing for TTVs in the fit. We use the uniform priors provided for the TTVs from \textsc{allesfitter}, which have upper and lower bounds of $\text{TTV}_{\text{mid}}\pm0.01$\,d (14.4 min). For this fit, we take the $t_0$ and $P_\mathrm{b}$ parameters from the linear fit (i.e. that discussed in Section~\ref{sec:analysis}) and fix them.
TTVs typically occur if another body is present within the system, and have especially large amplitudes if the planets are in a mean motion resonance \citep[MMR, see e.g.][]{trappist1, lammers2024, rivera2010}. 
While the only available photometric data for TOI-6478\,b consists of 3 transits, our fit does not show evidence of significant Transit Timing Variations (TTVs) (see Figure~\ref{fig:ttvs}). Since there is significant time between the 3 transits ($\sim$4 years between transit 1 and 2, and $\sim$ 1 year between transit 2 and 3), thus none are sequential, it is difficult to draw firm conclusions about whether there could be a second object in the system inducing TTVs.}

\subsection{Future Observations}

\subsubsection{Constraining the mass with radial velocities}

With \textcolor{black}{9} RV data points from MAROON-X we were able to compute an upper limit to the mass of TOI-6478\,b to be \textcolor{black}{$M_b\leq9.9\,\text{M}_{\oplus}$}. With the values from our fit, and assuming the orbit is circular, we can calculate the expected semi-amplitude for a given mass. We also attempt to estimate the smallest mass this planet could likely physically have, and find this value is $4.5\,\text{M}_{\oplus}$, based on the current known population of planets in this radius range; planets below $4.5\,\text{M}_{\oplus}$ are typically rocky, whereas a planet of radius $4.6\,\text{R}_\oplus$ is expected to be gaseous. This mass translates to a semi-amplitude signal of $2.12\,\text{m\,s}^{-1}$. Knowing the uncertainty per measurement from MAROON-X to be \textcolor{black}{$2.03\,\rm{m\,s}^{-1}$}, we estimate a further 60 spectra are needed in order to confidently constrain the mass of TOI-6478\,b.

\subsubsection{JWST Atmospheric Characterisation}
TOI-6478\,b represents a convenient opportunity to investigate cool, H$_2$-rich atmospheres. Its cold \textcolor{black}{($T_{\text{eq}}=204.4$\,K)} temperature is rivaled by only nine transiting planets\footnote{Obtained from the composite NASA exoplanet archive \textcolor{black}{(\url{https://exoplanetarchive.ipac.caltech.edu/cgi-bin/TblView/nph-tblView?app=ExoTbls&config=PSCompPars})} with the condition that planets must be transiting.} with radii > 2$\,\text{R}_{\oplus}$, \textcolor{black}{see Figure~\ref{fig:mass-rad} \textit{left}}. However TOI-6478\,b's orbital period is over $3\times$ shorter than the next coldest planet's, Kepler-309\,c \citep{rowe2014}, which has $P=105.3$\,d and $T_{\text{eq}}=204$\,K. The majority of these planets with similar $T_{\text{eq}}$ have $P>1\,\text{yr}$, posing obvious issues for detailed atmospheric characterisation and follow-up observations, whereas TOI-6478\,b transits almost monthly. Additionally, due to the under-dense nature of the planet, its estimated minimum transmission spectroscopy metric \citep[TSM,][]{tsm2018} is \textcolor{black}{$\sim 230$} (based on the mass upper limit of \textcolor{black}{$9.9\,\rm{M}_{\oplus}$}), which is significantly high compared to the other nine known colder transiting planets, which all have TSM < 7\footnote{The TSM is adapted for planets with $R_\text{p}<10\,\text{R}_\oplus$, and two of the cold planets have radii of $R_\text{p}=10.2\,\text{R}_\oplus$ and $R_\text{p}=12.44\,\text{R}_\oplus$. We use the scale factor for planets with $4.0 < R_{\text{p}} < 10\,\text{R}_{\oplus}$ but note that these TSM values may not be fully comparable.}. 

\begin{figure}
    \centering
    \includegraphics[width=\linewidth]{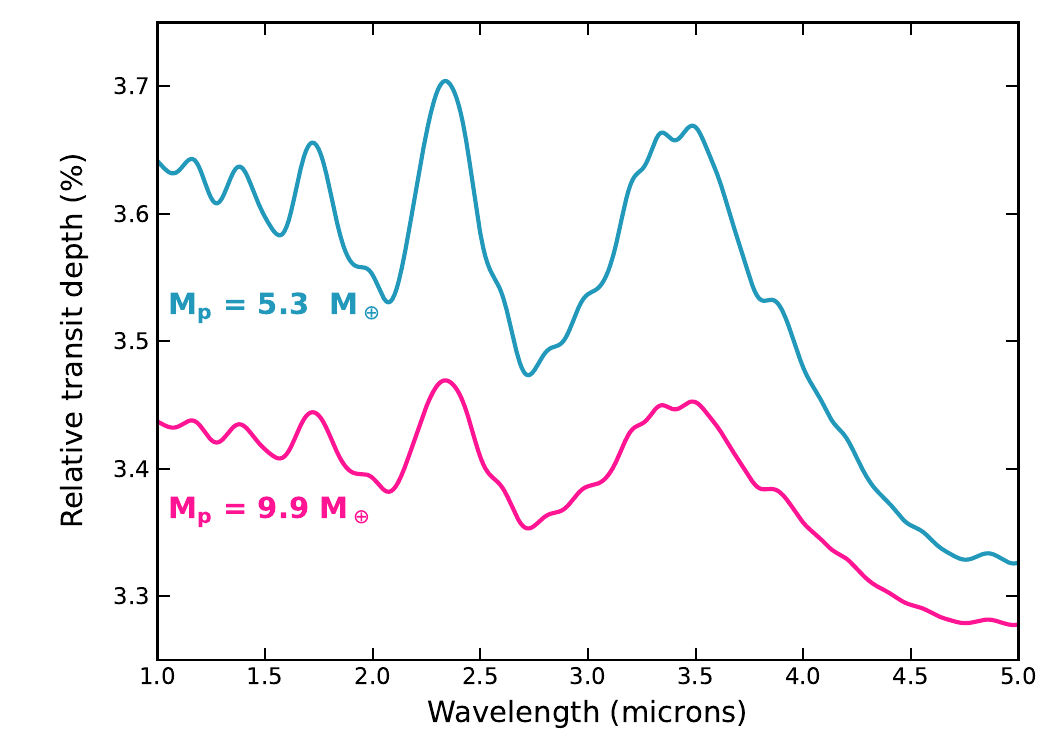}
    \caption{Model transmission spectra for TOI-6478~b corresponding to the upper mass limit \textcolor{black}{(9.9~M$_\oplus$, pink)} and the mass which results in a super-puff density of $0.3\,\text{g}\,\text{cm}^{-3}$ (5.3~M$_\oplus$, blue). Both models assume a 100$\times$ solar atmospheric metallicity, and a simple aerosol prescription equivalent to 10$^4$ times H$_2$ Rayleigh scattering. The spectrum is dominated by CH$_4$ features at 1.4, 1.7, 2.3 and 3.3 $\mu$m.}
    \label{fig:transmission_spec}
\end{figure}

We model the transmission spectrum of TOI-6478~b to demonstrate its observability (Figure \ref{fig:transmission_spec}). The synthetic spectra are generated using the \texttt{Genesis} atmospheric model \citep{Gandhi2017, Piette2020,Piette2023}, coupled with the \texttt{FastChem Cond} equilibrium chemistry code \citep{Stock2018,Kitzmann2024}. We assume a 100$\times$ solar elemental abundance, motivated by the  metallicities of Neptune and Uranus' atmospheres \citep{Atreya2018}, and include the effects of rainout condensation. Given the $\sim$205~K temperature of this atmosphere, condensation is a key process and can significantly deplete the H$_2$O abundance in the upper atmosphere. The remaining atmospheric composition is dominated by H$_2$, He and CH$_4$, with smaller contributions from NH$_3$. Given the prevalence of atmospheric clouds and hazes in Neptune-sized exoplanets \citep[e.g.][]{Brande2024}, we include a simple aerosol prescription corresponding to H$_2$ Rayleigh scattering boosted by a factor of 10$^4$. Despite the high atmospheric metallicity and presence of aerosols, the near-infrared CH$_4$ features in these models are large, spanning several 100s ppm, even in the case of the upper mass limit.

\section*{Acknowledgements}
\textcolor{black}{The authors would like to acknowledge the MAROON-X team for their work on the reduction and analysis of the radial velocity data used in this work. Their thorough data analysis, clear communication and willingness to assist with any inquiries have been greatly appriciated. We are grateful for their collaboration and support.}

Funding for the TESS mission is provided by NASA's Science Mission Directorate. We acknowledge the use of public TESS data from pipelines at the TESS Science Office and at the TESS Science Processing Operations Center. KAC and CNW acknowledge support from the TESS mission via subaward s3449 from MIT.

This research has made use of the Exoplanet Follow-up Observation Program (ExoFOP; DOI: 10.26134/ExoFOP5) website, which is operated by the California Institute of Technology, under contract with the National Aeronautics and Space Administration under the Exoplanet Exploration Program.

This paper made use of data collected by the TESS mission and are publicly available from the Mikulski Archive for Space Telescopes (MAST) operated by the Space Telescope Science Institute (STScI). Resources supporting this work were provided by the NASA High-End Computing (HEC) Program through the NASA Advanced Supercomputing (NAS) Division at Ames Research Center for the production of the SPOC data products.

Some of the observations in this paper made use of the High-Resolution Imaging instrument Zorro and were obtained under Gemini LLP Proposal Number: GN/S-2021A-LP-105. Zorro was funded by the NASA Exoplanet Exploration Program and built at the NASA Ames Research Center by Steve B. Howell, Nic Scott, Elliott P. Horch, and Emmett Quigley. Zorro was mounted on the Gemini South telescope of the international Gemini Observatory, a program of NSF’s OIR Lab, which is managed by the Association of Universities for Research in Astronomy (AURA) under a cooperative agreement with the National Science Foundation. on behalf of the Gemini partnership: the National Science Foundation (United States), National Research Council (Canada), Agencia Nacional de Investigación y Desarrollo (Chile), Ministerio de Ciencia, Tecnología e Innovación (Argentina), Ministério da Ciência, Tecnologia, Inovações e Comunicações (Brazil), and Korea Astronomy and Space Science Institute (Republic of Korea).

This work makes use of observations from the LCOGT network. Part of the LCOGT telescope time was granted by NOIRLab through the Mid-Scale Innovations Program (MSIP). MSIP is funded by NSF.

This paper is based on observations made with the MuSCAT instruments, developed by the Astrobiology Center (ABC) in Japan, the University of Tokyo, and Las Cumbres Observatory (LCOGT). MuSCAT3 was developed with financial support by JSPS KAKENHI (JP18H05439) and JST PRESTO (JPMJPR1775), and is located at the Faulkes Telescope North on Maui, HI (USA), operated by LCOGT. MuSCAT4 was developed with financial support provided by the Heising-Simons Foundation (grant 2022-3611), JST grant number JPMJCR1761, and the ABC in Japan, and is located at the Faulkes Telescope South at Siding Spring Observatory (Australia), operated by LCOGT.
This work is partly supported by JSPS KAKENHI Grant Number JP24H00017, JP24K00689 and JSPS Bilateral Program Number JPJSBP120249910.

The paper is based on observations made with the Kast spectrograph on the Shane 3m telescope at Lick Observatory.
A major upgrade of the Kast spectrograph was made possible through
generous gifts from the Heising-Simons Foundation as well
as William and Marina Kast. 
We acknowledge that Lick Observatory sits on the unceded ancestral homelands of the Chochenyo and Tamyen Ohlone peoples, including the Alson and Socostac tribes, who were the original inhabitants of the area that includes Mt. Hamilton.

Visiting Astronomer at the Infrared Telescope Facility, which is operated by the University of Hawaii under contract 80HQTR24DA010 with the National Aeronautics and Space Administration.

This material is based upon work supported by the National Aeronautics and Space Administration under Agreement No.\ 80NSSC21K0593 for the program ``Alien Earths''.
The results reported herein benefitted from collaborations and/or information exchange within NASA’s Nexus for Exoplanet System Science (NExSS) research coordination network sponsored by NASA’s Science Mission Directorate.

The postdoctoral fellowship of KB is funded by F.R.S.-FNRS grant T.0109.20 and by the Francqui Foundation.

This work was enabled by observations made from the Gemini North telescope, located within the Maunakea Science Reserve and adjacent to the summit of Maunakea. We are grateful for the privilege of observing the Universe from a place that is unique in both its astronomical quality and its cultural significance.
Based on observations obtained at the international Gemini Observatory, a program of NSF NOIRLab, which is managed by the Association of Universities for Research in Astronomy (AURA) under a cooperative agreement with the U.S. National Science Foundation on behalf of the Gemini Observatory partnership: the U.S. National Science Foundation (United States), National Research Council (Canada), Agencia Nacional de Investigaci\'{o}n y Desarrollo (Chile), Ministerio de Ciencia, Tecnolog\'{i}a e Innovaci\'{o}n (Argentina), Minist\'{e}rio da Ci\^{e}ncia, Tecnologia, Inova\c{c}\~{o}es e Comunica\c{c}\~{o}es (Brazil), and Korea Astronomy and Space Science Institute (Republic of Korea). Programme ID: GN-2024A-DD-102.

MGS acknowledges support from the UK Science and Technology Facilities Council (STFC).

For the purpose of open access, the author has applied a Creative Commons Attribution (CC BY) licence to the Author Accepted Manuscript version arising from this submission.

\section*{Data Availability}

\tess data products are available via the MAST portal at \url{https://mast.stsci.edu/portal/Mashup/Clients/Mast/Portal.html}.



\bibliographystyle{mnras}
\bibliography{TOI-6478} 

\begin{thebibliography}{}
\makeatletter
\relax
\def\mn@urlcharsother{\let\do\@makeother \do\$\do\&\do\#\do\^\do\_\do\%\do\~}
\def\mn@doi{\begingroup\mn@urlcharsother \@ifnextchar [ {\mn@doi@} {\mn@doi@[]}}
\def\mn@doi@[#1]#2{\def\@tempa{#1}\ifx\@tempa\@empty \href {http://dx.doi.org/#2} {doi:#2}\else \href {http://dx.doi.org/#2} {#1}\fi \endgroup}
\def\mn@eprint#1#2{\mn@eprint@#1:#2::\@nil}
\def\mn@eprint@arXiv#1{\href {http://arxiv.org/abs/#1} {{\tt arXiv:#1}}}
\def\mn@eprint@dblp#1{\href {http://dblp.uni-trier.de/rec/bibtex/#1.xml} {dblp:#1}}
\def\mn@eprint@#1:#2:#3:#4\@nil{\def\@tempa {#1}\def\@tempb {#2}\def\@tempc {#3}\ifx \@tempc \@empty \let \@tempc \@tempb \let \@tempb \@tempa \fi \ifx \@tempb \@empty \def\@tempb {arXiv}\fi \@ifundefined {mn@eprint@\@tempb}{\@tempb:\@tempc}{\expandafter \expandafter \csname mn@eprint@\@tempb\endcsname \expandafter{\@tempc}}}

\bibitem[\protect\citeauthoryear{{Atreya}, {Crida}, {Guillot}, {Lunine}, {Madhusudhan}  \& {Mousis}}{{Atreya} et~al.}{2018}]{Atreya2018}
{Atreya} S.~K.,  {Crida} A.,  {Guillot} T.,  {Lunine} J.~I.,  {Madhusudhan} N.,   {Mousis} O.,  2018, in {Baines} K.~H.,  {Flasar} F.~M.,  {Krupp} N.,   {Stallard} T.,  eds, , Saturn in the 21st Century.
pp 5--43, \mn@doi{10.1017/9781316227220.002}

\bibitem[\protect\citeauthoryear{{Bailer-Jones}, {Rybizki}, {Fouesneau}, {Demleitner}  \& {Andrae}}{{Bailer-Jones} et~al.}{2021}]{BJdist}
{Bailer-Jones} C.~A.~L.,  {Rybizki} J.,  {Fouesneau} M.,  {Demleitner} M.,   {Andrae} R.,  2021, \mn@doi [\aj] {10.3847/1538-3881/abd806}, \href {https://ui.adsabs.harvard.edu/abs/2021AJ....161..147B} {161, 147}

\bibitem[\protect\citeauthoryear{{Baraffe}, {Homeier}, {Allard}  \& {Chabrier}}{{Baraffe} et~al.}{2015}]{Baraffe2015}
{Baraffe} I.,  {Homeier} D.,  {Allard} F.,   {Chabrier} G.,  2015, \mn@doi [\aap] {10.1051/0004-6361/201425481}, \href {http://adsabs.harvard.edu/abs/2015A%26A...577A..42B} {577, A42}

\bibitem[\protect\citeauthoryear{{Barkaoui} et~al.,}{{Barkaoui} et~al.}{2023}]{Barkaoui2023}
{Barkaoui} K.,  et~al., 2023, \mn@doi [\aap] {10.1051/0004-6361/202346838}, \href {https://ui.adsabs.harvard.edu/abs/2023A&A...677A..38B} {677, A38}

\bibitem[\protect\citeauthoryear{{Barstow}, {Aigrain}, {Irwin}, {Kendrew}  \& {Fletcher}}{{Barstow} et~al.}{2015}]{2015MNRAS.448.2546B}
{Barstow} J.~K.,  {Aigrain} S.,  {Irwin} P.~G.~J.,  {Kendrew} S.,   {Fletcher} L.~N.,  2015, \mn@doi [\mnras] {10.1093/mnras/stv186}, \href {https://ui.adsabs.harvard.edu/abs/2015MNRAS.448.2546B} {448, 2546}

\bibitem[\protect\citeauthoryear{{Benneke} et~al.,}{{Benneke} et~al.}{2019}]{benneke2019}
{Benneke} B.,  et~al., 2019, \mn@doi [\apjl] {10.3847/2041-8213/ab59dc}, \href {https://ui.adsabs.harvard.edu/abs/2019ApJ...887L..14B} {887, L14}

\bibitem[\protect\citeauthoryear{{Bennett} \& {Bovy}}{{Bennett} \& {Bovy}}{2019}]{bennett2019}
{Bennett} M.,  {Bovy} J.,  2019, \mn@doi [\mnras] {10.1093/mnras/sty2813}, \href {https://ui.adsabs.harvard.edu/abs/2019MNRAS.482.1417B} {482, 1417}

\bibitem[\protect\citeauthoryear{{Bensby}, {Feltzing}  \& {Lundstr{\"o}m}}{{Bensby} et~al.}{2003}]{2003AA...410..527B}
{Bensby} T.,  {Feltzing} S.,   {Lundstr{\"o}m} I.,  2003, \mn@doi [\aap] {10.1051/0004-6361:20031213}, \href {http://adsabs.harvard.edu/abs/2003A%26A...410..527B} {410, 527}

\bibitem[\protect\citeauthoryear{{Bochanski}, {West}, {Hawley}  \& {Covey}}{{Bochanski} et~al.}{2007}]{2007AJ....133..531B}
{Bochanski} J.~J.,  {West} A.~A.,  {Hawley} S.~L.,   {Covey} K.~R.,  2007, \mn@doi [\aj] {10.1086/510240}, \href {http://adsabs.harvard.edu/abs/2007AJ....133..531B} {133, 531}

\bibitem[\protect\citeauthoryear{{Boisse}, {Bouchy}, {H{\'e}brard}, {Bonfils}, {Santos}  \& {Vauclair}}{{Boisse} et~al.}{2011}]{2011IAUS..273..281B}
{Boisse} I.,  {Bouchy} F.,  {H{\'e}brard} G.,  {Bonfils} X.,  {Santos} N.,   {Vauclair} S.,  2011, in {Prasad Choudhary} D.,  {Strassmeier} K.~G.,  eds,  IAU Symposium Vol. 273, Physics of Sun and Star Spots. pp 281--285 (\mn@eprint {arXiv} {1012.1452}), \mn@doi{10.1017/S1743921311015389}

\bibitem[\protect\citeauthoryear{{Bovy}}{{Bovy}}{2015a}]{galpy}
{Bovy} J.,  2015a, \mn@doi [\apjs] {10.1088/0067-0049/216/2/29}, \href {https://ui.adsabs.harvard.edu/abs/2015ApJS..216...29B} {216, 29}

\bibitem[\protect\citeauthoryear{{Bovy}}{{Bovy}}{2015b}]{bovy2015}
{Bovy} J.,  2015b, \mn@doi [\apjs] {10.1088/0067-0049/216/2/29}, \href {https://ui.adsabs.harvard.edu/abs/2015ApJS..216...29B} {216, 29}

\bibitem[\protect\citeauthoryear{{Brande} et~al.,}{{Brande} et~al.}{2024}]{Brande2024}
{Brande} J.,  et~al., 2024, \mn@doi [\apjl] {10.3847/2041-8213/ad1b5c}, \href {https://ui.adsabs.harvard.edu/abs/2024ApJ...961L..23B} {961, L23}

\bibitem[\protect\citeauthoryear{{Brown} et~al.,}{{Brown} et~al.}{2013}]{Brown:2013}
{Brown} T.~M.,  et~al., 2013, \mn@doi [\pasp] {10.1086/673168}, \href {https://ui.adsabs.harvard.edu/abs/2013PASP..125.1031B} {125, 1031}

\bibitem[\protect\citeauthoryear{{Burgasser} \& {Splat Development Team}}{{Burgasser} \& {Splat Development Team}}{2017}]{splat}
{Burgasser} A.~J.,  {Splat Development Team} 2017, in Astronomical Society of India Conference Series. pp 7--12 (\mn@eprint {arXiv} {1707.00062})

\bibitem[\protect\citeauthoryear{{Casagrande}, {Flynn}  \& {Bessell}}{{Casagrande} et~al.}{2008}]{2008MNRAS.389..585C}
{Casagrande} L.,  {Flynn} C.,   {Bessell} M.,  2008, \mn@doi [\mnras] {10.1111/j.1365-2966.2008.13573.x}, \href {https://ui.adsabs.harvard.edu/abs/2008MNRAS.389..585C} {389, 585}

\bibitem[\protect\citeauthoryear{{Chabrier}}{{Chabrier}}{2003}]{chabrier2003}
{Chabrier} G.,  2003, \mn@doi [\pasp] {10.1086/376392}, \href {https://ui.adsabs.harvard.edu/abs/2003PASP..115..763C} {115, 763}

\bibitem[\protect\citeauthoryear{{Chen} \& {Kipping}}{{Chen} \& {Kipping}}{2017}]{2017ApJ...834...17C}
{Chen} J.,  {Kipping} D.,  2017, \mn@doi [\apj] {10.3847/1538-4357/834/1/17}, \href {https://ui.adsabs.harvard.edu/abs/2017ApJ...834...17C} {834, 17}

\bibitem[\protect\citeauthoryear{{Ciardi}, {Beichman}, {Horch}  \& {Howell}}{{Ciardi} et~al.}{2015}]{2015ApJ...805...16C}
{Ciardi} D.~R.,  {Beichman} C.~A.,  {Horch} E.~P.,   {Howell} S.~B.,  2015, \mn@doi [\apj] {10.1088/0004-637X/805/1/16}, \href {https://ui.adsabs.harvard.edu/abs/2015ApJ...805...16C} {805, 16}

\bibitem[\protect\citeauthoryear{{Cochran} et~al.,}{{Cochran} et~al.}{2011}]{2011ApJS..197....7C}
{Cochran} W.~D.,  et~al., 2011, \mn@doi [\apjs] {10.1088/0067-0049/197/1/7}, \href {https://ui.adsabs.harvard.edu/abs/2011ApJS..197....7C} {197, 7}

\bibitem[\protect\citeauthoryear{{Collins}}{{Collins}}{2019}]{collins2019}
{Collins} K.,  2019, in American Astronomical Society Meeting Abstracts \#233. p. 140.05

\bibitem[\protect\citeauthoryear{{Collins}, {Kielkopf}, {Stassun}  \& {Hessman}}{{Collins} et~al.}{2017}]{Collins:2017}
{Collins} K.~A.,  {Kielkopf} J.~F.,  {Stassun} K.~G.,   {Hessman} F.~V.,  2017, \mn@doi [\aj] {10.3847/1538-3881/153/2/77}, \href {http://adsabs.harvard.edu/abs/2017AJ....153...77C} {153, 77}

\bibitem[\protect\citeauthoryear{{Cushing}, {Vacca}  \& {Rayner}}{{Cushing} et~al.}{2004}]{Cushing2004}
{Cushing} M.~C.,  {Vacca} W.~D.,   {Rayner} J.~T.,  2004, \mn@doi [\pasp] {10.1086/382907}, \href {https://ui.adsabs.harvard.edu/abs/2004PASP..116..362C} {116, 362}

\bibitem[\protect\citeauthoryear{{Cushing}, {Rayner}  \& {Vacca}}{{Cushing} et~al.}{2005}]{Cushing2005}
{Cushing} M.~C.,  {Rayner} J.~T.,   {Vacca} W.~D.,  2005, \mn@doi [\apj] {10.1086/428040}, \href {https://ui.adsabs.harvard.edu/abs/2005ApJ...623.1115C} {623, 1115}

\bibitem[\protect\citeauthoryear{{Cutri} et~al.,}{{Cutri} et~al.}{2003}]{2masscat}
{Cutri} R.~M.,  et~al., 2003, VizieR Online Data Catalog, \href {https://ui.adsabs.harvard.edu/abs/2003yCat.2246....0C} {p. II/246}

\bibitem[\protect\citeauthoryear{{Cutri} et~al.,}{{Cutri} et~al.}{2021}]{wisecat}
{Cutri} R.~M.,  et~al., 2021, VizieR Online Data Catalog, \href {https://ui.adsabs.harvard.edu/abs/2014yCat.2328....0C} {p. II/328}

\bibitem[\protect\citeauthoryear{{Davis} et~al.,}{{Davis} et~al.}{2024}]{davis2024}
{Davis} Y.~T.,  et~al., 2024, \mn@doi [\mnras] {10.1093/mnras/stae842}, \href {https://ui.adsabs.harvard.edu/abs/2024MNRAS.530.2565D} {530, 2565}

\bibitem[\protect\citeauthoryear{{Delrez} et~al.,}{{Delrez} et~al.}{2022}]{Delrez2022}
{Delrez} L.,  et~al., 2022, \mn@doi [\aap] {10.1051/0004-6361/202244041}, \href {https://ui.adsabs.harvard.edu/abs/2022A&A...667A..59D} {667, A59}

\bibitem[\protect\citeauthoryear{{Dittmann} et~al.,}{{Dittmann} et~al.}{2017}]{lhs1140b}
{Dittmann} J.~A.,  et~al., 2017, \mn@doi [\nat] {10.1038/nature22055}, \href {https://ui.adsabs.harvard.edu/abs/2017Natur.544..333D} {544, 333}

\bibitem[\protect\citeauthoryear{{Duck} et~al.,}{{Duck} et~al.}{2023}]{duck2023}
{Duck} A.,  et~al., 2023, \mn@doi [\mnras] {10.1093/mnras/stad452}, \href {https://ui.adsabs.harvard.edu/abs/2023MNRAS.521.6305D} {521, 6305}

\bibitem[\protect\citeauthoryear{{Eisner} et~al.,}{{Eisner} et~al.}{2021}]{eisner2021}
{Eisner} N.~L.,  et~al., 2021, \mn@doi [\mnras] {10.1093/mnras/staa3739}, \href {https://ui.adsabs.harvard.edu/abs/2021MNRAS.501.4669E} {501, 4669}

\bibitem[\protect\citeauthoryear{{Foreman-Mackey}, {Hogg}, {Lang}  \& {Goodman}}{{Foreman-Mackey} et~al.}{2013}]{emcee}
{Foreman-Mackey} D.,  {Hogg} D.~W.,  {Lang} D.,   {Goodman} J.,  2013, \mn@doi [\pasp] {10.1086/670067}, \href {https://ui.adsabs.harvard.edu/abs/2013PASP..125..306F} {125, 306}

\bibitem[\protect\citeauthoryear{{Foreman-Mackey}, {Agol}, {Ambikasaran}  \& {Angus}}{{Foreman-Mackey} et~al.}{2017}]{celerite}
{Foreman-Mackey} D.,  {Agol} E.,  {Ambikasaran} S.,   {Angus} R.,  2017, {celerite: Scalable 1D Gaussian Processes in C++, Python, and Julia} (\mn@eprint {ascl} {1709.008})

\bibitem[\protect\citeauthoryear{{Furlan} \& {Howell}}{{Furlan} \& {Howell}}{2017}]{2017AJ....154...66F}
{Furlan} E.,  {Howell} S.~B.,  2017, \mn@doi [\aj] {10.3847/1538-3881/aa7b70}, \href {https://ui.adsabs.harvard.edu/abs/2017AJ....154...66F} {154, 66}

\bibitem[\protect\citeauthoryear{{Furlan} \& {Howell}}{{Furlan} \& {Howell}}{2020}]{2020ApJ...898...47F}
{Furlan} E.,  {Howell} S.~B.,  2020, \mn@doi [\apj] {10.3847/1538-4357/ab9c9c}, \href {https://ui.adsabs.harvard.edu/abs/2020ApJ...898...47F} {898, 47}

\bibitem[\protect\citeauthoryear{{Gaia Collaboration}}{{Gaia Collaboration}}{2022}]{gaiaDR3cat}
{Gaia Collaboration} 2022, VizieR Online Data Catalog, \href {https://ui.adsabs.harvard.edu/abs/2022yCat.1355....0G} {p. I/355}

\bibitem[\protect\citeauthoryear{{Gaia Collaboration} et~al.,}{{Gaia Collaboration} et~al.}{2016}]{gaia2016}
{Gaia Collaboration} et~al., 2016, \mn@doi [\aap] {10.1051/0004-6361/201629272}, \href {https://ui.adsabs.harvard.edu/abs/2016A&A...595A...1G} {595, A1}

\bibitem[\protect\citeauthoryear{{Gaia Collaboration} et~al.,}{{Gaia Collaboration} et~al.}{2021}]{2021AA...649A...6G}
{Gaia Collaboration} et~al., 2021, \mn@doi [\aap] {10.1051/0004-6361/202039498}, \href {https://ui.adsabs.harvard.edu/abs/2021A&A...649A...6G} {649, A6}

\bibitem[\protect\citeauthoryear{{Gaia Collaboration} et~al.,}{{Gaia Collaboration} et~al.}{2023}]{gaiadr3}
{Gaia Collaboration} et~al., 2023, \mn@doi [\aap] {10.1051/0004-6361/202243940}, \href {https://ui.adsabs.harvard.edu/abs/2023A&A...674A...1G} {674, A1}

\bibitem[\protect\citeauthoryear{{Gandhi} \& {Madhusudhan}}{{Gandhi} \& {Madhusudhan}}{2017}]{Gandhi2017}
{Gandhi} S.,  {Madhusudhan} N.,  2017, \mn@doi [\mnras] {10.1093/mnras/stx1601}, \href {https://ui.adsabs.harvard.edu/abs/2017MNRAS.472.2334G} {472, 2334}

\bibitem[\protect\citeauthoryear{{Gao} \& {Zhang}}{{Gao} \& {Zhang}}{2020}]{gao&zhang2020}
{Gao} P.,  {Zhang} X.,  2020, \mn@doi [\apj] {10.3847/1538-4357/ab6a9b}, \href {https://ui.adsabs.harvard.edu/abs/2020ApJ...890...93G} {890, 93}

\bibitem[\protect\citeauthoryear{{Ghachoui} et~al.,}{{Ghachoui} et~al.}{2023}]{Ghachoui2023}
{Ghachoui} M.,  et~al., 2023, \mn@doi [\aap] {10.1051/0004-6361/202347040}, \href {https://ui.adsabs.harvard.edu/abs/2023A&A...677A..31G} {677, A31}

\bibitem[\protect\citeauthoryear{{Giacalone} \& {Dressing}}{{Giacalone} \& {Dressing}}{2020}]{2020ascl.soft02004G}
{Giacalone} S.,  {Dressing} C.~D.,  2020, {triceratops: Candidate exoplanet rating tool}, Astrophysics Source Code Library, record ascl:2002.004

\bibitem[\protect\citeauthoryear{{Giacalone} et~al.,}{{Giacalone} et~al.}{2021}]{triceratops}
{Giacalone} S.,  et~al., 2021, \mn@doi [\aj] {10.3847/1538-3881/abc6af}, \href {https://ui.adsabs.harvard.edu/abs/2021AJ....161...24G} {161, 24}

\bibitem[\protect\citeauthoryear{{Gillon} et~al.,}{{Gillon} et~al.}{2017}]{trappist1}
{Gillon} M.,  et~al., 2017, \mn@doi [\nat] {10.1038/nature21360}, \href {https://ui.adsabs.harvard.edu/abs/2017Natur.542..456G} {542, 456}

\bibitem[\protect\citeauthoryear{{Gillon} et~al.,}{{Gillon} et~al.}{2024}]{Gillon2024}
{Gillon} M.,  et~al., 2024, \mn@doi [Nature Astronomy] {10.1038/s41550-024-02271-2}, \href {https://ui.adsabs.harvard.edu/abs/2024NatAs...8..865G} {8, 865}

\bibitem[\protect\citeauthoryear{{Gravity Collaboration} et~al.,}{{Gravity Collaboration} et~al.}{2019}]{gravity2019}
{Gravity Collaboration} et~al., 2019, \mn@doi [\aap] {10.1051/0004-6361/201935656}, \href {https://ui.adsabs.harvard.edu/abs/2019A&A...625L..10G} {625, L10}

\bibitem[\protect\citeauthoryear{{Guerrero} et~al.,}{{Guerrero} et~al.}{2021a}]{guerrero2021}
{Guerrero} N.~M.,  et~al., 2021a, \mn@doi [\apjs] {10.3847/1538-4365/abefe1}, \href {https://ui.adsabs.harvard.edu/abs/2021ApJS..254...39G} {254, 39}

\bibitem[\protect\citeauthoryear{{Guerrero} et~al.,}{{Guerrero} et~al.}{2021b}]{toi}
{Guerrero} N.~M.,  et~al., 2021b, \mn@doi [\apjs] {10.3847/1538-4365/abefe1}, \href {https://ui.adsabs.harvard.edu/abs/2021ApJS..254...39G} {254, 39}

\bibitem[\protect\citeauthoryear{{G{\"u}nther} \& {Daylan}}{{G{\"u}nther} \& {Daylan}}{2019}]{allesfitter-code}
{G{\"u}nther} M.~N.,  {Daylan} T.,  2019, {Allesfitter: Flexible Star and Exoplanet Inference From Photometry and Radial Velocity}, Astrophysics Source Code Library (\mn@eprint {ascl} {1903.003})

\bibitem[\protect\citeauthoryear{{G{\"u}nther} \& {Daylan}}{{G{\"u}nther} \& {Daylan}}{2021}]{allesfitter-paper}
{G{\"u}nther} M.~N.,  {Daylan} T.,  2021, \mn@doi [\apjs] {10.3847/1538-4365/abe70e}, \href {https://ui.adsabs.harvard.edu/abs/2021ApJS..254...13G} {254, 13}

\bibitem[\protect\citeauthoryear{{Hallatt} \& {Lee}}{{Hallatt} \& {Lee}}{2024}]{hallatt2024arXiv240809319H}
{Hallatt} T.,  {Lee} E.~J.,  2024, \mn@doi [arXiv e-prints] {10.48550/arXiv.2408.09319}, \href {https://ui.adsabs.harvard.edu/abs/2024arXiv240809319H} {p. arXiv:2408.09319}

\bibitem[\protect\citeauthoryear{{Hardegree-Ullman}, {Apai}, {Bergsten}, {Pascucci}  \& {L{\'o}pez-Morales}}{{Hardegree-Ullman} et~al.}{2023}]{2023AJ....165..267H}
{Hardegree-Ullman} K.~K.,  {Apai} D.,  {Bergsten} G.~J.,  {Pascucci} I.,   {L{\'o}pez-Morales} M.,  2023, \mn@doi [\aj] {10.3847/1538-3881/acd1ec}, \href {https://ui.adsabs.harvard.edu/abs/2023AJ....165..267H} {165, 267}

\bibitem[\protect\citeauthoryear{Henry, Jao, Subasavage, Beaulieu, Ianna, Costa  \& Méndez}{Henry et~al.}{2006}]{henry2006}
Henry T.~J.,  Jao W.,  Subasavage J.~P.,  Beaulieu T.~D.,  Ianna P.~A.,  Costa E.,   Méndez R.~A.,  2006, \mn@doi [The Astronomical Journal] {10.1086/508233}, 132, 2360–2371

\bibitem[\protect\citeauthoryear{{Howell}, {Everett}, {Sherry}, {Horch}  \& {Ciardi}}{{Howell} et~al.}{2011}]{2011AJ....142...19H}
{Howell} S.~B.,  {Everett} M.~E.,  {Sherry} W.,  {Horch} E.,   {Ciardi} D.~R.,  2011, \mn@doi [\aj] {10.1088/0004-6256/142/1/19}, \href {https://ui.adsabs.harvard.edu/abs/2011AJ....142...19H} {142, 19}

\bibitem[\protect\citeauthoryear{{Howell}, {Everett}, {Horch}, {Winters}, {Hirsch}, {Nusdeo}  \& {Scott}}{{Howell} et~al.}{2016}]{2016ApJ...829L...2H}
{Howell} S.~B.,  {Everett} M.~E.,  {Horch} E.~P.,  {Winters} J.~G.,  {Hirsch} L.,  {Nusdeo} D.,   {Scott} N.~J.,  2016, \mn@doi [\apjl] {10.3847/2041-8205/829/1/L2}, \href {https://ui.adsabs.harvard.edu/abs/2016ApJ...829L...2H} {829, L2}

\bibitem[\protect\citeauthoryear{{Huang} et~al.,}{{Huang} et~al.}{2020a}]{QLP2020a}
{Huang} C.~X.,  et~al., 2020a, \mn@doi [Research Notes of the American Astronomical Society] {10.3847/2515-5172/abca2e}, \href {https://ui.adsabs.harvard.edu/abs/2020RNAAS...4..204H} {4, 204}

\bibitem[\protect\citeauthoryear{{Huang} et~al.,}{{Huang} et~al.}{2020b}]{QLP2020b}
{Huang} C.~X.,  et~al., 2020b, \mn@doi [Research Notes of the American Astronomical Society] {10.3847/2515-5172/abca2d}, \href {https://ui.adsabs.harvard.edu/abs/2020RNAAS...4..206H} {4, 206}

\bibitem[\protect\citeauthoryear{{Husser}, {Wende-von Berg}, {Dreizler}, {Homeier}, {Reiners}, {Barman}  \& {Hauschildt}}{{Husser} et~al.}{2013a}]{Husser:2013}
{Husser} T.~O.,  {Wende-von Berg} S.,  {Dreizler} S.,  {Homeier} D.,  {Reiners} A.,  {Barman} T.,   {Hauschildt} P.~H.,  2013a, \mn@doi [\aap] {10.1051/0004-6361/201219058}, \href {https://ui.adsabs.harvard.edu/abs/2013A&A...553A...6H} {553, A6}

\bibitem[\protect\citeauthoryear{{Husser}, {Wende-von Berg}, {Dreizler}, {Homeier}, {Reiners}, {Barman}  \& {Hauschildt}}{{Husser} et~al.}{2013b}]{phoenix}
{Husser} T.~O.,  {Wende-von Berg} S.,  {Dreizler} S.,  {Homeier} D.,  {Reiners} A.,  {Barman} T.,   {Hauschildt} P.~H.,  2013b, \mn@doi [\aap] {10.1051/0004-6361/201219058}, \href {https://ui.adsabs.harvard.edu/abs/2013AAA...553A...6H} {553, A6}

\bibitem[\protect\citeauthoryear{{Jenkins}}{{Jenkins}}{2002}]{jenkins2002}
{Jenkins} J.~M.,  2002, \mn@doi [\apj] {10.1086/341136}, \href {http://adsabs.harvard.edu/abs/2002ApJ...575..493J} {575, 493}

\bibitem[\protect\citeauthoryear{{Jenkins} et~al.,}{{Jenkins} et~al.}{2010}]{jenkins2010}
{Jenkins} J.~M.,  et~al., 2010, in {Radziwill} N.~M.,  {Bridger} A.,  eds,  Society of Photo-Optical Instrumentation Engineers (SPIE) Conference Series Vol. 7740, Software and Cyberinfrastructure for Astronomy. p. 77400D, \mn@doi{10.1117/12.856764}

\bibitem[\protect\citeauthoryear{{Jenkins} et~al.,}{{Jenkins} et~al.}{2016}]{SPOC}
{Jenkins} J.~M.,  et~al., 2016, in {Chiozzi} G.,  {Guzman} J.~C.,  eds,  Society of Photo-Optical Instrumentation Engineers (SPIE) Conference Series Vol. 9913, Software and Cyberinfrastructure for Astronomy IV. p. 99133E, \mn@doi{10.1117/12.2233418}

\bibitem[\protect\citeauthoryear{{Jenkins}, {Tenenbaum}, {Seader}, {Burke}, {McCauliff}, {Smith}, {Twicken}  \& {Chandrasekaran}}{{Jenkins} et~al.}{2020}]{jenkins2020}
{Jenkins} J.~M.,  {Tenenbaum} P.,  {Seader} S.,  {Burke} C.~J.,  {McCauliff} S.~D.,  {Smith} J.~C.,  {Twicken} J.~D.,   {Chandrasekaran} H.,  2020, {Kepler Data Processing Handbook: Transiting Planet Search}, Kepler Science Document KSCI-19081-003

\bibitem[\protect\citeauthoryear{{Jensen}}{{Jensen}}{2013}]{Jensen:2013}
{Jensen} E.,  2013, {Tapir: A web interface for transit/eclipse observability}, Astrophysics Source Code Library (\mn@eprint {ascl} {1306.007})

\bibitem[\protect\citeauthoryear{{Kanodia} et~al.,}{{Kanodia} et~al.}{2024}]{kanodia2024}
{Kanodia} S.,  et~al., 2024, \mn@doi [\aj] {10.3847/1538-3881/ad7796}, \href {https://ui.adsabs.harvard.edu/abs/2024AJ....168..235K} {168, 235}

\bibitem[\protect\citeauthoryear{Kass \& Raftery}{Kass \& Raftery}{1995}]{BayesFactor}
Kass R.~E.,  Raftery A.~E.,  1995, \mn@doi [Journal of the American Statistical Association] {10.1080/01621459.1995.10476572}, 90, 773

\bibitem[\protect\citeauthoryear{{Kempton} et~al.,}{{Kempton} et~al.}{2018}]{tsm2018}
{Kempton} E. M.~R.,  et~al., 2018, \mn@doi [\pasp] {10.1088/1538-3873/aadf6f}, \href {https://ui.adsabs.harvard.edu/abs/2018PASP..130k4401K} {130, 114401}

\bibitem[\protect\citeauthoryear{{Kempton} et~al.,}{{Kempton} et~al.}{2023}]{kempton2023}
{Kempton} E. M.~R.,  et~al., 2023, \mn@doi [\nat] {10.1038/s41586-023-06159-5}, \href {https://ui.adsabs.harvard.edu/abs/2023Natur.620...67K} {620, 67}

\bibitem[\protect\citeauthoryear{{Kesseli}, {Muirhead}, {Mann}  \& {Mace}}{{Kesseli} et~al.}{2018}]{2018AJ....155..225K}
{Kesseli} A.~Y.,  {Muirhead} P.~S.,  {Mann} A.~W.,   {Mace} G.,  2018, \mn@doi [\aj] {10.3847/1538-3881/aabccb}, \href {https://ui.adsabs.harvard.edu/abs/2018AJ....155..225K} {155, 225}

\bibitem[\protect\citeauthoryear{{Kipping}}{{Kipping}}{2013}]{kippingldcs}
{Kipping} D.~M.,  2013, \mn@doi [\mnras] {10.1093/mnras/stt1435}, \href {https://ui.adsabs.harvard.edu/abs/2013MNRAS.435.2152K} {435, 2152}

\bibitem[\protect\citeauthoryear{{Kitzmann}, {Stock}  \& {Patzer}}{{Kitzmann} et~al.}{2024}]{Kitzmann2024}
{Kitzmann} D.,  {Stock} J.~W.,   {Patzer} A. B.~C.,  2024, \mn@doi [\mnras] {10.1093/mnras/stad3515}, \href {https://ui.adsabs.harvard.edu/abs/2024MNRAS.527.7263K} {527, 7263}

\bibitem[\protect\citeauthoryear{{Kunimoto} \& {Daylan}}{{Kunimoto} \& {Daylan}}{2021}]{kunimoto2021}
{Kunimoto} M.,  {Daylan} T.,  2021, in Posters from the TESS Science Conference II (TSC2). p.~62, \mn@doi{10.5281/zenodo.5125527}

\bibitem[\protect\citeauthoryear{{Lammers} \& {Winn}}{{Lammers} \& {Winn}}{2024}]{lammers2024}
{Lammers} C.,  {Winn} J.~N.,  2024, \mn@doi [\apjl] {10.3847/2041-8213/ad50d2}, \href {https://ui.adsabs.harvard.edu/abs/2024ApJ...968L..12L} {968, L12}

\bibitem[\protect\citeauthoryear{{Lasker}, {Doggett}, {McLean}, {Sturch}, {Djorgovski}, {de Carvalho}  \& {Reid}}{{Lasker} et~al.}{1996}]{lasker1996}
{Lasker} B.~M.,  {Doggett} J.,  {McLean} B.,  {Sturch} C.,  {Djorgovski} S.,  {de Carvalho} R.~R.,   {Reid} I.~N.,  1996, in {Jacoby} G.~H.,  {Barnes} J.,  eds,  Astronomical Society of the Pacific Conference Series Vol. 101, Astronomical Data Analysis Software and Systems V. p.~88

\bibitem[\protect\citeauthoryear{{L{\'e}pine}, {Rich}  \& {Shara}}{{L{\'e}pine} et~al.}{2003}]{2003AJ....125.1598L}
{L{\'e}pine} S.,  {Rich} R.~M.,   {Shara} M.~M.,  2003, \mn@doi [\aj] {10.1086/345972}, \href {http://ads.ari.uni-heidelberg.de/abs/2003AJ....125.1598L} {125, 1598}

\bibitem[\protect\citeauthoryear{{L{\'e}pine}, {Rich}  \& {Shara}}{{L{\'e}pine} et~al.}{2007}]{2007ApJ...669.1235L}
{L{\'e}pine} S.,  {Rich} R.~M.,   {Shara} M.~M.,  2007, \mn@doi [\apj] {10.1086/521614}, \href {http://ads.ari.uni-heidelberg.de/abs/2007ApJ...669.1235L} {669, 1235}

\bibitem[\protect\citeauthoryear{{L{\'e}pine}, {Hilton}, {Mann}, {Wilde}, {Rojas-Ayala}, {Cruz}  \& {Gaidos}}{{L{\'e}pine} et~al.}{2013}]{2013AJ....145..102L}
{L{\'e}pine} S.,  {Hilton} E.~J.,  {Mann} A.~W.,  {Wilde} M.,  {Rojas-Ayala} B.,  {Cruz} K.~L.,   {Gaidos} E.,  2013, \mn@doi [\aj] {10.1088/0004-6256/145/4/102}, \href {http://adsabs.harvard.edu/abs/2013AJ....145..102L} {145, 102}

\bibitem[\protect\citeauthoryear{{Lester} et~al.,}{{Lester} et~al.}{2021}]{2021AJ....162...75L}
{Lester} K.~V.,  et~al., 2021, \mn@doi [\aj] {10.3847/1538-3881/ac0d06}, \href {https://ui.adsabs.harvard.edu/abs/2021AJ....162...75L} {162, 75}

\bibitem[\protect\citeauthoryear{{Li}, {Tenenbaum}, {Twicken}, {Burke}, {Jenkins}, {Quintana}, {Rowe}  \& {Seader}}{{Li} et~al.}{2019}]{li2019}
{Li} J.,  {Tenenbaum} P.,  {Twicken} J.~D.,  {Burke} C.~J.,  {Jenkins} J.~M.,  {Quintana} E.~V.,  {Rowe} J.~F.,   {Seader} S.~E.,  2019, \mn@doi [\pasp] {10.1088/1538-3873/aaf44d}, \href {https://ui.adsabs.harvard.edu/abs/2019PASP..131b4506L} {131, 024506}

\bibitem[\protect\citeauthoryear{{Madhusudhan}, {Nixon}, {Welbanks}, {Piette}  \& {Booth}}{{Madhusudhan} et~al.}{2020}]{madhu2020}
{Madhusudhan} N.,  {Nixon} M.~C.,  {Welbanks} L.,  {Piette} A. A.~A.,   {Booth} R.~A.,  2020, \mn@doi [\apjl] {10.3847/2041-8213/ab7229}, \href {https://ui.adsabs.harvard.edu/abs/2020ApJ...891L...7M} {891, L7}

\bibitem[\protect\citeauthoryear{{Mann}, {Brewer}, {Gaidos}, {L{\'e}pine}  \& {Hilton}}{{Mann} et~al.}{2013}]{Mann2013}
{Mann} A.~W.,  {Brewer} J.~M.,  {Gaidos} E.,  {L{\'e}pine} S.,   {Hilton} E.~J.,  2013, \mn@doi [\aj] {10.1088/0004-6256/145/2/52}, \href {https://ui.adsabs.harvard.edu/abs/2013AJ....145...52M} {145, 52}

\bibitem[\protect\citeauthoryear{{Mann}, {Feiden}, {Gaidos}, {Boyajian}  \& {von Braun}}{{Mann} et~al.}{2016}]{Mann:2016}
{Mann} A.~W.,  {Feiden} G.~A.,  {Gaidos} E.,  {Boyajian} T.,   {von Braun} K.,  2016, \mn@doi [\apj] {10.3847/0004-637X/819/1/87}, \href {https://ui.adsabs.harvard.edu/abs/2016ApJ...819...87M} {819, 87}

\bibitem[\protect\citeauthoryear{{Martioli} et~al.,}{{Martioli} et~al.}{2024}]{martioli2024}
{Martioli} E.,  et~al., 2024, \mn@doi [\aap] {10.1051/0004-6361/202450334}, \href {https://ui.adsabs.harvard.edu/abs/2024A&A...690A.312M} {690, A312}

\bibitem[\protect\citeauthoryear{{Massey} \& {Gronwall}}{{Massey} \& {Gronwall}}{1990}]{1990ApJ...358..344M}
{Massey} P.,  {Gronwall} C.,  1990, \mn@doi [\apj] {10.1086/168991}, \href {https://ui.adsabs.harvard.edu/abs/1990ApJ...358..344M} {358, 344}

\bibitem[\protect\citeauthoryear{{Maxted}}{{Maxted}}{2016}]{ellc}
{Maxted} P.~F.~L.,  2016, \mn@doi [\aap] {10.1051/0004-6361/201628579}, \href {https://ui.adsabs.harvard.edu/abs/2016AAA...591A.111M} {591, A111}

\bibitem[\protect\citeauthoryear{{Maxted}, {Triaud}  \& {Martin}}{{Maxted} et~al.}{2023}]{maxted2023}
{Maxted} P. F.~L.,  {Triaud} A. H.~M.~J.,   {Martin} D.~V.,  2023, \mn@doi [Universe] {10.3390/universe9120498}, \href {https://ui.adsabs.harvard.edu/abs/2023Univ....9..498M} {9, 498}

\bibitem[\protect\citeauthoryear{{McCully}, {Volgenau}, {Harbeck}, {Lister}, {Saunders}, {Turner}, {Siiverd}  \& {Bowman}}{{McCully} et~al.}{2018}]{McCully:2018}
{McCully} C.,  {Volgenau} N.~H.,  {Harbeck} D.-R.,  {Lister} T.~A.,  {Saunders} E.~S.,  {Turner} M.~L.,  {Siiverd} R.~J.,   {Bowman} M.,  2018, in \procspie. p. 107070K (\mn@eprint {arXiv} {1811.04163}), \mn@doi{10.1117/12.2314340}

\bibitem[\protect\citeauthoryear{{McMillan}}{{McMillan}}{2017}]{mcmillan2017}
{McMillan} P.~J.,  2017, \mn@doi [\mnras] {10.1093/mnras/stw2759}, \href {https://ui.adsabs.harvard.edu/abs/2017MNRAS.465...76M} {465, 76}

\bibitem[\protect\citeauthoryear{{Miller} \& {Stone}}{{Miller} \& {Stone}}{1994}]{kastspectrograph}
{Miller} J.~S.,  {Stone} R.~P.~S.,  1994, Technical Report~66, The Kast Double Spectrograph.
University of California Lick Observatory Technical Reports

\bibitem[\protect\citeauthoryear{{Moya}, {Zuccarino}, {Chaplin}  \& {Davies}}{{Moya} et~al.}{2018}]{2018ApJS..237...21M}
{Moya} A.,  {Zuccarino} F.,  {Chaplin} W.~J.,   {Davies} G.~R.,  2018, \mn@doi [\apjs] {10.3847/1538-4365/aacdae}, \href {https://ui.adsabs.harvard.edu/abs/2018ApJS..237...21M} {237, 21}

\bibitem[\protect\citeauthoryear{{Narita} et~al.,}{{Narita} et~al.}{2020}]{Narita:2020}
{Narita} N.,  et~al., 2020, in Society of Photo-Optical Instrumentation Engineers (SPIE) Conference Series. p. 114475K, \mn@doi{10.1117/12.2559947}

\bibitem[\protect\citeauthoryear{{Nettelmann}, {Kramm}, {Redmer}  \& {Neuh{\"a}user}}{{Nettelmann} et~al.}{2010}]{2010A&A...523A..26N}
{Nettelmann} N.,  {Kramm} U.,  {Redmer} R.,   {Neuh{\"a}user} R.,  2010, \mn@doi [\aap] {10.1051/0004-6361/200911985}, \href {https://ui.adsabs.harvard.edu/abs/2010A&A...523A..26N} {523, A26}

\bibitem[\protect\citeauthoryear{{Oke}}{{Oke}}{1990}]{1990AJ.....99.1621O}
{Oke} J.~B.,  1990, \mn@doi [\aj] {10.1086/115444}, \href {https://ui.adsabs.harvard.edu/abs/1990AJ.....99.1621O} {99, 1621}

\bibitem[\protect\citeauthoryear{{Parviainen} \& {Aigrain}}{{Parviainen} \& {Aigrain}}{2015}]{pyldtk}
{Parviainen} H.,  {Aigrain} S.,  2015, \mn@doi [\mnras] {10.1093/mnras/stv1857}, \href {https://ui.adsabs.harvard.edu/abs/2015MNRAS.453.3821P} {453, 3821}

\bibitem[\protect\citeauthoryear{{Pecaut} \& {Mamajek}}{{Pecaut} \& {Mamajek}}{2013}]{2013ApJS..208....9P}
{Pecaut} M.~J.,  {Mamajek} E.~E.,  2013, \mn@doi [\apjs] {10.1088/0067-0049/208/1/9}, \href {http://adsabs.harvard.edu/abs/2013ApJS..208....9P} {208, 9}

\bibitem[\protect\citeauthoryear{{Pickles}}{{Pickles}}{1998}]{1998PASP..110..863P}
{Pickles} A.~J.,  1998, \mn@doi [\pasp] {10.1086/316197}, \href {https://ui.adsabs.harvard.edu/abs/1998PASP..110..863P} {110, 863}

\bibitem[\protect\citeauthoryear{{Piette} \& {Madhusudhan}}{{Piette} \& {Madhusudhan}}{2020}]{Piette2020}
{Piette} A. A.~A.,  {Madhusudhan} N.,  2020, \mn@doi [\apj] {10.3847/1538-4357/abbfb1}, \href {https://ui.adsabs.harvard.edu/abs/2020ApJ...904..154P} {904, 154}

\bibitem[\protect\citeauthoryear{{Piette}, {Gao}, {Brugman}, {Shahar}, {Lichtenberg}, {Miozzi}  \& {Driscoll}}{{Piette} et~al.}{2023}]{Piette2023}
{Piette} A. A.~A.,  {Gao} P.,  {Brugman} K.,  {Shahar} A.,  {Lichtenberg} T.,  {Miozzi} F.,   {Driscoll} P.,  2023, \mn@doi [\apj] {10.3847/1538-4357/acdef2}, \href {https://ui.adsabs.harvard.edu/abs/2023ApJ...954...29P} {954, 29}

\bibitem[\protect\citeauthoryear{{Piro} \& {Vissapragada}}{{Piro} \& {Vissapragada}}{2020}]{2020AJ....159..131P}
{Piro} A.~L.,  {Vissapragada} S.,  2020, \mn@doi [\aj] {10.3847/1538-3881/ab7192}, \href {https://ui.adsabs.harvard.edu/abs/2020AJ....159..131P} {159, 131}

\bibitem[\protect\citeauthoryear{{Pollacco} et~al.,}{{Pollacco} et~al.}{2006}]{2006PASP..118.1407P}
{Pollacco} D.~L.,  et~al., 2006, \mn@doi [\pasp] {10.1086/508556}, \href {http://adsabs.harvard.edu/abs/2006PASP..118.1407P} {118, 1407}

\bibitem[\protect\citeauthoryear{{Rayner}, {Toomey}, {Onaka}, {Denault}, {Stahlberger}, {Vacca}, {Cushing}  \& {Wang}}{{Rayner} et~al.}{2003}]{Rayner2003}
{Rayner} J.~T.,  {Toomey} D.~W.,  {Onaka} P.~M.,  {Denault} A.~J.,  {Stahlberger} W.~E.,  {Vacca} W.~D.,  {Cushing} M.~C.,   {Wang} S.,  2003, \mn@doi [\pasp] {10.1086/367745}, \href {https://ui.adsabs.harvard.edu/abs/2003PASP..115..362R} {115, 362}

\bibitem[\protect\citeauthoryear{{Rayner}, {Cushing}  \& {Vacca}}{{Rayner} et~al.}{2009}]{Rayner2009}
{Rayner} J.~T.,  {Cushing} M.~C.,   {Vacca} W.~D.,  2009, \mn@doi [\apjs] {10.1088/0067-0049/185/2/289}, \href {https://ui.adsabs.harvard.edu/abs/2009ApJS..185..289R} {185, 289}

\bibitem[\protect\citeauthoryear{{Rebassa-Mansergas} et~al.,}{{Rebassa-Mansergas} et~al.}{2023}]{2023MNRAS.526.4787R}
{Rebassa-Mansergas} A.,  et~al., 2023, \mn@doi [\mnras] {10.1093/mnras/stad3050}, \href {https://ui.adsabs.harvard.edu/abs/2023MNRAS.526.4787R} {526, 4787}

\bibitem[\protect\citeauthoryear{{Reefe} et~al.,}{{Reefe} et~al.}{2022}]{2022AJ....163..269R}
{Reefe} M.~A.,  et~al., 2022, \mn@doi [\aj] {10.3847/1538-3881/ac658b}, \href {https://ui.adsabs.harvard.edu/abs/2022AJ....163..269R} {163, 269}

\bibitem[\protect\citeauthoryear{{Reid} et~al.,}{{Reid} et~al.}{1991}]{reid1991}
{Reid} I.~N.,  et~al., 1991, \mn@doi [\pasp] {10.1086/132866}, \href {https://ui.adsabs.harvard.edu/abs/1991PASP..103..661R} {103, 661}

\bibitem[\protect\citeauthoryear{{Reyl{\'e}}, {Jardine}, {Fouqu{\'e}}, {Caballero}, {Smart}  \& {Sozzetti}}{{Reyl{\'e}} et~al.}{2021}]{reyle2021}
{Reyl{\'e}} C.,  {Jardine} K.,  {Fouqu{\'e}} P.,  {Caballero} J.~A.,  {Smart} R.~L.,   {Sozzetti} A.,  2021, \mn@doi [\aap] {10.1051/0004-6361/202140985}, \href {https://ui.adsabs.harvard.edu/abs/2021A&A...650A.201R} {650, A201}

\bibitem[\protect\citeauthoryear{{Ricker} et~al.,}{{Ricker} et~al.}{2015}]{tess}
{Ricker} G.~R.,  et~al., 2015, \mn@doi [Journal of Astronomical Telescopes, Instruments, and Systems] {10.1117/1.JATIS.1.1.014003}, \href {https://ui.adsabs.harvard.edu/abs/2015JATIS...1a4003R} {1, 014003}

\bibitem[\protect\citeauthoryear{{Rivera}, {Laughlin}, {Butler}, {Vogt}, {Haghighipour}  \& {Meschiari}}{{Rivera} et~al.}{2010}]{rivera2010}
{Rivera} E.~J.,  {Laughlin} G.,  {Butler} R.~P.,  {Vogt} S.~S.,  {Haghighipour} N.,   {Meschiari} S.,  2010, \mn@doi [\apj] {10.1088/0004-637X/719/1/890}, \href {https://ui.adsabs.harvard.edu/abs/2010ApJ...719..890R} {719, 890}

\bibitem[\protect\citeauthoryear{{Rojas-Ayala}, {Covey}, {Muirhead}  \& {Lloyd}}{{Rojas-Ayala} et~al.}{2012}]{Rojas-Ayala2012}
{Rojas-Ayala} B.,  {Covey} K.~R.,  {Muirhead} P.~S.,   {Lloyd} J.~P.,  2012, \mn@doi [\apj] {10.1088/0004-637X/748/2/93}, \href {https://ui.adsabs.harvard.edu/abs/2012ApJ...748...93R} {748, 93}

\bibitem[\protect\citeauthoryear{{Rowe} et~al.,}{{Rowe} et~al.}{2014}]{rowe2014}
{Rowe} J.~F.,  et~al., 2014, \mn@doi [\apj] {10.1088/0004-637X/784/1/45}, \href {https://ui.adsabs.harvard.edu/abs/2014ApJ...784...45R} {784, 45}

\bibitem[\protect\citeauthoryear{{Santerne} et~al.,}{{Santerne} et~al.}{2019}]{2019arXiv191107355S}
{Santerne} A.,  et~al., 2019, \mn@doi [arXiv e-prints] {10.48550/arXiv.1911.07355}, \href {https://ui.adsabs.harvard.edu/abs/2019arXiv191107355S} {p. arXiv:1911.07355}

\bibitem[\protect\citeauthoryear{{Sch{\"o}nrich}, {Binney}  \& {Dehnen}}{{Sch{\"o}nrich} et~al.}{2010}]{schonrich2010}
{Sch{\"o}nrich} R.,  {Binney} J.,   {Dehnen} W.,  2010, \mn@doi [\mnras] {10.1111/j.1365-2966.2010.16253.x}, \href {http://adsabs.harvard.edu/abs/2010MNRAS.403.1829S} {403, 1829}

\bibitem[\protect\citeauthoryear{{Scott} et~al.,}{{Scott} et~al.}{2021}]{2021FrASS...8..138S}
{Scott} N.~J.,  et~al., 2021, \mn@doi [Frontiers in Astronomy and Space Sciences] {10.3389/fspas.2021.716560}, \href {https://ui.adsabs.harvard.edu/abs/2021FrASS...8..138S} {8, 138}

\bibitem[\protect\citeauthoryear{{Seager} \& {Deming}}{{Seager} \& {Deming}}{2010}]{2010ARA&A..48..631S}
{Seager} S.,  {Deming} D.,  2010, \mn@doi [\araa] {10.1146/annurev-astro-081309-130837}, \href {https://ui.adsabs.harvard.edu/abs/2010ARA&A..48..631S} {48, 631}

\bibitem[\protect\citeauthoryear{{Sebastian} et~al.,}{{Sebastian} et~al.}{2021}]{speculoos}
{Sebastian} D.,  et~al., 2021, \mn@doi [\aap] {10.1051/0004-6361/202038827}, \href {https://ui.adsabs.harvard.edu/abs/2021A&A...645A.100S} {645, A100}

\bibitem[\protect\citeauthoryear{{Seifahrt}, {Bean}, {St{\"u}rmer}, {Gers}, {Grobler}, {Reed}  \& {Jones}}{{Seifahrt} et~al.}{2016}]{maroonx2016}
{Seifahrt} A.,  {Bean} J.~L.,  {St{\"u}rmer} J.,  {Gers} L.,  {Grobler} D.~S.,  {Reed} T.,   {Jones} D.~J.,  2016, in {Evans} C.~J.,  {Simard} L.,   {Takami} H.,  eds,  Society of Photo-Optical Instrumentation Engineers (SPIE) Conference Series Vol. 9908, Ground-based and Airborne Instrumentation for Astronomy VI. p. 990818 (\mn@eprint {arXiv} {1606.07140}), \mn@doi{10.1117/12.2232069}

\bibitem[\protect\citeauthoryear{{Seifahrt} et~al.,}{{Seifahrt} et~al.}{2020}]{maroonx2020}
{Seifahrt} A.,  et~al., 2020, in {Evans} C.~J.,  {Bryant} J.~J.,   {Motohara} K.,  eds,  Society of Photo-Optical Instrumentation Engineers (SPIE) Conference Series Vol. 11447, Ground-based and Airborne Instrumentation for Astronomy VIII. p. 114471F (\mn@eprint {arXiv} {2106.02157}), \mn@doi{10.1117/12.2561564}

\bibitem[\protect\citeauthoryear{{Seifahrt} et~al.,}{{Seifahrt} et~al.}{2022}]{maroonx2022}
{Seifahrt} A.,  et~al., 2022, in {Evans} C.~J.,  {Bryant} J.~J.,   {Motohara} K.,  eds,  Society of Photo-Optical Instrumentation Engineers (SPIE) Conference Series Vol. 12184, Ground-based and Airborne Instrumentation for Astronomy IX. p. 121841G (\mn@eprint {arXiv} {2210.06563}), \mn@doi{10.1117/12.2629428}

\bibitem[\protect\citeauthoryear{{Smith} et~al.,}{{Smith} et~al.}{2012}]{smith2012}
{Smith} J.~C.,  et~al., 2012, \mn@doi [\pasp] {10.1086/667697}, \href {http://adsabs.harvard.edu/abs/2012PASP..124.1000S} {124, 1000}

\bibitem[\protect\citeauthoryear{{Spada}, {Demarque}, {Kim}  \& {Sills}}{{Spada} et~al.}{2013}]{2013ApJ...776...87S}
{Spada} F.,  {Demarque} P.,  {Kim} Y.~C.,   {Sills} A.,  2013, \mn@doi [\apj] {10.1088/0004-637X/776/2/87}, \href {https://ui.adsabs.harvard.edu/abs/2013ApJ...776...87S} {776, 87}

\bibitem[\protect\citeauthoryear{{Speagle}}{{Speagle}}{2020}]{dynesty}
{Speagle} J.~S.,  2020, \mn@doi [\mnras] {10.1093/mnras/staa278}, \href {https://ui.adsabs.harvard.edu/abs/2020MNRAS.493.3132S} {493, 3132}

\bibitem[\protect\citeauthoryear{{Spitoni}, {Matteucci}, {Gratton}, {Ratcliffe}, {Minchev}  \& {Cescutti}}{{Spitoni} et~al.}{2024}]{spitoni2024}
{Spitoni} E.,  {Matteucci} F.,  {Gratton} R.,  {Ratcliffe} B.,  {Minchev} I.,   {Cescutti} G.,  2024, \mn@doi [\aap] {10.1051/0004-6361/202450754}, \href {https://ui.adsabs.harvard.edu/abs/2024A&A...690A.208S} {690, A208}

\bibitem[\protect\citeauthoryear{{Stassun} \& {Torres}}{{Stassun} \& {Torres}}{2016}]{Stassun:2016}
{Stassun} K.~G.,  {Torres} G.,  2016, \mn@doi [\aj] {10.3847/0004-6256/152/6/180}, \href {https://ui.adsabs.harvard.edu/abs/2016AJ....152..180S} {152, 180}

\bibitem[\protect\citeauthoryear{{Stassun} \& {Torres}}{{Stassun} \& {Torres}}{2021}]{StassunTorres:2021}
{Stassun} K.~G.,  {Torres} G.,  2021, \mn@doi [\apjl] {10.3847/2041-8213/abdaad}, \href {https://ui.adsabs.harvard.edu/abs/2021ApJ...907L..33S} {907, L33}

\bibitem[\protect\citeauthoryear{{Stassun}, {Collins}  \& {Gaudi}}{{Stassun} et~al.}{2017}]{Stassun:2017}
{Stassun} K.~G.,  {Collins} K.~A.,   {Gaudi} B.~S.,  2017, \mn@doi [\aj] {10.3847/1538-3881/aa5df3}, \href {https://ui.adsabs.harvard.edu/abs/2017AJ....153..136S} {153, 136}

\bibitem[\protect\citeauthoryear{{Stassun}, {Corsaro}, {Pepper}  \& {Gaudi}}{{Stassun} et~al.}{2018}]{Stassun:2018}
{Stassun} K.~G.,  {Corsaro} E.,  {Pepper} J.~A.,   {Gaudi} B.~S.,  2018, \mn@doi [\aj] {10.3847/1538-3881/aa998a}, \href {https://ui.adsabs.harvard.edu/abs/2018AJ....155...22S} {155, 22}

\bibitem[\protect\citeauthoryear{{Stassun} et~al.,}{{Stassun} et~al.}{2019}]{TICv8}
{Stassun} K.~G.,  et~al., 2019, \mn@doi [\aj] {10.3847/1538-3881/ab3467}, \href {https://ui.adsabs.harvard.edu/abs/2019AJ....158..138S} {158, 138}

\bibitem[\protect\citeauthoryear{{Stock}, {Kitzmann}, {Patzer}  \& {Sedlmayr}}{{Stock} et~al.}{2018}]{Stock2018}
{Stock} J.~W.,  {Kitzmann} D.,  {Patzer} A. B.~C.,   {Sedlmayr} E.,  2018, \mn@doi [\mnras] {10.1093/mnras/sty1531}, \href {https://ui.adsabs.harvard.edu/abs/2018MNRAS.479..865S} {479, 865}

\bibitem[\protect\citeauthoryear{{Stumpe} et~al.,}{{Stumpe} et~al.}{2012}]{stumpe2012}
{Stumpe} M.~C.,  et~al., 2012, \mn@doi [\pasp] {10.1086/667698}, \href {https://ui.adsabs.harvard.edu/abs/2012PASP..124..985S} {124, 985}

\bibitem[\protect\citeauthoryear{{Stumpe}, {Smith}, {Catanzarite}, {Van Cleve}, {Jenkins}, {Twicken}  \& {Girouard}}{{Stumpe} et~al.}{2014}]{stumpe2014}
{Stumpe} M.~C.,  {Smith} J.~C.,  {Catanzarite} J.~H.,  {Van Cleve} J.~E.,  {Jenkins} J.~M.,  {Twicken} J.~D.,   {Girouard} F.~R.,  2014, \mn@doi [\pasp] {10.1086/674989}, \href {http://adsabs.harvard.edu/abs/2014PASP..126..100S} {126, 100}

\bibitem[\protect\citeauthoryear{{Swayne} et~al.,}{{Swayne} et~al.}{2024}]{swayne2024}
{Swayne} M.~I.,  et~al., 2024, \mn@doi [\mnras] {10.1093/mnras/stad3866}, \href {https://ui.adsabs.harvard.edu/abs/2024MNRAS.528.5703S} {528, 5703}

\bibitem[\protect\citeauthoryear{{The JWST Transiting Exoplanet Community Early Release Science Team} et~al.,}{{The JWST Transiting Exoplanet Community Early Release Science Team} et~al.}{2022}]{jwstERS}
{The JWST Transiting Exoplanet Community Early Release Science Team} et~al., 2022, arXiv e-prints, \href {https://ui.adsabs.harvard.edu/abs/2022arXiv220811692T} {p. arXiv:2208.11692}

\bibitem[\protect\citeauthoryear{{Timmermans} et~al.,}{{Timmermans} et~al.}{2024}]{Timmermans2024}
{Timmermans} M.,  et~al., 2024, \mn@doi [\aap] {10.1051/0004-6361/202347981}, \href {https://ui.adsabs.harvard.edu/abs/2024A&A...687A..48T} {687, A48}

\bibitem[\protect\citeauthoryear{{Torres}, {Andersen}  \& {Gim{\'e}nez}}{{Torres} et~al.}{2010}]{2010A&ARv..18...67T}
{Torres} G.,  {Andersen} J.,   {Gim{\'e}nez} A.,  2010, \mn@doi [\aapr] {10.1007/s00159-009-0025-1}, \href {https://ui.adsabs.harvard.edu/abs/2010A&ARv..18...67T} {18, 67}

\bibitem[\protect\citeauthoryear{{Triaud}}{{Triaud}}{2021}]{triaud2021_smallstars}
{Triaud} A. H.~M.~J.,  2021, in {Madhusudhan} N.,  ed., , ExoFrontiers; Big Questions in Exoplanetary Science.
pp~6--1, \mn@doi{10.1088/2514-3433/abfa8fch6}

\bibitem[\protect\citeauthoryear{{Triaud} et~al.,}{{Triaud} et~al.}{2011}]{Triaud2011}
{Triaud} A.~H.~M.~J.,  et~al., 2011, \mn@doi [\aap] {10.1051/0004-6361/201016367}, \href {https://ui.adsabs.harvard.edu/abs/2011A&A...531A..24T} {531, A24}

\bibitem[\protect\citeauthoryear{{Triaud} et~al.,}{{Triaud} et~al.}{2013}]{eblm2013}
{Triaud} A.~H.~M.~J.,  et~al., 2013, \mn@doi [\aap] {10.1051/0004-6361/201219643}, \href {https://ui.adsabs.harvard.edu/abs/2013A&A...549A..18T} {549, A18}

\bibitem[\protect\citeauthoryear{{Triaud} et~al.,}{{Triaud} et~al.}{2023}]{Triaud2023}
{Triaud} A. H.~M.~J.,  et~al., 2023, \mn@doi [\mnras] {10.1093/mnrasl/slad097}, \href {https://ui.adsabs.harvard.edu/abs/2023MNRAS.525L..98T} {525, L98}

\bibitem[\protect\citeauthoryear{{Twicken} et~al.,}{{Twicken} et~al.}{2018}]{twicken2018}
{Twicken} J.~D.,  et~al., 2018, \mn@doi [\pasp] {10.1088/1538-3873/aab694}, \href {http://adsabs.harvard.edu/abs/2018PASP..130f4502T} {130, 064502}

\bibitem[\protect\citeauthoryear{{Vanderburg}, {Plavchan}, {Johnson}, {Ciardi}, {Swift}  \& {Kane}}{{Vanderburg} et~al.}{2016}]{2016MNRAS.459.3565V}
{Vanderburg} A.,  {Plavchan} P.,  {Johnson} J.~A.,  {Ciardi} D.~R.,  {Swift} J.,   {Kane} S.~R.,  2016, \mn@doi [\mnras] {10.1093/mnras/stw863}, \href {https://ui.adsabs.harvard.edu/abs/2016MNRAS.459.3565V} {459, 3565}

\bibitem[\protect\citeauthoryear{{Weeks} et~al.,}{{Weeks} et~al.}{2024}]{weeks2024}
{Weeks} A.,  et~al., 2024, \mn@doi [arXiv e-prints] {10.48550/arXiv.2411.17358}, \href {https://ui.adsabs.harvard.edu/abs/2024arXiv241117358W} {p. arXiv:2411.17358}

\bibitem[\protect\citeauthoryear{{West}, {Hawley}, {Bochanski}, {Covey}, {Reid}, {Dhital}, {Hilton}  \& {Masuda}}{{West} et~al.}{2008}]{2008AJ....135..785W}
{West} A.~A.,  {Hawley} S.~L.,  {Bochanski} J.~J.,  {Covey} K.~R.,  {Reid} I.~N.,  {Dhital} S.,  {Hilton} E.~J.,   {Masuda} M.,  2008, \mn@doi [\aj] {10.1088/0004-6256/135/3/785}, \href {http://adsabs.harvard.edu/abs/2008AJ....135..785W} {135, 785}

\bibitem[\protect\citeauthoryear{{Winters} et~al.,}{{Winters} et~al.}{2022}]{winters2022}
{Winters} J.~G.,  et~al., 2022, \mn@doi [\aj] {10.3847/1538-3881/ac50a9}, \href {https://ui.adsabs.harvard.edu/abs/2022AJ....163..168W} {163, 168}

\bibitem[\protect\citeauthoryear{{Zechmeister} et~al.,}{{Zechmeister} et~al.}{2018}]{serval}
{Zechmeister} M.,  et~al., 2018, \mn@doi [\aap] {10.1051/0004-6361/201731483}, \href {https://ui.adsabs.harvard.edu/abs/2018A&A...609A..12Z} {609, A12}

\bibitem[\protect\citeauthoryear{{Zeng}, {Sasselov}  \& {Jacobsen}}{{Zeng} et~al.}{2016}]{zeng2016}
{Zeng} L.,  {Sasselov} D.~D.,   {Jacobsen} S.~B.,  2016, \mn@doi [\apj] {10.3847/0004-637X/819/2/127}, \href {https://ui.adsabs.harvard.edu/abs/2016ApJ...819..127Z} {819, 127}

\bibitem[\protect\citeauthoryear{{Zink} et~al.,}{{Zink} et~al.}{2023}]{zink2023}
{Zink} J.~K.,  et~al., 2023, \mn@doi [\aj] {10.3847/1538-3881/acd24c}, \href {https://ui.adsabs.harvard.edu/abs/2023AJ....165..262Z} {165, 262}

\makeatother
\end{thebibliography}




\appendix
\textcolor{black}
{
\section{RV DATA}
In Table~\ref{tab:rvs} we provide the 10 RV data points obtained with the MAROON-X spectrograph in Hawai'i.}

\begin{table*}
\caption{\textcolor{black}{MAROON-X observations of TOI-6478, where the red and blue columns denote the red and blue arms of the spectrograph respectively. 
\newline $^*$This data point is omitted from our analysis as it was flagged as `usable' rather than `pass', and we wanted to remain conservative when obtaining an upper mass limit.}}
    \begin{tabular}{c|c|c|c|c|c|c}
    \hline
    BJD (day) & \multicolumn{3}{c|}{Red} & \multicolumn{3}{c}{Blue} \\
    \hline
    & RV (m\,s$^{-1}$) & $\sigma$ (m\,s$^{-1}$) & SNR  & RV (m\,s$^{-1}$) & $\sigma$ (m\,s$^{-1}$) & SNR \\
    \hline
  2460398.89216 & 2.12                               & 2.15                                                  & 50.0 & 0.96                               & 3.99                                                  & 20.0 \\
    2460399.87039 & 2.18                               & 2.23                                                  & 48.0 & 6.31                               & 4.13                                                  & 19.0 \\
    2460401.90117$^*$  & -1.35                              & 2.00                                                  & 54.0 & -6.15                              & 4.66                                                  & 19.0 \\
    2460401.91435 & 3.34                               & 1.94                                                  & 54.0 & -1.72                              & 3.59                                                  & 22.0 \\
    2460403.81640   & 0.26                               & 2.34                                                  & 46.0 & -5.44                              & 4.34                                                  & 18.0 \\
    2460410.81273 & -3.06                              & 1.82                                                  & 58.0 & -2.92                              & 3.28                                                  & 24.0 \\
    2460417.88471 & -1.79                              & 2.53                                                  & 41.0 & -8.56                              & 4.73                                                  & 17.0 \\
    2460418.77670 & -1.33                              & 1.78                                                  & 56.0 & -1.61                              & 3.33                                                  & 23.0 \\
    2460421.83060  & 3.39                               & 2.04                                                  & 49.0 & 9.18                               & 4.22                                                  & 19.0 \\
    2460423.74191 & -2.68                              & 1.60                                                  & 61.0 & -2.60                              & 3.00                                                  & 25.0
    \label{tab:rvs}
    \end{tabular}
\end{table*}



\bsp	
\label{lastpage}
\end{document}